\documentclass[longauth]{aa_2016}
\usepackage[varg]{txfonts}
\usepackage{graphicx}
\usepackage{graphics}
\usepackage{amssymb}
\usepackage{amsmath}
\usepackage[T1]{fontenc}
\usepackage[utf8]{inputenc}
\usepackage{url,hyperref}
\usepackage{footmisc}
\usepackage[switch]{lineno}
\usepackage{multicol}
\def\beq{\begin{eqnarray}}
\def\eeq{\end{eqnarray}}
\def\bdm{\begin{displaymath}}
\def\edm{\end{displaymath}}
\def\be {\begin{equation}}
\def\ee {\end{equation}}
\newcommand{\stat}{_{\rm stat}}
\newcommand{\sys}{_{\rm sys}}
\newcommand{\hess}{H.E.S.S.}
\newcommand{\fermi}{{\it Fermi}-LAT}
\newcommand{\source}{PKS~1510$-$089}
\newcommand{\g}{\ensuremath{\gamma}}
\newcommand{\E}[1]{\times 10^{#1}}
\newcommand{\as}{^{\ast}}
\newcommand{\p}{^{\prime}}
\newcommand{\est}[3]{\left( \frac{#1}{#2} \right)^{#3}}
\makeatletter
\renewcommand*{\@fnsymbol}[1]{\ifcase#1\or*\or$\dagger$\or$\ddagger$\or**\or$\dagger\dagger$\or$\ddagger\ddagger$\fi}
\renewcommand*\aa@pageof{, page \thepage{} of \pageref*{LastPage}} 
\makeatother
\begin{document}
\title{Observation of a sudden cessation of a very-high-energy \g-ray flare in PKS~1510$-$089 with H.E.S.S. and MAGIC in May 2016}
\titlerunning{The VHE \g-ray outburst of PKS~1510$-$089 in May 2016}
\authorrunning{H.E.S.S. \& MAGIC collaborations et al.}
\author{
\small
H.E.S.S. Collaboration
\and H.~Abdalla \inst{\ref{NWU}}
\and R.~Adam \inst{\ref{LLR}}
\and F.~Aharonian \inst{\ref{MPIK},\ref{DIAS},\ref{RAU}}
\and F.~Ait~Benkhali \inst{\ref{MPIK}}
\and E.O.~Ang\"uner \inst{\ref{CPPM}}
\and C.~Arcaro \inst{\ref{NWU}}
\and C.~Armand \inst{\ref{LAPP}}
\and T.~Armstrong \inst{\ref{Oxford}}
\and H.~Ashkar \inst{\ref{IRFU}}
\and M.~Backes \inst{\ref{UNAM},\ref{NWU}}
\and V.~Baghmanyan \inst{\ref{IFJPAN}}
\and V.~Barbosa~Martins \inst{\ref{DESY}}
\and A.~Barnacka \inst{\ref{UJK}}
\and M.~Barnard \inst{\ref{NWU}}
\and Y.~Becherini \inst{\ref{Linnaeus}}
\and D.~Berge \inst{\ref{DESY}}
\and K.~Bernl\"ohr \inst{\ref{MPIK}}
\and B.~Bi \inst{\ref{IAAT}}
\and M.~B\"ottcher \inst{\ref{NWU}}
\and C.~Boisson \inst{\ref{LUTH}}
\and J.~Bolmont \inst{\ref{LPNHE}}
\and S.~Bonnefoy \inst{\ref{DESY}}
\and M.~de~Bony~de~Lavergne \inst{\ref{LAPP}}
\and J.~Bregeon \inst{\ref{LUPM}}
\and M.~Breuhaus \inst{\ref{MPIK}}
\and F.~Brun \inst{\ref{IRFU}}
\and P.~Brun \inst{\ref{IRFU}}
\and M.~Bryan \inst{\ref{GRAPPA}}
\and M.~B\"{u}chele \inst{\ref{ECAP}}
\and T.~Bulik \inst{\ref{UWarsaw}}
\and T.~Bylund \inst{\ref{Linnaeus}}
\and S.~Caroff \inst{\ref{LPNHE}}
\and A.~Carosi \inst{\ref{LAPP}}
\and S.~Casanova \inst{\ref{IFJPAN},\ref{MPIK}}
\and T.~Chand \inst{\ref{NWU}}
\and S.~Chandra \inst{\ref{NWU}}
\and A.~Chen \inst{\ref{WITS}}
\and G.~Cotter \inst{\ref{Oxford}}
\and M.~Cury{\l}o \inst{\ref{UWarsaw}}
\and J.~Damascene~Mbarubucyeye \inst{\ref{DESY}}
\and I.D.~Davids \inst{\ref{UNAM}}
\and J.~Davies \inst{\ref{Oxford}}
\and C.~Deil \inst{\ref{MPIK}}
\and J.~Devin \inst{\ref{CENBG}}
\and P.~deWilt \inst{\ref{Adelaide}}
\and L.~Dirson \inst{\ref{HH}}
\and A.~Djannati-Ata\"i \inst{\ref{APC}}
\and A.~Dmytriiev \inst{\ref{LUTH}}
\and A.~Donath \inst{\ref{MPIK}}
\and V.~Doroshenko \inst{\ref{IAAT}}
\and J.~Dyks \inst{\ref{NCAC}}
\and K.~Egberts \inst{\ref{UP}}
\and F.~Eichhorn \inst{\ref{ECAP}}
\and S.~Einecke \inst{\ref{Adelaide}}
\and G.~Emery \inst{\ref{LPNHE}}
\and J.-P.~Ernenwein \inst{\ref{CPPM}}
\and K.~Feijen \inst{\ref{Adelaide}}
\and S.~Fegan \inst{\ref{LLR}}
\and A.~Fiasson \inst{\ref{LAPP}}
\and G.~Fichet~de~Clairfontaine \inst{\ref{LUTH}}
\and M.~Filipovic \inst{\ref{Adelaide}}
\and G.~Fontaine \inst{\ref{LLR}}
\and S.~Funk \inst{\ref{ECAP}}
\and M.~F\"u{\ss}ling \inst{\ref{DESY}}
\and S.~Gabici \inst{\ref{APC}}
\and Y.A.~Gallant \inst{\ref{LUPM}}
\and G.~Giavitto \inst{\ref{DESY}}
\and L.~Giunti \inst{\ref{APC}, \ref{IRFU}}
\and D.~Glawion \inst{\ref{LSW}}
\and J.F.~Glicenstein \inst{\ref{IRFU}}
\and D.~Gottschall \inst{\ref{IAAT}}
\and M.-H.~Grondin \inst{\ref{CENBG}}
\and J.~Hahn \inst{\ref{MPIK}}
\and M.~Haupt \inst{\ref{DESY}}
\and G.~Hermann \inst{\ref{MPIK}}
\and J.A.~Hinton \inst{\ref{MPIK}}
\and W.~Hofmann \inst{\ref{MPIK}}
\and C.~Hoischen \inst{\ref{UP}}
\and T.~L.~Holch \inst{\ref{HUB}}
\and M.~Holler \inst{\ref{LFUI}}
\and M.~H\"{o}rbe \inst{\ref{Oxford}}
\and D.~Horns \inst{\ref{HH}}
\and D.~Huber \inst{\ref{LFUI}}
\and M.~Jamrozy \inst{\ref{UJK}}
\and D.~Jankowsky \inst{\ref{ECAP}}
\and F.~Jankowsky \inst{\ref{LSW}}
\and A.~Jardin-Blicq \inst{\ref{MPIK}}
\and V.~Joshi \inst{\ref{ECAP}}
\and I.~Jung-Richardt \inst{\ref{ECAP}}
\and M.A.~Kastendieck \inst{\ref{HH}}
\and K.~Katarzy{\'n}ski \inst{\ref{NCUT}}
\and U.~Katz \inst{\ref{ECAP}}
\and D.~Khangulyan \inst{\ref{Rikkyo}}
\and B.~Kh\'elifi \inst{\ref{APC}}
\and S.~Klepser \inst{\ref{DESY}}
\and W.~Klu\'{z}niak \inst{\ref{NCAC}}
\and Nu.~Komin \inst{\ref{WITS}}
\and R.~Konno \inst{\ref{DESY}}
\and K.~Kosack \inst{\ref{IRFU}}
\and D.~Kostunin \inst{\ref{DESY}} 
\and M.~Kreter \inst{\ref{NWU}}
\and G.~Lamanna \inst{\ref{LAPP}}
\and A.~Lemi\`ere \inst{\ref{APC}}
\and M.~Lemoine-Goumard \inst{\ref{CENBG}}
\and J.-P.~Lenain \inst{\ref{LPNHE}}
\and C.~Levy \inst{\ref{LPNHE}}
\and T.~Lohse \inst{\ref{HUB}}
\and I.~Lypova \inst{\ref{DESY}}
\and J.~Mackey \inst{\ref{DIAS}}
\and J.~Majumdar \inst{\ref{DESY}}
\and D.~Malyshev \inst{\ref{IAAT}}
\and D.~Malyshev \inst{\ref{ECAP}}
\and V.~Marandon \inst{\ref{MPIK}}
\and P.~Marchegiani \inst{\ref{WITS}}
\and A.~Marcowith \inst{\ref{LUPM}}
\and A.~Mares \inst{\ref{CENBG}}
\and G.~Mart\'i-Devesa \inst{\ref{LFUI}}
\and R.~Marx \inst{\ref{LSW}, \ref{MPIK}}
\and G.~Maurin \inst{\ref{LAPP}}
\and P.J.~Meintjes \inst{\ref{UFS}}
\and M.~Meyer \inst{\ref{ECAP}}
\and A.M.W.~Mitchell \inst{\ref{MPIK}, \ref{Zurich}}
\and R.~Moderski \inst{\ref{NCAC}}
\and M.~Mohamed \inst{\ref{LSW}}
\and L.~Mohrmann \inst{\ref{ECAP}}
\and A.~Montanari \inst{\ref{IRFU}}
\and C.~Moore \inst{\ref{Leicester}}
\and P.~Morris \inst{\ref{Oxford}}
\and E.~Moulin \inst{\ref{IRFU}}
\and J.~Muller \inst{\ref{LLR}}
\and T.~Murach \inst{\ref{DESY}}
\and K.~Nakashima \inst{\ref{ECAP}}
\and A.~Nayerhoda \inst{\ref{IFJPAN}}
\and M.~de~Naurois \inst{\ref{LLR}}
\and H.~Ndiyavala  \inst{\ref{NWU}}
\and F.~Niederwanger \inst{\ref{LFUI}}
\and J.~Niemiec \inst{\ref{IFJPAN}}
\and L.~Oakes \inst{\ref{HUB}}
\and P.~O'Brien \inst{\ref{Leicester}}
\and H.~Odaka \inst{\ref{Tokyo}}
\and S.~Ohm \inst{\ref{DESY}}
\and L.~Olivera-Nieto \inst{\ref{MPIK}}
\and E.~de~Ona~Wilhelmi \inst{\ref{DESY}}
\and M.~Ostrowski \inst{\ref{UJK}}
\and M.~Panter \inst{\ref{MPIK}}
\and S.~Panny \inst{\ref{LFUI}}
\and R.D.~Parsons \inst{\ref{HUB}}
\and G.~Peron \inst{\ref{MPIK}}
\and B.~Peyaud \inst{\ref{IRFU}}
\and Q.~Piel \inst{\ref{LAPP}}
\and S.~Pita \inst{\ref{APC}}
\and V.~Poireau \inst{\ref{LAPP}}
\and A.~Priyana~Noel \inst{\ref{UJK}}
\and D.A.~Prokhorov \inst{\ref{WITS},\ref{GRAPPA}}
\and H.~Prokoph \inst{\ref{DESY}}
\and G.~P\"uhlhofer \inst{\ref{IAAT}}
\and M.~Punch \inst{\ref{APC},\ref{Linnaeus}}
\and A.~Quirrenbach \inst{\ref{LSW}}
\and S.~Raab \inst{\ref{ECAP}}
\and R.~Rauth \inst{\ref{LFUI}}
\and P.~Reichherzer \inst{\ref{IRFU}}
\and A.~Reimer \inst{\ref{LFUI}}
\and O.~Reimer \inst{\ref{LFUI}}
\and Q.~Remy \inst{\ref{MPIK}}
\and M.~Renaud \inst{\ref{LUPM}}
\and F.~Rieger \inst{\ref{MPIK}}
\and L.~Rinchiuso \inst{\ref{IRFU}}
\and C.~Romoli \inst{\ref{MPIK}}
\and G.~Rowell \inst{\ref{Adelaide}}
\and B.~Rudak \inst{\ref{NCAC}}
\and E.~Ruiz-Velasco \inst{\ref{MPIK}}
\and V.~Sahakian \inst{\ref{YPI}}
\and S.~Sailer \inst{\ref{MPIK}}
\and D.A.~Sanchez \inst{\ref{LAPP}} \protect\footnotemark[1]
\and A.~Santangelo \inst{\ref{IAAT}}
\and M.~Sasaki \inst{\ref{ECAP}}
\and M.~Scalici \inst{\ref{IAAT}}
\and F.~Sch\"ussler \inst{\ref{IRFU}}
\and H.M.~Schutte \inst{\ref{NWU}}
\and U.~Schwanke \inst{\ref{HUB}}
\and S.~Schwemmer \inst{\ref{LSW}}
\and M.~Seglar-Arroyo \inst{\ref{IRFU}}
\and M.~Senniappan \inst{\ref{Linnaeus}}
\and A.S.~Seyffert \inst{\ref{NWU}}
\and N.~Shafi \inst{\ref{WITS}}
\and K.~Shiningayamwe \inst{\ref{UNAM}}
\and R.~Simoni \inst{\ref{GRAPPA}}
\and A.~Sinha \inst{\ref{APC}}
\and H.~Sol \inst{\ref{LUTH}}
\and A.~Specovius \inst{\ref{ECAP}}
\and S.~Spencer \inst{\ref{Oxford}}
\and M.~Spir-Jacob \inst{\ref{APC}}
\and {\L.}~Stawarz \inst{\ref{UJK}}
\and L.~Sun \inst{\ref{GRAPPA}}
\and R.~Steenkamp \inst{\ref{UNAM}}
\and C.~Stegmann \inst{\ref{UP},\ref{DESY}}
\and S.~Steinmassl \inst{\ref{MPIK}}
\and C.~Steppa \inst{\ref{UP}}
\and T.~Takahashi  \inst{\ref{KAVLI}}
\and T.~Tavernier \inst{\ref{IRFU}}
\and A.M.~Taylor \inst{\ref{DESY}}
\and R.~Terrier \inst{\ref{APC}}
\and D.~Tiziani \inst{\ref{ECAP}}
\and M.~Tluczykont \inst{\ref{HH}}
\and L.~Tomankova \inst{\ref{ECAP}}
\and C.~Trichard \inst{\ref{LLR}}
\and M.~Tsirou \inst{\ref{LUPM}}
\and R.~Tuffs \inst{\ref{MPIK}}
\and Y.~Uchiyama \inst{\ref{Rikkyo}}
\and D.J.~van~der~Walt \inst{\ref{NWU}}
\and C.~van~Eldik \inst{\ref{ECAP}}
\and C.~van~Rensburg \inst{\ref{NWU}}
\and B.~van~Soelen \inst{\ref{UFS}}
\and G.~Vasileiadis \inst{\ref{LUPM}}
\and J.~Veh \inst{\ref{ECAP}}
\and C.~Venter \inst{\ref{NWU}}
\and P.~Vincent \inst{\ref{LPNHE}}
\and J.~Vink \inst{\ref{GRAPPA}}
\and H.J.~V\"olk \inst{\ref{MPIK}}
\and T.~Vuillaume \inst{\ref{LAPP}}
\and Z.~Wadiasingh \inst{\ref{NWU}}
\and S.J.~Wagner \inst{\ref{LSW}}
\and J.~Watson \inst{\ref{Oxford}}
\and F.~Werner \inst{\ref{MPIK}}
\and R.~White \inst{\ref{MPIK}}
\and A.~Wierzcholska \inst{\ref{IFJPAN},\ref{LSW}}
\and Yu Wun Wong \inst{\ref{ECAP}}
\and A.~Yusafzai \inst{\ref{ECAP}}
\and M.~Zacharias \inst{\ref{NWU}} \protect\footnotemark[1]
\and R.~Zanin \inst{\ref{MPIK}}
\and D.~Zargaryan \inst{\ref{DIAS}}
\and A.A.~Zdziarski \inst{\ref{NCAC}}
\and A.~Zech \inst{\ref{LUTH}}
\and S.J.~Zhu \inst{\ref{DESY}}
\and J.~Zorn \inst{\ref{MPIK}}
\and S.~Zouari \inst{\ref{APC}}
\and N.~\.Zywucka \inst{\ref{NWU}} \and
\\
%
MAGIC Collaboration
\and V.~A.~Acciari \inst{\ref{M1}}
\and S.~Ansoldi \inst{\ref{M2},\ref{M5}}
\and L.~A.~Antonelli \inst{\ref{M3}}
\and A.~Arbet Engels \inst{\ref{M4}}
\and K.~Asano \inst{\ref{M5}}
\and D.~Baack \inst{\ref{M6}}
\and A.~Babi\'c \inst{\ref{M7}}
\and A.~Baquero \inst{\ref{M8}}
\and U.~Barres de Almeida \inst{\ref{M9}}
\and J.~A.~Barrio \inst{\ref{M8}}
\and J.~Becerra Gonz\'alez \inst{\ref{M1}}
\and W.~Bednarek \inst{\ref{M10}}
\and L.~Bellizzi \inst{\ref{M11}}
\and E.~Bernardini \inst{\ref{DESY},\ref{M16}}
\and A.~Berti \inst{\ref{M13}}
\and J.~Besenrieder \inst{\ref{M14}}
\and W.~Bhattacharyya \inst{\ref{DESY}}
\and C.~Bigongiari \inst{\ref{M3}}
\and A.~Biland \inst{\ref{M4}}
\and O.~Blanch \inst{\ref{M15}}
\and G.~Bonnoli \inst{\ref{M11}}
\and \v{Z}.~Bo\v{s}njak \inst{\ref{M7}}
\and G.~Busetto \inst{\ref{M16}}
\and R.~Carosi \inst{\ref{M17}}
\and G.~Ceribella \inst{\ref{M14}}
\and M.~Cerruti \inst{\ref{M18}}
\and Y.~Chai \inst{\ref{M14}}
\and A.~Chilingarian \inst{\ref{M19}}
\and S.~Cikota \inst{\ref{M7}}
\and S.~M.~Colak \inst{\ref{M15}}
\and U.~Colin \inst{\ref{M14}}
\and E.~Colombo \inst{\ref{M1}}
\and J.~L.~Contreras \inst{\ref{M8}}
\and J.~Cortina \inst{\ref{M20}}
\and S.~Covino \inst{\ref{M3}}
\and G.~D'Amico \inst{\ref{M14}}
\and V.~D'Elia \inst{\ref{M3}}
\and P.~Da Vela \inst{\ref{M17},\ref{M26}}
\and F.~Dazzi \inst{\ref{M3}}
\and A.~De Angelis \inst{\ref{M16}}
\and B.~De Lotto \inst{\ref{M2}}
\and M.~Delfino \inst{\ref{M15},\ref{M27}}
\and J.~Delgado \inst{\ref{M15},\ref{M27}}
\and D.~Depaoli \inst{\ref{M13}}
\and F.~Di Pierro \inst{\ref{M13}}
\and L.~Di Venere \inst{\ref{M13}}
\and E.~Do Souto Espi\~neira \inst{\ref{M15}}
\and D.~Dominis Prester \inst{\ref{M7}}
\and A.~Donini \inst{\ref{M2}}
\and D.~Dorner \inst{\ref{M21}}
\and M.~Doro \inst{\ref{M16}}
\and D.~Elsaesser \inst{\ref{M6}}
\and V.~Fallah Ramazani \inst{\ref{M22}}
\and A.~Fattorini \inst{\ref{M6}}
\and G.~Ferrara \inst{\ref{M3}}
\and L.~Foffano \inst{\ref{M16}}
\and M.~V.~Fonseca \inst{\ref{M8}}
\and L.~Font \inst{\ref{M23}}
\and C.~Fruck \inst{\ref{M14}}
\and S.~Fukami \inst{\ref{M5}}
\and R.~J.~Garc\'ia L\'opez \inst{\ref{M1}}
\and M.~Garczarczyk \inst{\ref{DESY}}
\and S.~Gasparyan \inst{\ref{M19}}
\and M.~Gaug \inst{\ref{M23}}
\and N.~Giglietto \inst{\ref{M13}}
\and F.~Giordano \inst{\ref{M13}}
\and P.~Gliwny \inst{\ref{M10}}
\and N.~Godinovi\'c \inst{\ref{M7}}
\and D.~Green \inst{\ref{M14}}
\and D.~Hadasch \inst{\ref{M5}}
\and A.~Hahn \inst{\ref{M14}}
\and L.~Heckmann \inst{\ref{M14}}
\and J.~Herrera \inst{\ref{M1}}
\and J.~Hoang \inst{\ref{M8}}
\and D.~Hrupec \inst{\ref{M7}}
\and M.~H\"utten \inst{\ref{M14}}
\and T.~Inada \inst{\ref{M5}}
\and S.~Inoue \inst{\ref{M5}}
\and K.~Ishio \inst{\ref{M14}}
\and Y.~Iwamura \inst{\ref{M5}}
\and L.~Jouvin \inst{\ref{M15}}
\and Y.~Kajiwara \inst{\ref{M5}}
\and M.~Karjalainen \inst{\ref{M1}}
\and D.~Kerszberg \inst{\ref{M15}}
\and Y.~Kobayashi \inst{\ref{M5}}
\and H.~Kubo \inst{\ref{M5}}
\and J.~Kushida \inst{\ref{M5}}
\and A.~Lamastra \inst{\ref{M3}}
\and D.~Lelas \inst{\ref{M7}}
\and F.~Leone \inst{\ref{M3}}
\and E.~Lindfors \inst{\ref{M22}}
\and S.~Lombardi \inst{\ref{M3}}
\and F.~Longo \inst{\ref{M2},\ref{M28}}
\and M.~L\'opez \inst{\ref{M8}}
\and R.~L\'opez-Coto \inst{\ref{M16}}
\and A.~L\'opez-Oramas \inst{\ref{M1}}
\and S.~Loporchio \inst{\ref{M13}}
\and B.~Machado de Oliveira Fraga \inst{\ref{M9}}
\and C.~Maggio \inst{\ref{M23}}
\and P.~Majumdar \inst{\ref{M24}}
\and M.~Makariev \inst{\ref{M25}}
\and M.~Mallamaci \inst{\ref{M16}}
\and G.~Maneva \inst{\ref{M25}}
\and M.~Manganaro \inst{\ref{M7}}
\and K.~Mannheim \inst{\ref{M21}}
\and L.~Maraschi \inst{\ref{M3}}
\and M.~Mariotti \inst{\ref{M16}}
\and M.~Mart\'inez \inst{\ref{M15}}
\and D.~Mazin \inst{\ref{M14},\ref{M5}}
\and S.~Mender \inst{\ref{M6}}
\and S.~Mi\'canovi\'c \inst{\ref{M7}}
\and D.~Miceli \inst{\ref{M2}}
\and T.~Miener \inst{\ref{M8}}
\and M.~Minev \inst{\ref{M25}}
\and J.~M.~Miranda \inst{\ref{M11}}
\and R.~Mirzoyan \inst{\ref{M14}}
\and E.~Molina \inst{\ref{M18}}
\and A.~Moralejo \inst{\ref{M15}}
\and D.~Morcuende \inst{\ref{M8}}
\and V.~Moreno \inst{\ref{M23}}
\and E.~Moretti \inst{\ref{M15}}
\and P.~Munar-Adrover \inst{\ref{M23}}
\and V.~Neustroev \inst{\ref{M22}}
\and C.~Nigro \inst{\ref{M15}}
\and K.~Nilsson \inst{\ref{M22}}
\and D.~Ninci \inst{\ref{M15}}
\and K.~Nishijima \inst{\ref{M5}}
\and K.~Noda \inst{\ref{M5}}
\and S.~Nozaki \inst{\ref{M5}}
\and Y.~Ohtani \inst{\ref{M5}}
\and T.~Oka \inst{\ref{M5}}
\and J.~Otero-Santos \inst{\ref{M1}}
\and M.~Palatiello \inst{\ref{M2}}
\and D.~Paneque \inst{\ref{M14}}
\and R.~Paoletti \inst{\ref{M11}}
\and J.~M.~Paredes \inst{\ref{M18}}
\and L.~Pavleti\'c \inst{\ref{M7}}
\and P.~Pe\~nil \inst{\ref{M8}}
\and C.~Perennes \inst{\ref{M16}}
\and M.~Persic \inst{\ref{M2},\ref{M29}}
\and P.~G.~Prada Moroni \inst{\ref{M17}}
\and E.~Prandini \inst{\ref{M16}}
\and C.~Priyadarshi \inst{\ref{M15}}
\and I.~Puljak \inst{\ref{M7}}
\and W.~Rhode \inst{\ref{M6}}
\and M.~Rib\'o \inst{\ref{M18}}
\and J.~Rico \inst{\ref{M15}}
\and C.~Righi \inst{\ref{M3}}
\and A.~Rugliancich \inst{\ref{M17}}
\and L.~Saha \inst{\ref{M8}}
\and N.~Sahakyan \inst{\ref{M19}}
\and T.~Saito \inst{\ref{M5}}
\and S.~Sakurai \inst{\ref{M5}}
\and K.~Satalecka \inst{\ref{DESY}}
\and B.~Schleicher \inst{\ref{M21}}
\and K.~Schmidt \inst{\ref{M6}}
\and T.~Schweizer \inst{\ref{M14}}
\and J.~Sitarek \inst{\ref{M10}} \protect\footnotemark[1]
\and I.~\v{S}nidari\'c \inst{\ref{M7}}
\and D.~Sobczynska \inst{\ref{M10}}
\and A.~Spolon \inst{\ref{M16}}
\and A.~Stamerra \inst{\ref{M3}}
\and D.~Strom \inst{\ref{M14}}
\and M.~Strzys \inst{\ref{M5}}
\and Y.~Suda \inst{\ref{M14}}
\and T.~Suri\'c \inst{\ref{M7}}
\and M.~Takahashi \inst{\ref{M5}}
\and F.~Tavecchio \inst{\ref{M3}}
\and P.~Temnikov \inst{\ref{M25}}
\and T.~Terzi\'c \inst{\ref{M7}} \protect\footnotemark[1]
\and M.~Teshima \inst{\ref{M14},\ref{M5}}
\and N.~Torres-Alb\`a \inst{\ref{M18}}
\and L.~Tosti \inst{\ref{M13}}
\and S.~Truzzi \inst{\ref{M11}}
\and J.~van Scherpenberg \inst{\ref{M14}}
\and G.~Vanzo \inst{\ref{M1}}
\and M.~Vazquez Acosta \inst{\ref{M1}}
\and S.~Ventura \inst{\ref{M11}}
\and V.~Verguilov \inst{\ref{M25}}
\and C.~F.~Vigorito \inst{\ref{M13}}
\and V.~Vitale \inst{\ref{M13}}
\and I.~Vovk \inst{\ref{M5}}
\and M.~Will \inst{\ref{M14}}
\and D.~Zari\'c \inst{\ref{M7}} \and
\\
%
S.G.~Jorstad \inst{\ref{Boston},\ref{Petersburg}} 
\and A.P.~Marscher \inst{\ref{Boston}} 
\and B.~Boccardi \inst{\ref{Bonn}}
\and C.~Casadio \inst{\ref{Bonn},\ref{greece}}
\and J.~Hodgson \inst{\ref{Daejeon}}
\and J.-Y.~Kim \inst{\ref{Bonn}}
\and T.P.~Krichbaum \inst{\ref{Bonn}}
\and A.~L\"{a}hteenm\"{a}ki \inst{\ref{Mets},\ref{Aalto},\ref{Tartu}}
\and M.~Tornikoski \inst{\ref{Mets}}
\and E.~Traianou \inst{\ref{Bonn}}
\and Z.R.~Weaver \inst{\ref{Boston}}
}
\institute{
Centre for Space Research, North-West University, Potchefstroom 2520, South Africa \label{NWU} \and 
Universit\"at Hamburg, Institut f\"ur Experimentalphysik, Luruper Chaussee 149, D 22761 Hamburg, Germany \label{HH} \and 
Max-Planck-Institut f\"ur Kernphysik, P.O. Box 103980, D 69029 Heidelberg, Germany \label{MPIK} \and 
Dublin Institute for Advanced Studies, 31 Fitzwilliam Place, Dublin 2, Ireland \label{DIAS} \and 
High Energy Astrophysics Laboratory, RAU,  123 Hovsep Emin St  Yerevan 0051, Armenia \label{RAU} \and
Yerevan Physics Institute, 2 Alikhanian Brothers St., 375036 Yerevan, Armenia \label{YPI} \and
Institut f\"ur Physik, Humboldt-Universit\"at zu Berlin, Newtonstr. 15, D 12489 Berlin, Germany \label{HUB} \and
University of Namibia, Department of Physics, Private Bag 13301, Windhoek, Namibia, 12010 \label{UNAM} \and
GRAPPA, Anton Pannekoek Institute for Astronomy, University of Amsterdam,  Science Park 904, 1098 XH Amsterdam, The Netherlands \label{GRAPPA} \and
Department of Physics and Electrical Engineering, Linnaeus University,  351 95 V\"axj\"o, Sweden \label{Linnaeus} \and
Institut f\"ur Astro- und Teilchenphysik, Leopold-Franzens-Universit\"at Innsbruck, A-6020 Innsbruck, Austria \label{LFUI} \and
School of Physical Sciences, University of Adelaide, Adelaide 5005, Australia \label{Adelaide} \and
Laboratoire Univers et Théories, Observatoire de Paris, Université PSL, CNRS, Université de Paris, 92190 Meudon, France \label{LUTH} \and
Sorbonne Universit\'e, Universit\'e Paris Diderot, Sorbonne Paris Cit\'e, CNRS/IN2P3, Laboratoire de Physique Nucl\'eaire et de Hautes Energies, LPNHE, 4 Place Jussieu, F-75252 Paris, France \label{LPNHE} \and
Laboratoire Univers et Particules de Montpellier, Universit\'e Montpellier, CNRS/IN2P3,  CC 72, Place Eug\`ene Bataillon, F-34095 Montpellier Cedex 5, France \label{LUPM} \and
IRFU, CEA, Universit\'e Paris-Saclay, F-91191 Gif-sur-Yvette, France \label{IRFU} \and
Astronomical Observatory, The University of Warsaw, Al. Ujazdowskie 4, 00-478 Warsaw, Poland \label{UWarsaw} \and
Aix Marseille Universit\'e, CNRS/IN2P3, CPPM, Marseille, France \label{CPPM} \and
Instytut Fizyki J\c{a}drowej PAN, ul. Radzikowskiego 152, 31-342 Krak{\'o}w, Poland \label{IFJPAN} \and
School of Physics, University of the Witwatersrand, 1 Jan Smuts Avenue, Braamfontein, Johannesburg, 2050 South Africa \label{WITS} \and
Laboratoire d'Annecy de Physique des Particules, Univ. Grenoble Alpes, Univ. Savoie Mont Blanc, CNRS, LAPP, 74000 Annecy, France \label{LAPP} \and
Landessternwarte, Universit\"at Heidelberg, K\"onigstuhl, D 69117 Heidelberg, Germany \label{LSW} \and
Universit\'e Bordeaux, CNRS/IN2P3, Centre d'\'Etudes Nucl\'eaires de Bordeaux Gradignan, 33175 Gradignan, France \label{CENBG} \and
Institut f\"ur Astronomie und Astrophysik, Universit\"at T\"ubingen, Sand 1, D 72076 T\"ubingen, Germany \label{IAAT} \and
Laboratoire Leprince-Ringuet, École Polytechnique, CNRS, Institut Polytechnique de Paris, F-91128 Palaiseau, France \label{LLR} \and
Université de Paris, CNRS, Astroparticule et Cosmologie, F-75013 Paris, France \label{APC} \and
Department of Physics and Astronomy, The University of Leicester, University Road, Leicester, LE1 7RH, United Kingdom \label{Leicester} \and
Nicolaus Copernicus Astronomical Center, Polish Academy of Sciences, ul. Bartycka 18, 00-716 Warsaw, Poland \label{NCAC} \and
Institut f\"ur Physik und Astronomie, Universit\"at Potsdam,  Karl-Liebknecht-Strasse 24/25, D 14476 Potsdam, Germany \label{UP} \and
Friedrich-Alexander-Universit\"at Erlangen-N\"urnberg, Erlangen Centre for Astroparticle Physics, Erwin-Rommel-Str. 1, D 91058 Erlangen, Germany \label{ECAP} \and
DESY, D-15738 Zeuthen, Germany \label{DESY} \and
Obserwatorium Astronomiczne, Uniwersytet Jagiello{\'n}ski, ul. Orla 171, 30-244 Krak{\'o}w, Poland \label{UJK} \and
Institute of Astronomy, Faculty of Physics, Astronomy and Informatics, Nicolaus Copernicus University,  Grudziadzka 5, 87-100 Torun, Poland \label{NCUT} \and
Department of Physics, University of the Free State,  PO Box 339, Bloemfontein 9300, South Africa \label{UFS} \and
Department of Physics, Rikkyo University, 3-34-1 Nishi-Ikebukuro, Toshima-ku, Tokyo 171-8501, Japan \label{Rikkyo} \and
Kavli Institute for the Physics and Mathematics of the Universe (WPI), The University of Tokyo Institutes for Advanced Study (UTIAS), The University of Tokyo, 5-1-5 Kashiwa-no-Ha, Kashiwa, Chiba, 277-8583, Japan \label{KAVLI} \and
Department of Physics, The University of Tokyo, 7-3-1 Hongo, Bunkyo-ku, Tokyo 113-0033, Japan \label{Tokyo} \and
University of Oxford, Department of Physics, Denys Wilkinson Building, Keble Road, Oxford OX1 3RH, UK \label{Oxford} \and
Now at Physik Institut, Universit\"at Z\"urich, Winterthurerstrasse 190, CH-8057 Z\"urich, Switzerland \label{Zurich}
\and Inst. de Astrof\'isica de Canarias, E-38200 La Laguna, and Universidad de La Laguna, Dpto. Astrof\'isica, E-38206 La Laguna, Tenerife, Spain, \label{M1}
\and Universit\`a di Udine, and INFN Trieste, I-33100 Udine, Italy, \label{M2}
\and National Institute for Astrophysics (INAF), I-00136 Rome, Italy, \label{M3}
\and ETH Zurich, CH-8093 Zurich, Switzerland, \label{M4}
\and Japanese MAGIC Consortium: ICRR, The University of Tokyo, 277-8582 Chiba, Japan; Department of Physics, Kyoto University, 606-8502 Kyoto, Japan; Tokai University, 259-1292 Kanagawa, Japan; RIKEN, 351-0198 Saitama, Japan, \label{M5}
\and Technische Universit\"at Dortmund, D-44221 Dortmund, Germany, \label{M6}
\and Croatian Consortium: University of Rijeka, Department of Physics, 51000 Rijeka; University of Split - FESB, 21000 Split; University of Zagreb - FER, 10000 Zagreb; University of Osijek, 31000 Osijek; Rudjer Boskovic Institute, 10000 Zagreb, Croatia, \label{M7}
\and IPARCOS Institute and EMFTEL Department, Universidad Complutense de Madrid, E-28040 Madrid, Spain, \label{M8}
\and Centro Brasileiro de Pesquisas F\'isicas (CBPF), 22290-180 URCA, Rio de Janeiro (RJ), Brasil, \label{M9}
\and University of Lodz, Faculty of Physics and Applied Informatics, Department of Astrophysics, 90-236 Lodz, Poland, \label{M10}
\and Universit\`a di Siena and INFN Pisa, I-53100 Siena, Italy, \label{M11}
\and Istituto Nazionale Fisica Nucleare (INFN), 00044 Frascati (Roma) Italy, \label{M13}
\and Max-Planck-Institut f\"ur Physik, D-80805 M\"unchen, Germany, \label{M14}
\and Institut de F\'isica d'Altes Energies (IFAE), The Barcelona Institute of Science and Technology (BIST), E-08193 Bellaterra (Barcelona), Spain, \label{M15}
\and Universit\`a di Padova and INFN, I-35131 Padova, Italy, \label{M16}
\and Universit\`a di Pisa, and INFN Pisa, I-56126 Pisa, Italy, \label{M17}
\and Universitat de Barcelona, ICCUB, IEEC-UB, E-08028 Barcelona, Spain, \label{M18}
\and The Armenian Consortium: ICRANet-Armenia at NAS RA, A. Alikhanyan National Laboratory, \label{M19}
\and Centro de Investigaciones Energ\'eticas, Medioambientales y Tecnol\'ogicas, E-28040 Madrid, Spain, \label{M20}
\and Universit\"at W\"urzburg, D-97074 W\"urzburg, Germany, \label{M21}
\and Finnish MAGIC Consortium: Finnish Centre of Astronomy with ESO (FINCA), University of Turku, FI-20014 Turku, Finland; Astronomy Research Unit, University of Oulu, FI-90014 Oulu, Finland, \label{M22}
\and Departament de F\'isica, and CERES-IEEC, Universitat Aut\`onoma de Barcelona, E-08193 Bellaterra, Spain, \label{M23}
\and Saha Institute of Nuclear Physics, HBNI, 1/AF Bidhannagar, Salt Lake, Sector-1, Kolkata 700064, India, \label{M24}
\and Inst. for Nucl. Research and Nucl. Energy, Bulgarian Academy of Sciences, BG-1784 Sofia, Bulgaria, \label{M25}
\and now at University of Innsbruck, \label{M26}
\and also at Port d'Informaci\'o Cient\'ifica (PIC) E-08193 Bellaterra (Barcelona) Spain, \label{M27}
\and also at Dipartimento di Fisica, Universit\`a di Trieste, I-34127 Trieste, Italy, \label{M28}
\and also at INAF-Trieste and Dept. of Physics \& Astronomy, University of Bologna \label{M29}
\and Institute for Astrophysical Research, Boston University, 725 Commonwealth Avenue, Boston, MA 02215 \label{Boston}
\and Astronomical Institute, St. Petersburg University, Universitetskij Pr. 28, Petrodvorets, 198504 St. Petersburg, Russia \label{Petersburg}
\and Max-Planck-Institut f\"ur Radioastronomie, Auf dem H\"ugel 69, 53121 Bonn, Germany \label{Bonn}
\and Foundation for Research and Technology – Hellas, IESL \& Institute of Astrophysics, Voutes 7110, Heraklion, Greece \label{greece}
\and Korea Astronomy and Space Science Institute, 776 Daedeok-daero, Yuseong-gu, Daejeon 34055, Korea \label{Daejeon}
\and Aalto University Mets\"ahovi Radio Observatory Mets\"ahovintie 114, FIN-02540 Kylm\"al\"a, Finland \label{Mets}
\and Aalto University Dept of Electronics and Nanoengineering, P.O. BOX 15500, FI-00076, Aalto, Finland \label{Aalto}
\and Tartu Observatory, Observatooriumi 1, 61602 T\~{o}ravere, Estonia \label{Tartu}
}
\offprints{H.E.S.S.~collaboration,
\protect\\\email{\href{mailto:contact.hess@hess-experiment.eu}{contact.hess@hess-experiment.eu}}; 
MAGIC~collaboration,
\protect\\\email{\href{mailto:contact.magic@mppmu.mpg.de}{contact.magic@mppmu.mpg.de}}; 
\protect\\\protect\footnotemark[1] Corresponding authors
}
\date{Received ? / accepted ? }
\abstract{The flat spectrum radio quasar (FSRQ) PKS~1510$-$089 is known for its complex multiwavelength behavior, and is one of only a few FSRQs detected at very high energy (VHE, $E>100\,$GeV) $\gamma$-rays. VHE \g-ray observations with H.E.S.S. and MAGIC during late May and early June 2016 resulted in the detection of an unprecedented flare, which reveals for the first time VHE \g-ray intranight variability in this source. While a common variability timescale of $1.5\,$hr is found, there is a significant deviation near the end of the flare with a timescale of $\sim 20\,$min marking the cessation of the event. 
The peak flux is nearly two orders of magnitude above the low-level emission. For the first time, curvature is detected in the VHE \g-ray spectrum of \source, which is fully explained through absorption by the extragalactic background light. 
Optical R-band observations with ATOM reveal a counterpart of the \g-ray flare, even though the detailed flux evolution differs from the VHE \g-ray lightcurve. Interestingly, a steep flux decrease is observed at the same time as the cessation of the VHE \g-ray flare. In the high energy (HE, $E>100\,$MeV) \g-ray band only a moderate flux increase is observed with \fermi, while the HE \g-ray spectrum significantly hardens up to a photon index of $1.6$. A search for broad-line region (BLR) absorption features in the \g-ray spectrum indicates that the emission region is located outside of the BLR. 
Radio VLBI observations reveal a fast moving knot interacting with a standing jet feature around the time of the flare. As the standing feature is located $\sim 50\,$pc from the black hole, the emission region of the flare may have been located at a significant distance from the black hole. 
If this correlation is indeed true, VHE \g\ rays have been produced far down the jet where turbulent plasma crosses a standing shock. 
}
\maketitle
%
\makeatletter
\renewcommand*{\@fnsymbol}[1]{\ifcase#1\@arabic{#1}\fi}
\makeatother
%
%
\section{Introduction} \label{sec:intro}
Rapid flares in blazars -- active galactic nuclei with the relativistic jet aligned with the observer's line-of-sight \citep{br74} -- are among the most puzzling events in the Universe. These flares can exhibit flux variations on the order of a few minutes and have been observed in all blazar types \citep{gea96,aHea07,tea11,aFea16b}. The spectral energy distribution (SED) of blazars exhibits two broadband components, which are usually attributed to synchrotron and inverse-Compton emission by relativistic electrons \citep{ds93,sbr94,mk97,bsmm00,bea13}, or hadronic interactions \citep{bea13,m93,mea03}. Within the standard one-zone model, the bulk of the broadband radiation -- from far-infrared to VHE \g-ray energies -- is produced in a single emission region, while the radio emission is produced in the extended jet. The single emission region supposedly fills the cross section of the jet, and therefore fast flux variations and their inferred small emission regions are often suggested to originate from a region close to the base of the jet. Observations of fast-variable VHE \g-ray emission challenge this picture \citep[see e.g.][]{pks1222}. 

\source\ is a blazar at cosmological redshift $z=0.361$ \citep{tea96}, which corresponds to a luminosity distance of $1.9\,$Gpc assuming a cosmic expansion with Hubble parameter $H_0=70\,$km/s/Mpc, matter density $\Omega_m=0.3$, and dark energy density $\Omega_{\Lambda}=0.7$. \source\ exhibits broad emission lines in its optical spectrum leading to its classification as a flat spectrum radio quasar (FSRQ). The broad lines are produced by gas in the so-called broad-line region (BLR), which is in close proximity to the black hole. 
As BLR photons absorb VHE \g-ray photons through pair production, any VHE \g-ray emission detected from FSRQs is most likely produced beyond the BLR \citep[e.g.,][]{cea18,msb19}.

Multiwavelength monitoring observations of \source\ reveal a complex behavior \citep{b13,Foschini2013,sea13,sea15,kea16} with changing correlation patterns between different energy bands. In turn, the SED modeling has resulted in different interpretations, such as two distinct photon fields for the inverse-Compton process \citep{nea12} or two spatially separated emission zones \citep{bea14}. While these models are not exclusive, they already emphasize the complex nature of this blazar. In addition, one of the highest recorded speeds of apparent superluminal motion -- up to $\sim 46\,c$ -- has been detected in radio observations of the jet of \source\ \citep{jo05}.

\source\ was detected for the first time in VHE \g\ rays with H.E.S.S. during a bright high-energy (HE, $E>100\,$MeV) $\gamma$-ray flare in 2009 \citep{Hea13}. Subsequent observations were initially guided by multiwavelength flares leading to a detection with MAGIC during another high state of HE \g -ray emission in 2012 \citep{aMea14}. Only later, systematic monitoring efforts at VHE \g\ rays have commenced \citep{zea19}. They led to the detection of \source\ in VHE \g-rays with MAGIC during low HE $\gamma$-ray states \citep{acc18}. 

In 2016, the monitoring with \hess\ resulted in the detection of a strong flare at VHE \g\ rays that was not triggered by multiwavelength events, and which has been followed up with MAGIC. These observations allowed for the first time studies on sub-hour time scales at VHE \g\ rays in this source. The results of these observations are presented here along with detailed observations at HE $\gamma$-rays, and optical frequencies, as well as long-term radio very-long baseline interferometry (VLBI) observations (Section \ref{sec:ana}). The flare behavior is discussed in Section \ref{sec:corstd} including its variability, absorption of VHE \g\ rays, the evolution on large scales, and constraints on the emission region parameters. A summary and the conclusions are given in Section \ref{sec:sumcon}.

%
%
\section{Data analysis and results} \label{sec:ana}
The flare took place on JD~2457538 (May 29-30, 2016) and JD~2457539 (May 30-31, 2016; hereafter ``flare night''), and has been observed with H.E.S.S. and MAGIC in VHE \g\ rays, with {\it Fermi}-LAT in HE \g\ rays, and with ATOM in the optical R band. Radio VLBI monitoring data at $43\,$GHz and $86\,$GHz taken over several months complement the data set. 
\subsection{H.E.S.S.} \label{sec:hess}
\begin{figure}[t]
\centering
\includegraphics[width=0.48\textwidth]{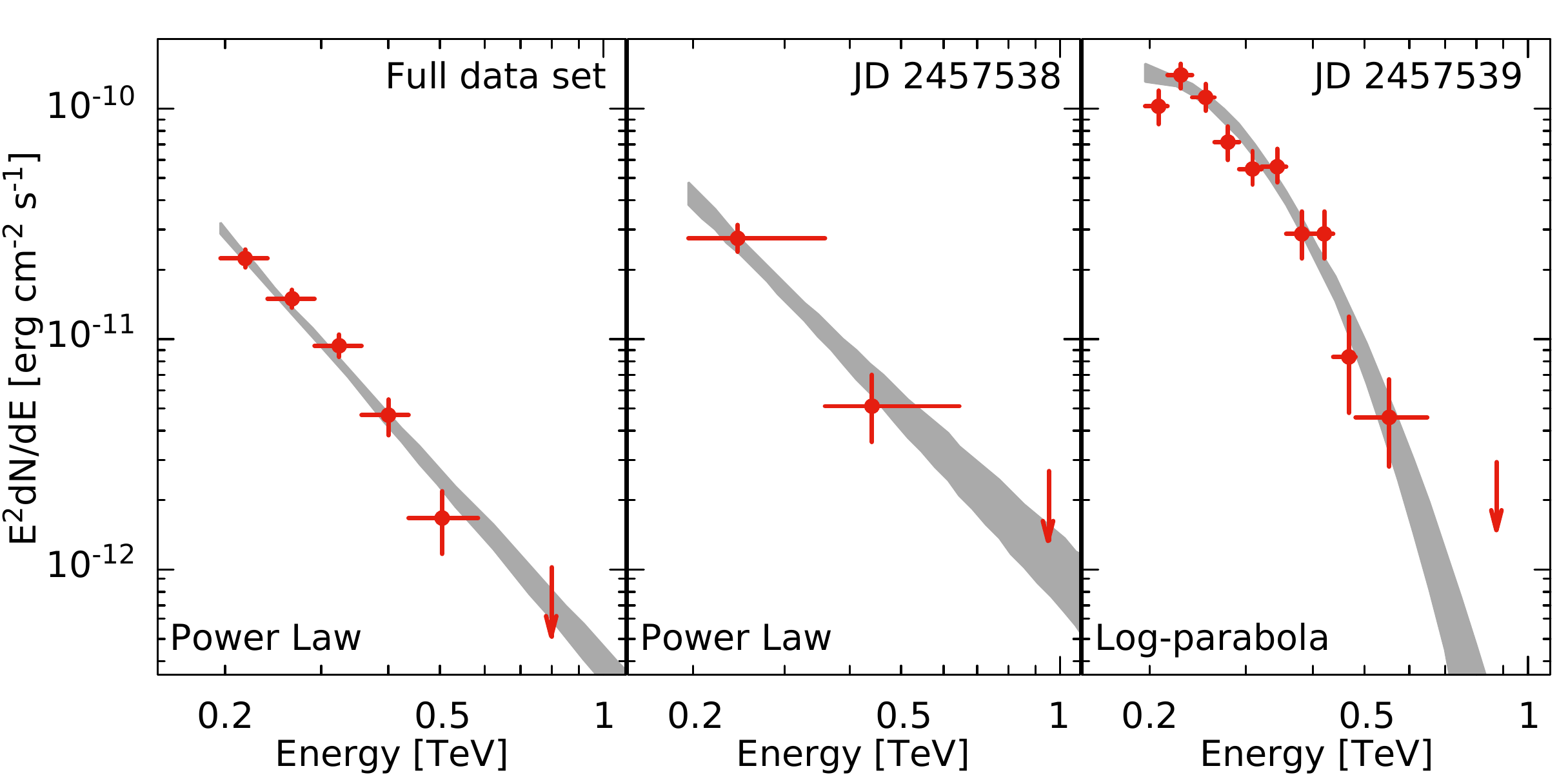}
\caption{Measured VHE \g-ray spectra from H.E.S.S. with spectral type and time frames as indicated. Points have a significance of at least $3\sigma$, while upper limits are at $99\%$ confidence level. The "Full data set" spectrum is the average spectrum of all observations from JD~2457536 -- 2457546.
}
\label{fig:spec_hess}
\end{figure}
H.E.S.S. is an array of five Imaging Atmospheric Cherenkov Telescopes (IACTs) located in the Khomas Highland in Namibia at an altitude of about $1800\,$m. The first phase of H.E.S.S. began in 2004 with four telescopes (CT1-4) with 107\,m$^2$ mirror area each, laid out in a square of 120\,m side length giving an optimal energy threshold of $\sim 100\,$GeV. In 2012, a fifth telescope (CT5) with 600\,m$^2$ mirror area was added to the center of the array, reducing the energy threshold to $\sim 50\,$GeV. 
Due to instrumental issues in the considered observation period (JD~2457536 -- 2457546, May 27 -- June 06 2016), the analysis only covers events recorded with CT2-4. The reduced number of telescopes, and observation conditions raise the energy threshold of this data set to $\gtrsim 190\,$GeV.

A total number of 31 runs (1 run lasts for about $28\,$min) passed the standard quality selection \citep{aHea06} resulting in an acceptance corrected observation time of $13.6\,$hr. The data set has been analyzed with the Model analysis chain using \textsc{loose cuts} \citep{dnr09}. The results have been cross-checked and verified using the independent reconstruction and analysis chain ImPACT \citep{ph14} giving consistent results.

In the entire observation period the source is detected with a significance of $29.5\sigma$ above an energy threshold of $E_{\rm thr} = 196\,$GeV. A power-law fit, 

\begin{align}
 F(E) = N(E_0)\times\left( \frac{E}{E_0} \right)^{-\Gamma} \label{eq:pwl},
\end{align}
to the measured spectrum results in a normalization of $N(0.265\,\text{TeV})=(1.12\pm 0.04\stat{}^{+0.7}_{-0.4}{}\sys)\times 10^{-10}\,$ph\,cm$^{-2}$s$^{-1}$TeV$^{-1}$ with an index of $\Gamma = 4.7\pm 0.1\stat\pm 0.15\sys$. The normalization is derived at the decorrelation energy $E_0$. The large systematic errors are a consequence of the energy uncertainty of 15\% \citep{aHea06} folded with the very soft spectrum of the source. Using a log-parabola function, 

\begin{align}
 F(E) = N(E_0)\times \left( \frac{E}{E_0} \right)^{-\alpha-\beta\log{(E/E_0)}} \label{eq:logp},
\end{align}
with normalization $N(0.232\,\text{TeV})=(2.0\pm 0.1\stat{}^{+1.2}_{-0.7}{}\sys)\times 10^{-10}\,$ph\,cm$^{-2}$s$^{-1}$TeV$^{-1}$ and parameters $\alpha = 3.8\pm 0.3\stat\pm 0.15\sys$ and $\beta = 1.9\pm 0.6\stat$ marginally improves the fit by $2.8\sigma$ for one additional free parameter. The measured spectrum of the entire observation period is shown in the left panel of Fig.~\ref{fig:spec_hess}. The average flux of the period above $200\,$GeV is $(2.1 \pm 0.1\stat{}^{+1.3}_{-0.7}{}\sys)\times 10^{-11}\,$ph\,cm$^{-2}$s$^{-1}$. 

\begin{figure}[t]
\centering
\includegraphics[width=0.48\textwidth]{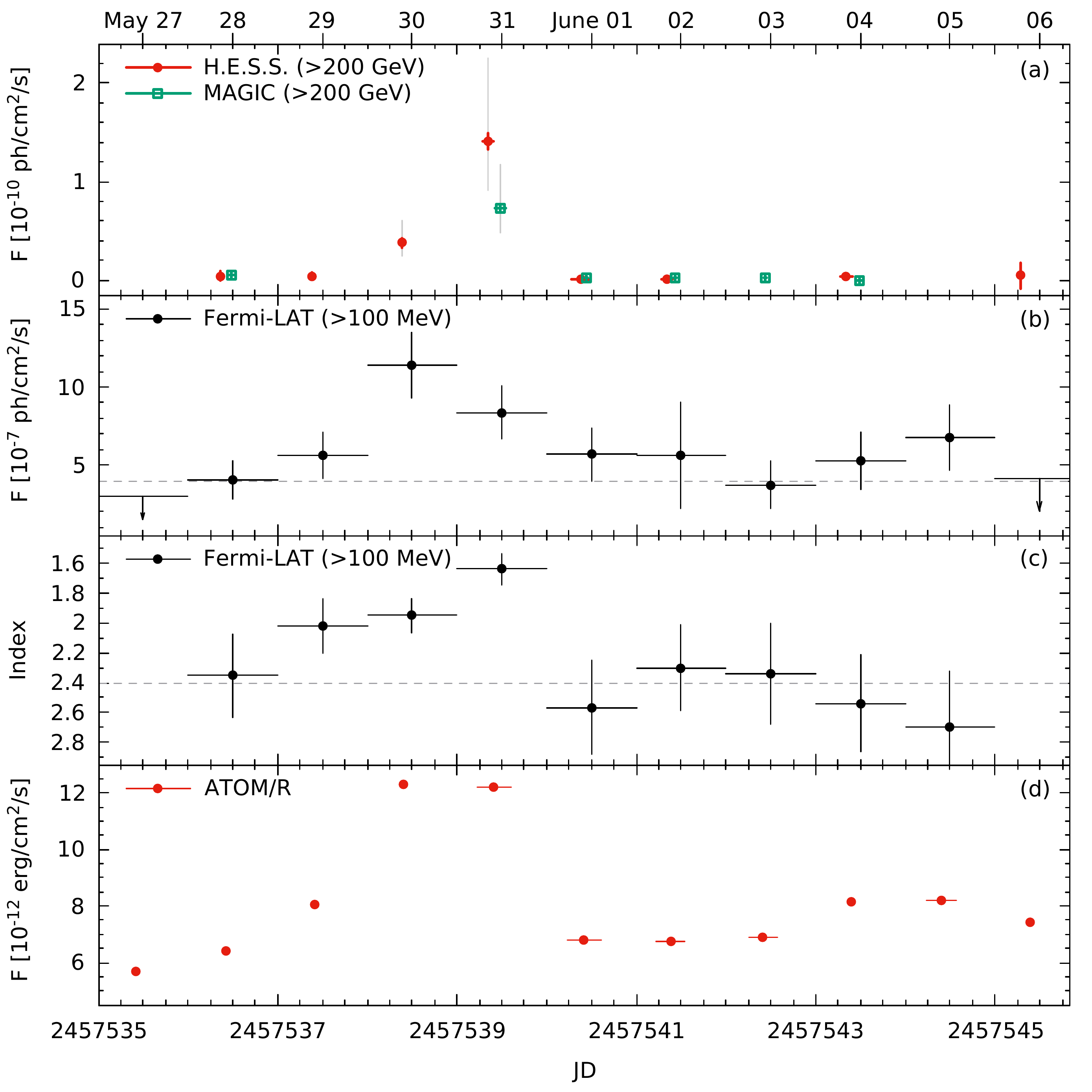}
\caption{Lightcurves of \source\ of the observation period. 
{\bf (a)} Nightly-averaged VHE \g-ray lightcurve from H.E.S.S. (red points) and MAGIC (green open squares). The grey bars mark the systematic uncertainties. 
{\bf (b)} $24\,$hr-average HE \g-ray lightcurve from \fermi. The dashed line marks the $11$ years average with $\bar{F} = 3.92\E{-7}\,$ph\,cm$^{-2}$s$^{-1}$. Arrows mark 95\% C.L. upper limits for bins with test statistics (TS) $<9.0$. 
{\bf (c)} $24\,$hr-binned HE \g-ray spectral index assuming a power-law spectrum. The dashed line marks the $11$ years average with ${\mbox{Index}}=2.402$. 
{\bf (d)} Nightly-averaged optical R-band lightcurve (de-reddened) from ATOM. The horizontal bars mark the observation time.
}
\label{fig:mwl_lc_all}
\end{figure}

The lightcurve with nightly averages and energy integrated above $200\,$GeV using the average spectrum of the full data set is shown in Fig. \ref{fig:mwl_lc_all}(a). A fit of a constant to the lightcurve is ruled out on a run-by-run binning with more than $10\sigma$. However, it is obvious that the source is not significantly detected apart from the two nights on JD~2457538 and 2457539, respectively. Given the high statistics of these two nights, they have been analyzed individually.

On JD~2457538 (May 29-30, 2016), H.E.S.S. observed \source\  for 3 runs and an acceptance-corrected observation time of $1.3\,$hr. The spectrum of this night is compatible with a power-law with normalization $N(0.274\,\text{TeV})=(1.9\pm 0.2\stat{}^{+1.1}_{-0.7}{}\sys)\times 10^{-10}\,$ph\,cm$^{-2}$s$^{-1}$TeV$^{-1}$ and an index of $\Gamma = 4.3\pm 0.3\stat\pm 0.15\sys$. A log-parabola, while not statistically preferred over the power-law, is described by a normalization of $N(0.240\,\text{TeV})=(2.5\pm 0.3\stat{}^{+1.5}_{-0.9}{}\sys)\times 10^{-10}\,$ph\,cm$^{-2}$s$^{-1}$TeV$^{-1}$ and parameters $\alpha = 3.4\pm 0.5\stat\pm 0.15\sys$ and $\beta = 3\pm 1\stat$. The spectrum is shown in the middle panel of Fig.~\ref{fig:spec_hess}. 
Deriving a power-law spectrum that accounts for the absorption by the extragalactic background light (EBL) using the model of \cite{frv08}, gives a normalization $N(0.274\,\text{TeV})=(5.2\pm 0.4\stat{}^{+3.1}_{-1.8}{}\sys)\times 10^{-10}\,$ph\,cm$^{-2}$s$^{-1}$TeV$^{-1}$ and an index of $\Gamma = 2.3\pm 0.3\stat\pm 0.15\sys$.
The average flux above $200\,$GeV is $(3.8 \pm 0.3\stat{}^{+2.3}_{-1.3}{}\sys)\times 10^{-11}\,$ph\,cm$^{-2}$s$^{-1}$ with no significant deviation from this value in the individual runs.

During the next night on JD~2457539 (May 30-31, 2016) the source was observed for 4 runs and an acceptance-corrected observation time of $1.8\,$hr. A power-law fit to the spectrum results in a normalization of $N(0.268\,\text{TeV})=(7.0\pm 0.3\stat{}^{+4.2}_{-2.5}{}\sys)\times 10^{-10}\,$ph\,cm$^{-2}$s$^{-1}$TeV$^{-1}$ and an index of $\Gamma = 4.8\pm 0.1\stat\pm 0.15\sys$. A log-parabola function with normalization $N(0.233\,\text{TeV})=(14.3\pm 0.7\stat{}^{+8.6}_{-5.0}{}\sys)\times 10^{-10}\,$ph\,cm$^{-2}$s$^{-1}$TeV$^{-1}$ and parameters $\alpha = 3.3\pm 0.4\stat\pm 0.15\sys$ and $\beta = 3.1\pm 0.7\stat$ improves the fit with respect to the power-law at a level of $4.4\sigma$. This is the first detection of significant curvature in the VHE \g-ray spectrum of \source. The spectrum is shown in the right panel of Fig.~\ref{fig:spec_hess}. Apparently, the spectral parameters $\alpha$ and $\beta$ do not change significantly between the two nights. They are also consistent within errors with the spectral parameters of the entire observation period. Analyzing the spectra for the individual runs does not give a significant deviation from the power-law parameters of the entire night. Applying the EBL model of \cite{frv08} to the spectrum of the whole night the curvature is much reduced, and the data are well described by a power-law with normalization of $N(0.268\,\text{TeV})=(19\pm 0.8\stat{}^{+11}_{-7}{}\sys)\times 10^{-10}\,$ph\,cm$^{-2}$s$^{-1}$TeV$^{-1}$ and an index of $\Gamma = 2.9\pm 0.2\stat\pm 0.15\sys$.

\begin{figure}[th]
\centering
\includegraphics[width=0.48\textwidth]{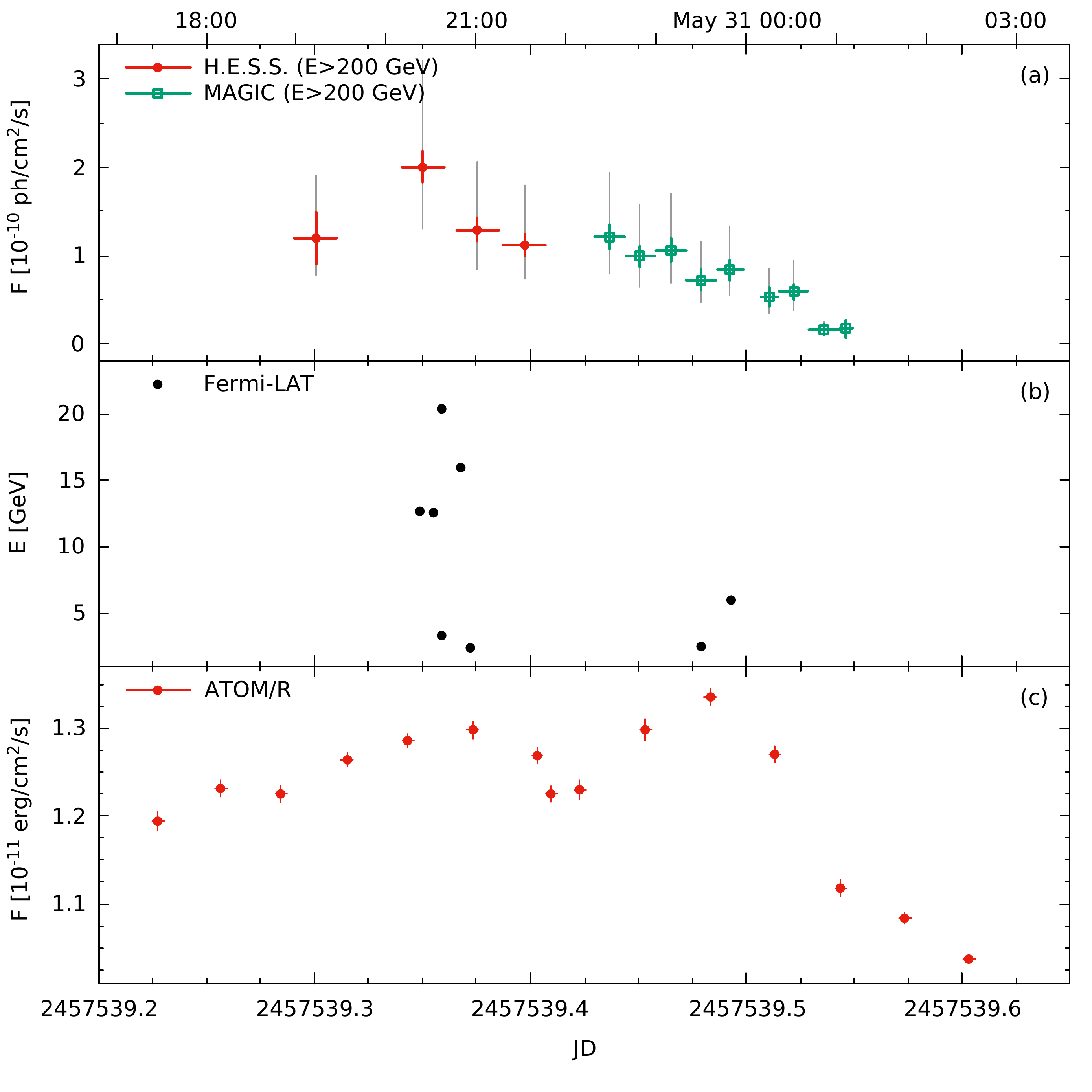}
\caption{Lightcurves of \source\ of the flare night, JD~2457539. 
{\bf (a)} VHE \g-ray lightcurve from H.E.S.S. (red points) and MAGIC (green open squares). The binning is 28\,min and 20\,min for the \hess\ and MAGIC lightcurves, respectively. The grey bars mark the systematic uncertainties.
{\bf (b)} Detected energies from \fermi\ above $1\,$GeV. 
The clustering of points in two bunches is a result of the orbital motion of the \textit{Fermi} satellite.
{\bf (c)} Optical R-band lightcurve from ATOM showing individual exposures of about $8\,$min duration each.
}
\label{fig:mwl_lc_57538}
\end{figure}

The lightcurve on JD 2457539 is shown in Fig.~\ref{fig:mwl_lc_57538}(a). The average flux of the night above $200\,$GeV amounts to $(14.3 \pm 0.6\stat{}^{+8.6}_{-5.0}{}\sys)\times 10^{-11}\,$cm$^{-2}$s$^{-1}$ or $\sim 56\%$ of the Crab nebula flux (or Crab Unit, C.U.) as measured with \hess\ above the same energy threshold. However, the flux deviates with $4.5\sigma$ from this average value on run-wise time-scales, clearly indicating for the first time VHE \g-ray intranight variability in \source. The peak flux of the night equals $(20 \pm 1\stat{}^{+12.0}_{-7.0}{}\sys)\times 10^{-11}\,$cm$^{-2}$s$^{-1}$ or $\sim 80\%$ C.U. above $200\,$GeV.
%

%
\subsection{MAGIC} \label{sec:magic}
\begin{figure}[th]
\centering
\includegraphics[width=0.48\textwidth]{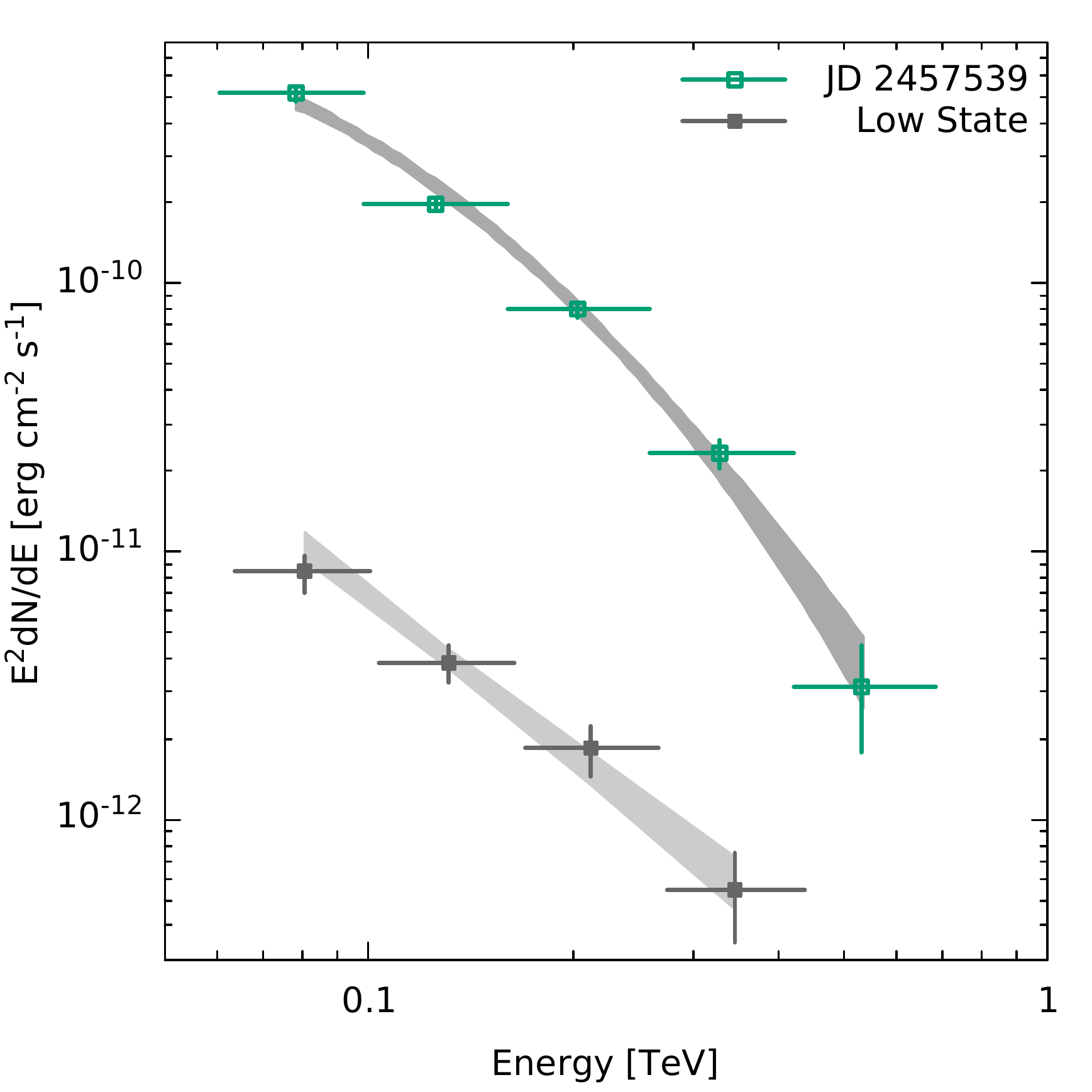}
\caption{Measured VHE \g-ray spectrum from MAGIC for JD~2457539 in green open squares, and the VHE \g-ray low state spectrum from \cite{acc18} in grey filled squares.
}
\label{fig:spec_magic}
\end{figure}
MAGIC is a system of two IACTs with reflectors of $17\,$m diameter each. 
They are located in the Canary Island of La Palma, at a height of 2200\,m a.s.l. \citep{al16a}.  
As \source\ is a southern source, only observable by MAGIC at zenith angle above $38^\circ$, the corresponding trigger threshold for a Crab-like spectrum is $\gtrsim 90$\,GeV \citep{al16b}, about 1.7 times larger than for the low zenith observations. 

\source\ has been regularly monitored by the MAGIC telescopes since 2013. 
The last observation before the flare night happened on JD~2457536.48, i.e. three days before.
During the flare night MAGIC observed \source\ for $2.53\,$hr. 
After the flare nightly follow-up observations were performed until JD~2457543.50, resulting in a total observation time of $7.5\,$hr. 
The data have been reduced and analyzed using MARS, the standard MAGIC analysis software \citep{za13,al16b}. 

During the investigated period, the only night in which a significant signal has been observed was the one on JD~2457539 (May 30-31, 2016). 
The average flux above 200 GeV in this night is
$(7.36\pm0.40\stat {}_{-2.6, {\rm sys}}^{+4.4})\times 10^{-11} \mathrm{ph\,cm^{-2}s^{-1}}$ corresponding to 
$(32.5\pm1.8\stat{}_{-11, {\rm sys}}^{+19})\%$ C.U. 
as measured with MAGIC above the same energy threshold\footnote{The integral Crab flux measured by MAGIC agrees within 7\% with the H.E.S.S. measurement.} \citep{al16b}. 
The flux shows an increase by a factor of $\gtrsim 5$ with respect to the flux upper limits measured on neighboring nights, as can be seen in Fig.~\ref{fig:mwl_lc_all}(a). 
The large systematic uncertainty is dominated by a $\sim 15\%$ uncertainty in the energy scale \citep{al16b} convolved with the very soft spectrum of \source.
The effect of the systematics in the energy scale on the integral flux was estimated by performing dedicated Monte Carlo simulations in which the signals obtained in all the camera's pixels were scaled by $+15\%$ or $-15\%$. 

The observed spectrum on the flare night was reconstructed between 60 and 700 GeV and is shown in Fig.~\ref{fig:spec_magic}. 
It exhibits clear curvature and can be described by a log-parabola, Eq.~(\ref{eq:logp}), with normalization $N(0.175\,\text{TeV})=(23.8\pm 1.4\stat{}_{-10}^{+11}{}\sys)\times 10^{-10}\,$ph\,cm$^{-2}$s$^{-1}$TeV$^{-1}$ 
and parameters $\alpha = 4.31\pm 0.12\stat\pm 0.15\sys$ and $\beta = 1.60\pm 0.41\stat$.
Correcting for the EBL absorption according to \cite{frv08}, the spectrum can be described by the simpler power-law form with normalization $N(0.175\,\text{TeV})=(35\pm 2\stat{}_{-15}^{+17}{}\sys)\times 10^{-10}\,$ph\,cm$^{-2}$s$^{-1}$TeV$^{-1}$ 
and an index of $\Gamma = 3.37\pm 0.09\stat\pm 0.15\sys$.
Similar to the case of the integral fluxes, the systematic uncertainties on the flux normalization are dominated by the 15\% uncertainty in the energy scale of MAGIC \citep{al16b}. 

The intranight light curve binned in 20\,min bins, as shown in Fig.~\ref{fig:mwl_lc_57538}(a), shows clear variability, since a constant fit can be excluded with more than $10\sigma$. 
The flux drops from $\sim50\%$ C.U. at the beginning of the observation down to $\sim 7.5\%$ C.U. at the end of the night. 

%
\subsection{\fermi} \label{sec:fermi}
{\it Fermi}-LAT monitors the HE $\gamma$-ray sky every $3\,$hr in the energy range from $20\,$MeV to beyond $300\,$GeV \citep{aFea09}. A binned analysis of the P8R3 SOURCE class events between energies of $100\,$MeV and $316\,$GeV was performed for a Region of Interest (ROI) of $15^{\circ}\times15^\circ$ centered at the position of \source.\footnote{The upper energy limit was chosen, because it corresponds to the energy binning of the instrumental response functions.} 
Sources within a region $30^{\circ}\times30^\circ$ centred on the source listed in the 4FGL catalog \citep{4fgl} have been accounted for in the likelihood analysis. 
In order to reduce contamination from the Earth Limb, a zenith angle cut of $90^{\circ}$ was applied. The analysis was performed with \textsc{fermipy}\footnote{\url{https://fermipy.readthedocs.io/}} v.0.17.4 and the \textsc{fermitools}\footnote{\url{https://fermi.gsfc.nasa.gov/ssc/data/analysis/}} v.1.0.0 software packages using the \textsc{P8R3\_SOURCE\_V2}\footnote{\url{http://fermi.gsfc.nasa.gov/ssc/data/analysis/documentation/Cicerone/Cicerone_LAT_IRFs/IRF_overview.html}} instrument response function \citep{Fermip8} and the \textsc{GLL\_IEM\_v07} and \textsc{ISO\_P8R3\_SOURCE\_V2\_v01} models\footnote{\url{http://fermi.gsfc.nasa.gov/ssc/data/access/lat/BackgroundModels.html}} for the Galactic and isotropic diffuse emission \citep{aFea16a}, respectively.

In a first step the average spectrum of \source\ is derived within the time window JD~2457201 -- 2457750, roughly 550 days centered around the VHE \g-ray flare. 
In the fit, all spectral parameters of sources within $3^\circ$ from \source\ are left free to vary along with the normalization of sources up to $10^\circ$ from the ROI center and of the background templates. 
All other source parameters are fixed to their respective 4FGL values. 
The normalizations of the background templates are left as additional free parameters. 
Additionally, spectral indices are frozen if a hint of emission from the source is detected with a significance $5<\mathrm{TS}<9$,\footnote{The TS value is defined as twice the difference of log-likelihood values of the optimized ROI model with and without the source included, $\mathrm{TS} = -2(\ln\mathcal{L}_1 - \ln\mathcal{L}_0)$~\citep{mattox1996}.} and all parameters are fixed if $\mathrm{TS}<5$ or the predicted number of photons is less than 0.01. 
Neither the residual nor TS maps show any particular hot spots above (or below) a significance at the $\sim2\,\sigma$ level. Therefore, the best-fit model describes the ROI well.
The spectrum of \source\ is described with a log-parabola function, Eq.~(\ref{eq:logp}), 
with normalization $N(743.5\,\text{MeV})=(9.6 \pm 0.2\stat)\times 10^{-11}\,$ph\,cm$^{-2}$s$^{-1}$MeV$^{-1}$ and parameters $\alpha = 2.33\pm 0.01\stat$ and $\beta = 0.041\pm 0.008\stat$. 

The best-fit ROI model is then used to derive a daily light curve of \source. 
In each time bin, the spectrum of \source\ is set to a simple power law and only the normalization of the source 4FGL~J1514.8-0949, associated with PMN~J1514-0948, a blazar of unknown type, located at a distance of only $0.9^\circ$ of \source, is allowed to vary. 
This is the brightest source within $5^\circ$ of \source, which corresponds approximately to the 68\,\% confidence radius of the point spread function of the LAT at 100\,MeV. 
The lightcurve bins have been centered around the VHE \g-ray observation windows. The resulting lightcurve is shown in Fig.~\ref{fig:mwl_lc_all}(b). 
The average flux of \source\ in the HE \g-ray energy band (in the $\sim500$\,days window chosen above) 
is $(7.04\pm0.14)\times 10^{-7}\,$ph\,cm$^{-2}$s$^{-1}$. The flux variation during the VHE \g-ray flare is not particularly strong, and by no means matches HE \g-ray outbursts already detected in this source \citep[e.g.][]{apm10,sea13,Foschini2013,oea13,aMea17,msb19}. 
The spectral index of the power-law fit to each $24\,$hr bin shows significant hardening of the HE \g-ray spectrum during the VHE \g-ray flare (see Fig.~\ref{fig:mwl_lc_all}(c)), reaching an index as hard as $1.6\pm 0.1$. 
During the \hess\ observation window, the spectral index further appears to harden to $1.4\pm0.2$, whereas during the MAGIC observations the spectrum marginally softens to $1.7\pm0.2$ (compatible results are obtained with the unbinned analysis pipeline of the \textsc{Fermitools}).\footnote{For these short observation times, we calculate the effective area in 5 bins of the the azimuthal spacecraft coordinates to account for this dependence on such short time scales. For longer time scales, the dependence is averaged over. See \url{https://fermi.gsfc.nasa.gov/ssc/data/analysis/scitools/help/gtltcube.txt}.}
This indicates that the flare mainly influenced the highest energies, while lower energy emission was not particularly enhanced. Hence, the peak frequency of the inverse Compton SED shifted significantly from $\lesssim 0.1\,$GeV on average to more than $10\,$GeV during the outburst. 
This is further supported by looking at the highest energy photons detected with \fermi\ during the \hess\ and MAGIC observation windows [see Fig.~\ref{fig:mwl_lc_57538}(b)].
During the peak VHE \g-ray flux, several photons above 10\,GeV are detected that originate from \source\ with a probability $>0.9$\footnote{The probabilities are calculated with the \texttt{gtsrcprob} tool of the standard \fermi\ software.}. The highest energy photon during the \hess\ (MAGIC) observations has an energy of $20.6\,$GeV ($6.0\,$GeV) and a probability of belonging to the source of 99.0\,\% (99.5\,\%).

%
\subsection{ATOM} \label{sec:atom}
The Automatic Telescope for Optical Monitoring (ATOM) is an optical telescope with 75\,cm aperture located on the H.E.S.S. site \citep{hea04}. Operating since 2005, ATOM provides optical monitoring of $\gamma$-ray sources and observed \source\ with high cadence in the R band. The data were analyzed using the fully automated ATOM Data Reduction and Analysis Software and their quality has been checked manually. 
The resulting flux was calculated via differential photometry using 5 custom-calibrated secondary standard stars in the same field-of-view.

After the detection of the beginning VHE \g-ray flare, the frequency of observations was increased to several exposures per night resulting in a detailed lightcurve. Fig.~\ref{fig:mwl_lc_all}(c) shows the nightly averaged optical lightcurve, while Fig.~\ref{fig:mwl_lc_57538}(c) displays individual exposures for the flare night, JD~2457539. It is apparent that the optical flux exhibited an outburst, even though the flux of the flare was several factors below the historical high state in July 2015 \citep{jea15,zea19}. 
Interestingly, the details between the VHE \g-ray and the optical lightcurve differ significantly during the flare night. While the VHE \g-ray flux shows a single peak, the optical flux exhibits a double peak structure. More detailed analyses regarding this fact are given in section \ref{sec:varia}.

%
\subsection{VLBA and GMVA} \label{sec:radio}
\begin{table*}[t]
\footnotesize
\caption{Map Information. \label{tab:Obs}}
\begin{tabular}{ccccccc}
Epoch&Beam&PA of Beam&$I_{peak}$&RMS&$f_{amp}$&M\\
(1)&(2)&(3)&(4)&(5)&(6)&(7) \\
\hline
2016/03/19&0.36$\times$0.14&$-$5.0&2991&0.70&1.21$\pm$0.07&10 \\
2016/04/23&0.36$\times$0.14&$-$2.6&2224&0.90&1.05$\pm$0.03&10 \\
2016/06/11&0.36$\times$0.14&$-$4.9&1992&0.65&1.01$\pm$0.05&10 \\
2016/07/05&0.41$\times$0.15&$-$8.5&2078&0.77&1.03$\pm$0.05&10 \\
2016/07/31&0.38$\times$0.15&$-$4.9&1629&1.45&1.05$\pm$0.05&10 \\
2016/09/05&0.57$\times$0.16&$-$15.0&2831&1.47&1.01$\pm$0.07&8, KP,SC \\
2016/10/06&0.85$\times$0.17&24.9&3042&1.50&1.20$\pm$0.11&9, MK \\
2016/10/23&0.34$\times$0.13&$-$1.8&2653&1.29&1.10$\pm$0.08&9, HN \\
2016/11/28&0.36$\times$0.13&$-$4.4&1799&1.00&1.50$\pm$0.10&10 \\
2016/12/24&0.38$\times$0.13&$-$7.0&2196&2.76&1.05$\pm$0.07&10 \\
2017/01/14&0.40$\times$0.14&$-$8.3&1700&1.10&1.02$\pm$0.05&10 \\
2017/02/04&0.41$\times$0.13&$-$7.0&1829&0.98&1.04$\pm$0.05&10 \\
2017/03/19&0.38$\times$0.14&$-$6.9&1180&0.47&1.31$\pm$0.09&10 \\
2017/04/16&0.39$\times$0.14&$-$8.3&1005&0.56&1.41$\pm$0.07&10 \\
2017/05/13&0.37$\times$0.14&$-$5.3&1680&1.10&1.01$\pm$0.05&10 \\
2017/06/09&0.46$\times$0.15&$-$13.5&1609&1.14&0.98$\pm$0.05&10 \\
\end{tabular}\\
{(1) - epoch of VLBA observation, year, month, and day; (2) - size of the restoring beam in mas; 
(3)- position angle of the beam in degree; (4) - total intensity map peak in mJy/beam, not corrected for $f_{amp}$; 
(5) - systematic noise level in mJy/beam estimated by {\it Difmap}, not corrected for $f_{amp}$; (6) - flux density 
correction factor, $f_{amp}$, see text; (7) - number of antennas, $M$, if $M<$10 the names of absent antennas (KP - Kitt Peak; SC - St. Croix; MK - Mauna Kea;
HN - Hancock, see \url{https://public.nrao.edu/telescopes/vlba/}) are indicated.}
\end{table*} 
\begin{table*}[t]
\footnotesize
\caption{Parameters of Knots. \label{tab:KParm}}
\begin{tabular}{llcccc}
\multicolumn{2}{l}{Parameter}&$K15$&$K16^{*}$&$A1$&$A2$\\
\multicolumn{2}{l}{(1)}&(2)&(3)&(4)&(5) \\
\hline
$\mu$&[mas~yr$^{-1}$]&0.24$\pm$0.03&0.48$\pm$0.11&$-$0.03$\pm$0.02&0.02$\pm$0.03 \\
$\beta_{\rm app}$&[$c$]&5.4$\pm$0.6&10.8$\pm$2.5& &  \\
$T_\circ$& [JD] &2457237$\pm$22&2457361$\pm$58& &  \\
$S_{\rm max}$& [Jy]&1.26$\pm$0.07&2.39$\pm$0.08&0.30$\pm$0.05&0.70$\pm$0.05\\
$\tau_{\rm var}$& [yr]&0.10$\pm$0.04&0.11$\pm$0.03& &  \\
$a$& [mas]&0.07$\pm$0.01&0.13$\pm$0.01&0.05$\pm$0.05&0.11$\pm$0.08\\
$\delta$& &34$\pm$3&43$\pm$5& &  \\
$\Gamma_{\rm b}$& &18$\pm$2&23$\pm$4& &  \\
$\langle R\rangle$& [mas]&0.2$\pm$0.1&0.3$\pm$0.1&0.18$\pm$0.05&0.47$\pm$0.06 \\
$\langle \Theta \rangle$& [deg]&37$\pm$21&$-$40$\pm$4&$-$41$\pm$14&$-$24$\pm$18 \\
$\Theta_\circ$& [deg]&0.5$\pm$0.1&0.6$\pm$0.3& &  \\
$N$& &8& 14&11&13 \\
\end{tabular}\\
{$\mu$ - proper motion and its 1$\sigma$ uncertainty; $\beta_{\rm app}$ - apparent speed and its 1$\sigma$ uncertainty (for K16 only the initial value is given); $T_\circ$ - time of ejection and its 1$\sigma$ uncertainty; $S_{\rm max}$ - maximum flux density and its 1$\sigma$ uncertainty; $\tau_{\rm var}$ - timescale of flux variability and its 1$\sigma$ uncertainty;
$a$ - average angular size of component and its standard deviation within 0.3~mas from the core for K15 and K16;  $\delta$ - Doppler factor and  its 1$\sigma$ uncertainty, $\Gamma_{\rm b}$ - Lorentz factor and its 1$\sigma$ uncertainty; $<R>$ - average distance from the core over epochs presented here and its standard deviation;  $<\Theta>$ - average position angle of component with respect to the core in projection on the plane of sky and its standard deviation; $\Theta_\circ$ - angle between velocity vector of component and line of sight and its 1$\sigma$ uncertainty; $N$ - number of epochs at which component was detected; * - the parameters of K16 are given for the period when the knot is within 0.5~mas from the core.}
\end{table*}
\begin{figure*}[th]
\centering
\includegraphics[width=0.33\textwidth]{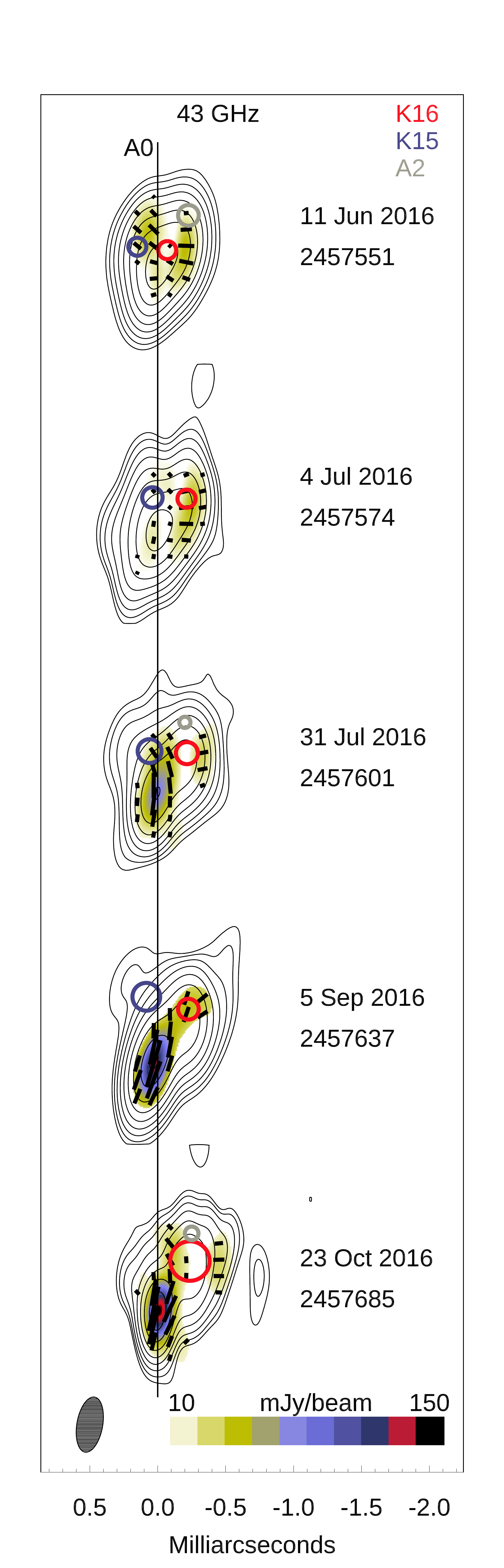}
\includegraphics[width=0.33\textwidth]{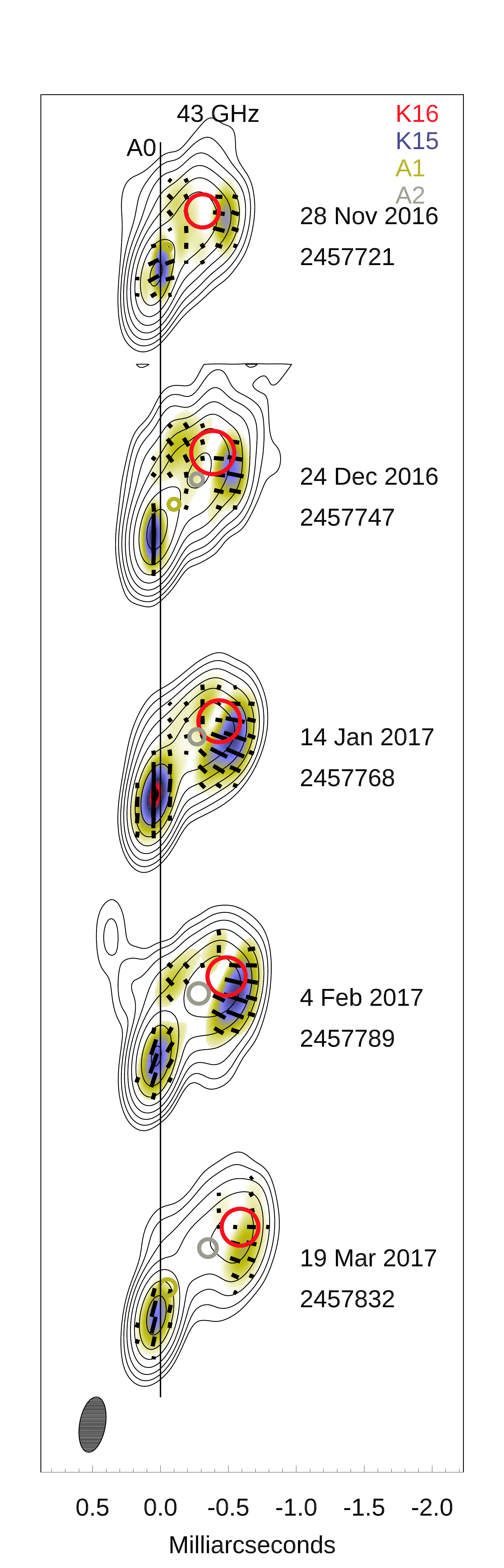}
\caption{Sequence of total (contours) and polarized (color scale) intensity images of \source\ at 43 GHz, 
convolved with a beam of FWHM dimensions 0.38$\times$0.15 mas$^2$ along PA=-10$^\circ$. The global total 
intensity peak is 2677 mJy/beam and the global polarized intensity peak is 147 mJy/beam. Black line segments 
within each image show the direction of the polarization electric vector; the length of each segment is proportional to the polarized intensity value; the black vertical line indicates the position of the core, A0; circles within the images mark positions of superluminal knots K16 (red)  as well as K15 (blue), and quasi-stationary features A1 (yellow) and A2 (gray), according to modeling.} 
\label{fig:vlba16}
\end{figure*}
\begin{figure}[th]
\centering
\includegraphics[width=0.33\textwidth]{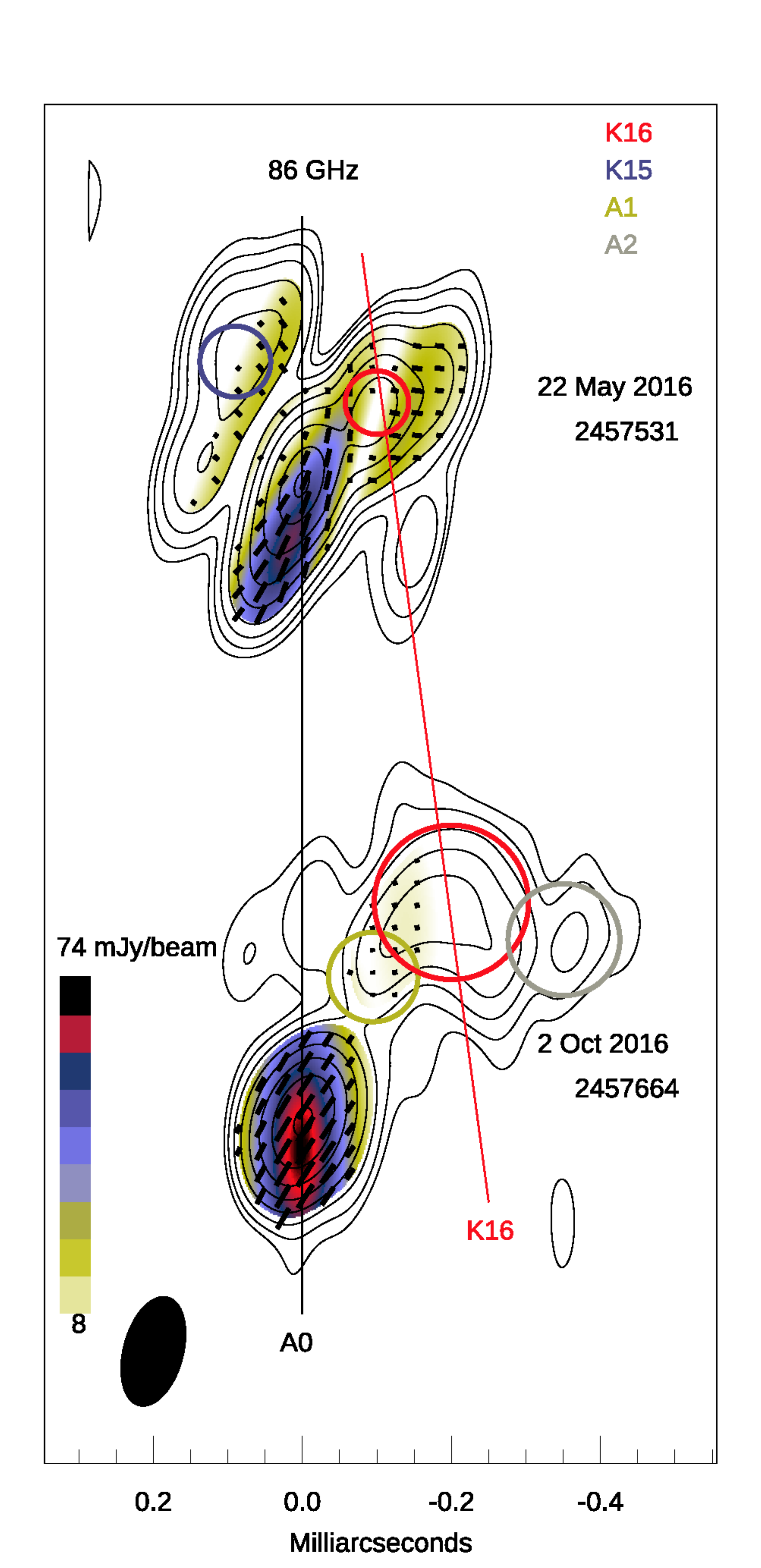}
\caption{A sequence of total (contours) and polarized (color scale) intensity images of \source\ at 86 GHz, with a beam of FWHM dimensions 0.10$\times$0.07 mas$^2$ along PA=-15$^\circ$. The global total intensity peak is 1137 mJy/beam and the global polarized intensity peak is 74 mJy/beam; the contours are 0.5\%, 1\%, 2\%,\ldots 64\% and 95\% of the global intensity peak. 
Black line segments within each image show the direction of the polarization electric vector; the length of the segment is proportional to the polarized intensity values; the black vertical line indicates the position of the core, A0; the blue, red, yellow, and grey circles mark positions of knots K15, K16, A2, and A1, respectively, according to modeling.} 
\label{fig:vlba3mm}
\end{figure}
\begin{figure*}[th]
\begin{minipage}[t]{0.49\textwidth}
\centering
\includegraphics[width=0.99\textwidth,trim=30 0 20 25,clip]{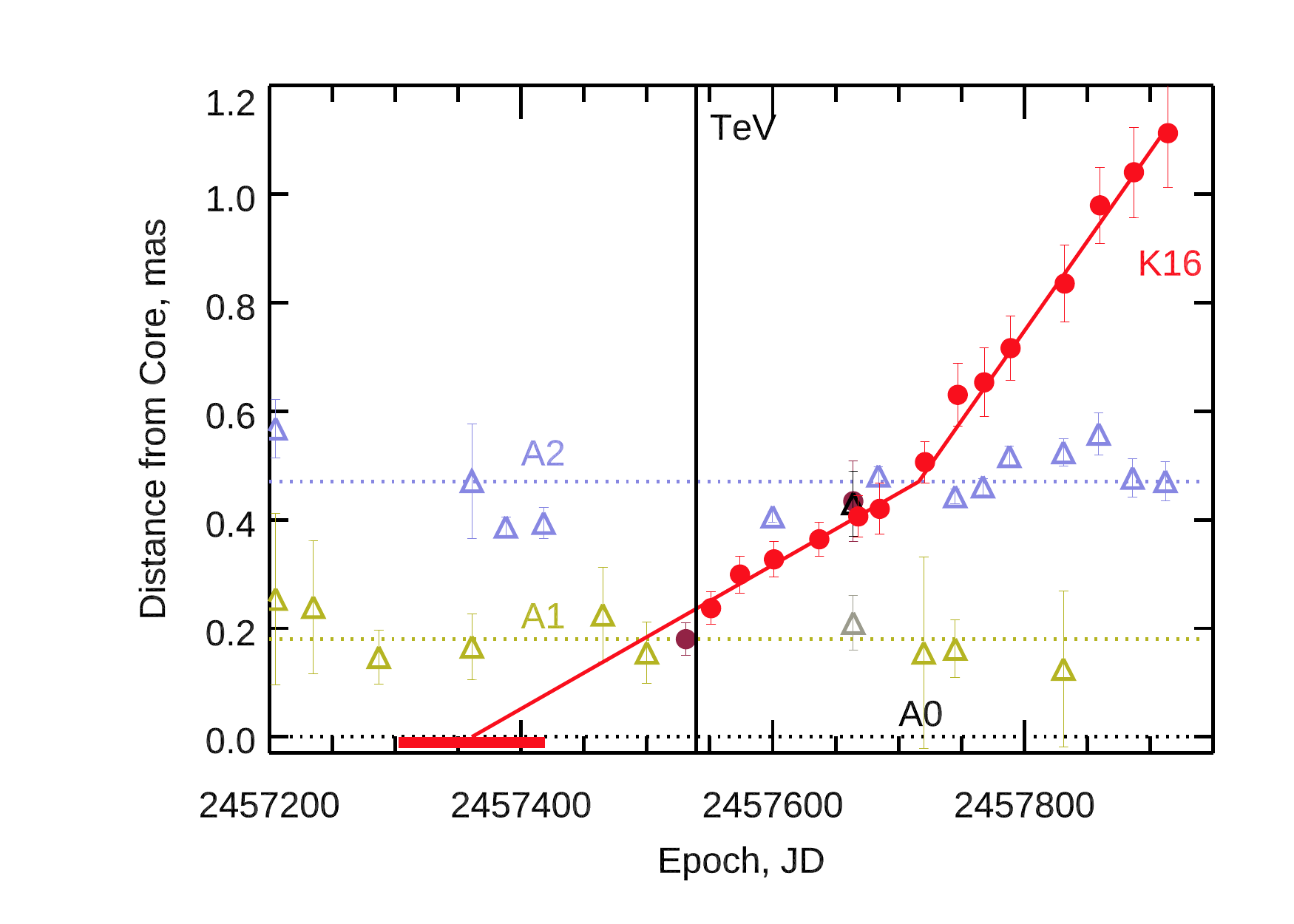}
\end{minipage}
~
\begin{minipage}[t]{0.49\textwidth}
\centering
\includegraphics[width=0.99\textwidth,trim=30 0 20 25,clip]{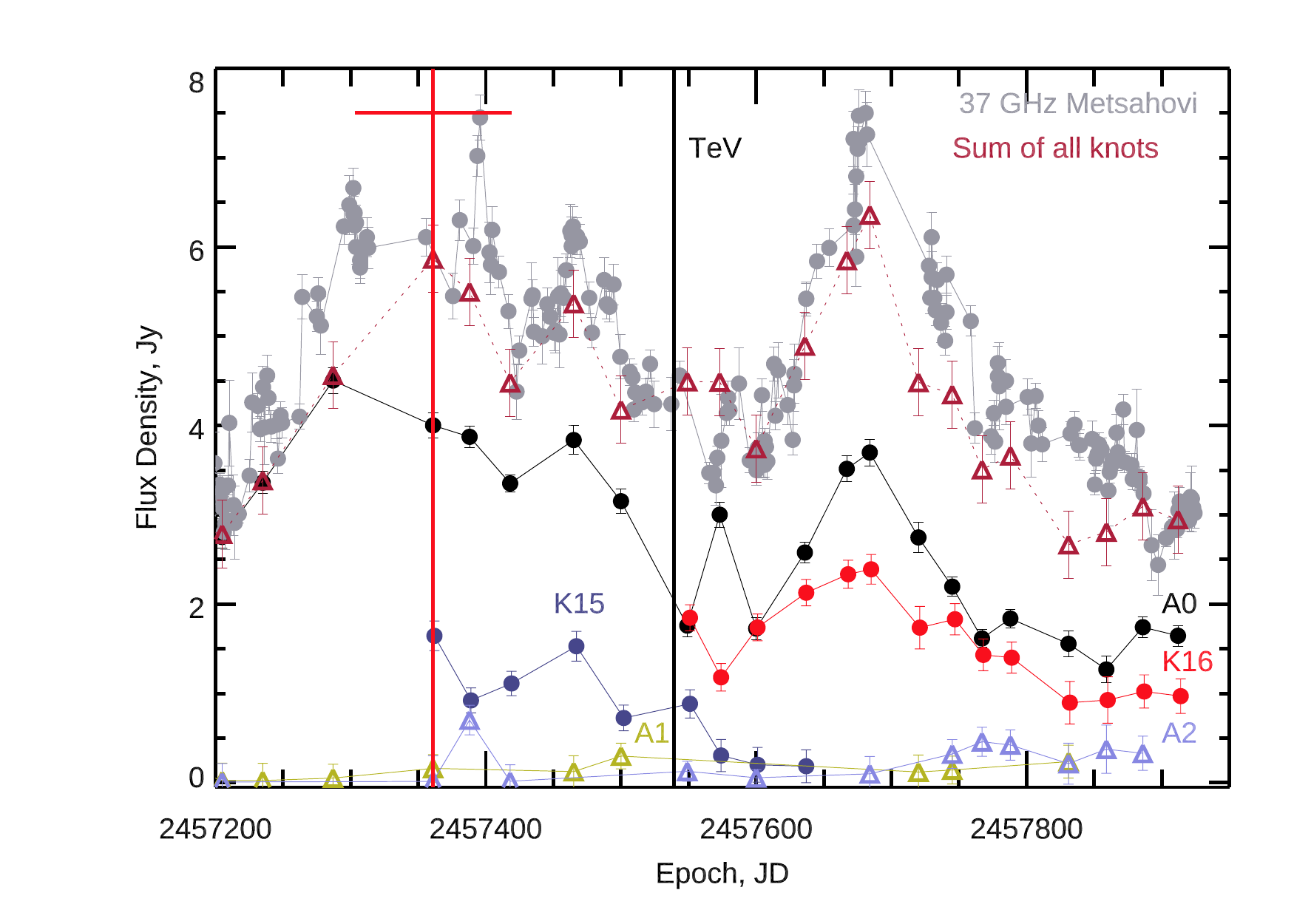}
\end{minipage}
\caption{{\it Left:} Separation of knots K16 (red/brown circles at 7~mm/3~mm, respectively) from the core A0 (black dotted line) according to the 43~GHz and 86 GHz maps; the red lines approximate the motion of K16; dotted yellow and blue lines show the average positions of A1 (yellow/gray triangles at 7~mm/3~mm) and A2 (cyan/black triangles at 7~mm/3~mm); the red line segment at the position of $A0$ indicates the 1$\sigma$ uncertainty of the ejection time of K16. 
{\it Right:} Light curves of the core A0 (black), A1 (yellow), A2 (cyan), K15 (blue), K16 (red), the sum of all components (magenta) at 43~GHz, and from the entire source at 37~GHz (gray). The red vertical line indicates the ejection time (and its $1\sigma$ uncertainty) of K16 from A0. 
In both panels, the black vertical line marks the VHE \g-ray flare.}  
\label{fig:knots}
\end{figure*}
\begin{figure}[th]
\centering
\includegraphics[width=0.48\textwidth]{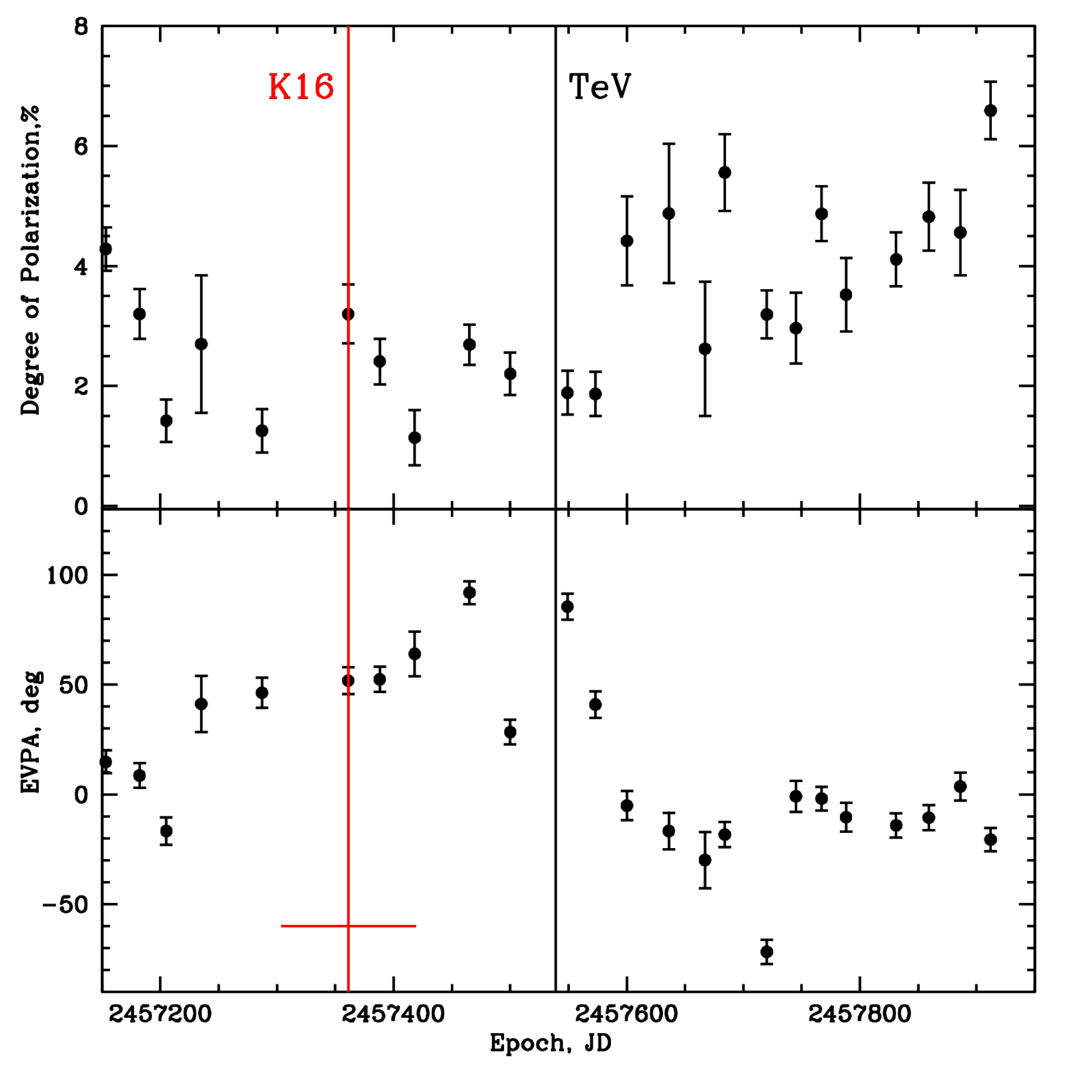}
\caption{{\it Top:} Degree of linear polarization of the core, A0, computed from the VLBA data at 43 GHz. 
{\it Bottom:} Position angle of polarization of the core. The red vertical line marks the time of ejection of K16 with its 1$\sigma$ uncertainty. The black vertical line marks the VHE \g-ray flare.} 
\label{fig:PolA0}
\end{figure}
\begin{figure*}[th]
\begin{minipage}[t]{0.36\textwidth}
\centering
\includegraphics[width=0.99\textwidth]{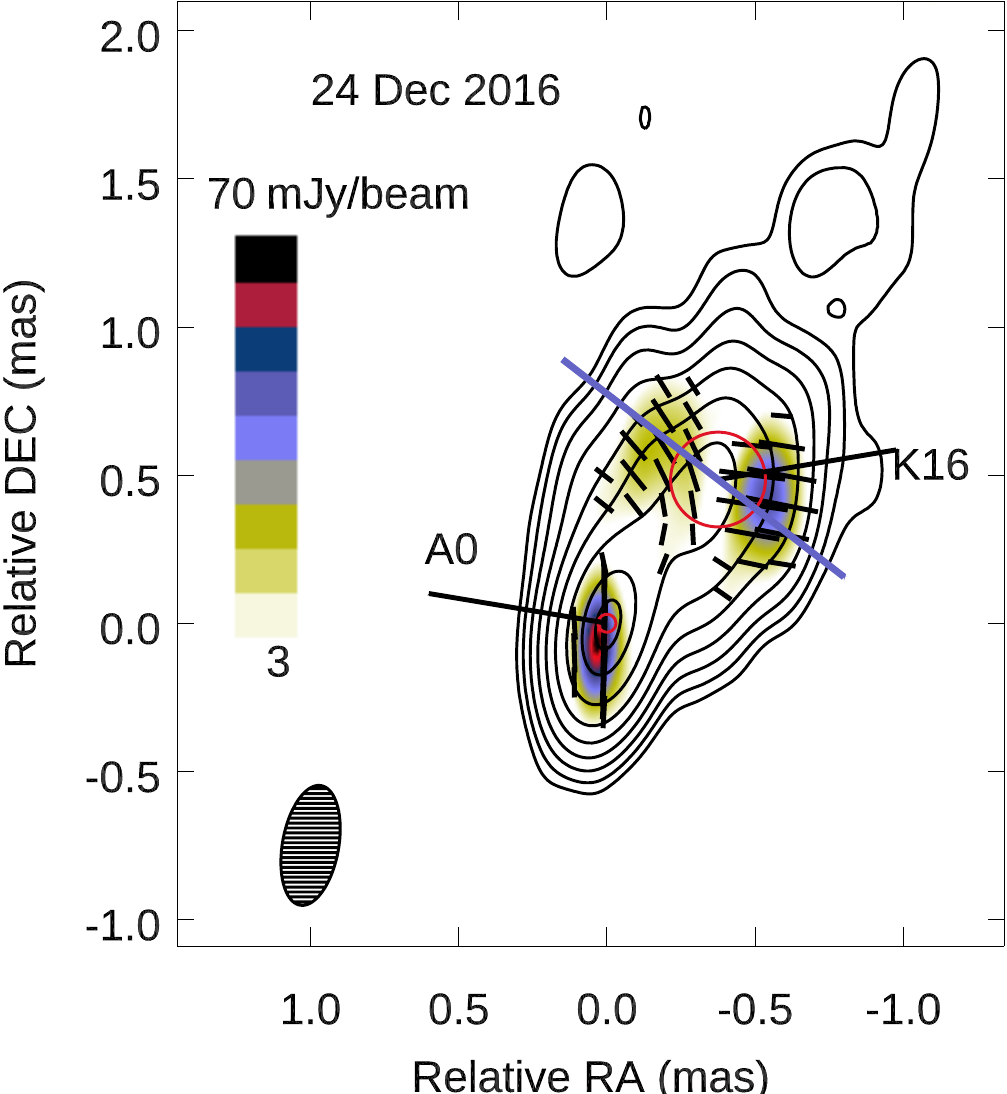}
\end{minipage}
~
\begin{minipage}[t]{0.61\textwidth}
\centering
\includegraphics[width=0.99\textwidth]{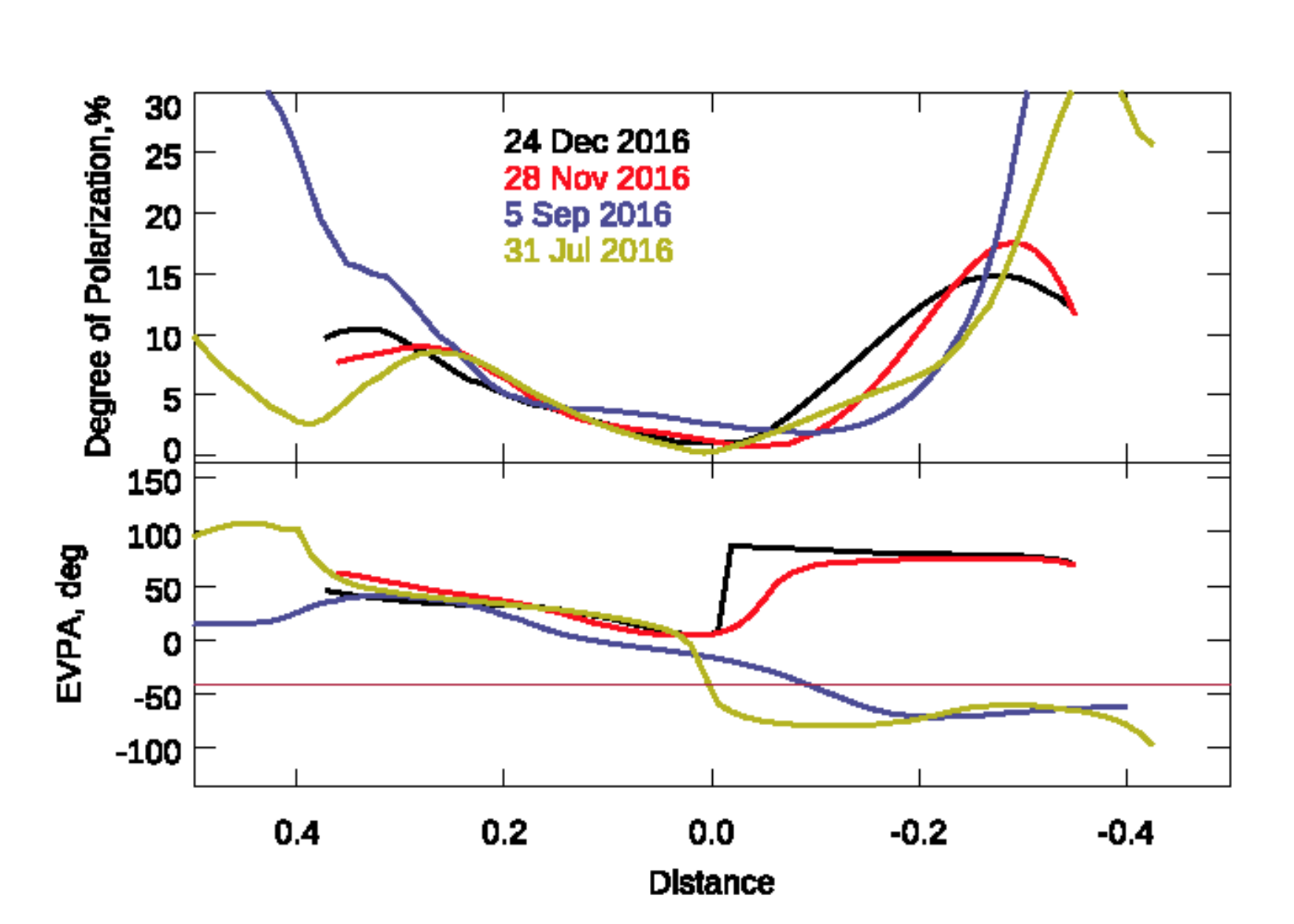}
\end{minipage}
\caption{{\it Left:} Total (contours) and polarized (color scale) intensity images of \source\ at 43 GHz. The total intensity peak is 2357 mJy/beam and the polarized intensity peak is 69 mJy/beam; the contours are 0.5\%, 1\%, 2\%,\ldots 64\% and 95\% of the total intensity peak.
The red circle indicates the position of K16 according to modeling; the blue line across K16 and perpendicular to the jet direction shows the profile line for calculating polarization parameters. 
{\it Right:} Profiles of degree ({\it top}) and position angle ({\it bottom}) of polarization at different epochs; the magenta horizontal line indicates the average jet direction. 
}
\label{fig:PolK16}
\end{figure*} 
%
\source\ has been monitored monthly with the Very Long Baseline Array (VLBA) at 43\,GHz (7\,mm) under the VLBA-BU-Blazar program\footnote{\url{http://www.bu.edu/blazars/VLBAproject.html}} since June 2007. The resulting total and polarized intensity images of the FSRQ have been analyzed at 16 epochs from March 2016 to June 2017, the period relevant to the VHE \g-ray activity reported here.
The data reduction for the VLBA analysis was performed using the Astronomical Image Process System (AIPS, version 31DEC17) and {\it Difmap} (version 2.4) software packages in the manner described by \cite{J17}. Table~\ref{tab:Obs} gives information about the observations at each epoch: number of antennas, restoring beam dimensions and position angle, and flux density correction factor, $f_{amp}$. The flux density correction factor at a given epoch was derived by calculating the total flux densities in the images of sources that have weak emission outside the angular size range of the VLBA images at 43~GHz (PKS 0235+164, S5 0716+71, PKS 1055+01, B2 1156+29, and PKS 1749+096). These sources were observed alongside \source\ in the program. The value for each source was compared with the total flux density measured at 37~GHz at the Mets\"ahovi Radio Observatory (Aalto University, Finland) within 1-2 days of the VLBA observation. The final factor, $f_{amp}$, which is the average of correction factors over the sources, was applied to the \source\ images to adjust the flux density scale.

The total intensity images were modeled by components with circular Gaussian brightness distributions, with each component (knot) characterized by the following parameters: flux density, $S$, distance from the core, $r$, position angle with respect to the core, $\Theta$, and angular size of the component, $a$ (FWHM of the Gaussian). The parameters used correspond to the best fit between the model and data with respect to the $\chi^2$ value provided by {\it Difmap}. The uncertainties of the parameters were calculated using the formalism given in \cite{J17}, which relates the uncertainties to the brightness temperature of the respective knot. The analysis of polarized intensity images was carried out with the {\it Interactive Data Language} (IDL, version 8.6.1)  software, using I, Q, and U Stokes parameter maps produced via {\it Difmap} to calculate the degree of polarization and electric vector position angle (EVPA). 

Fig.~\ref{fig:vlba16} shows a sequence of the VLBA images representing the jet behavior following the VHE \g-ray activity. 
There are five main components in the jet, designated in Fig.~\ref{fig:vlba16} as A0 (the core), A1, A2, K15, and K16. The core, A1 and A2 are presumably stationary features of the jet, while knots K15 and K16 move with respect to them. Table~\ref{tab:KParm} gives the parameters of the knots. The appearance of knot K15 and its possible association with the previous VHE \g-ray outburst of \source\ in May 2015, was discussed in \cite{aMea17}. The parameters of K15 listed in Table~\ref{tab:KParm} are based on additional epochs to those in \cite{aMea17}; they confirm the properties of K15 reported previously.

\source\ was also observed with the Global Millimeter-VLBI Array\footnote{\url{www.bu.edu/blazars/vlbi3mm/index.html}} (GMVA) at 3\,mm (86 GHz) at two epochs close to the VHE \g-ray event (May 22 and October 2, 2016), with a resolution of $\sim$70\,$\mu$as. The data reduction was performed in the same manner as described in \cite{Carolina19}, which resulted in total and polarized intensity images shown in Fig.~\ref{fig:vlba3mm}. 
There are two contemporaneous images at different frequencies (3\,mm on May 22, 2016, and 7\,mm on June 11, 2016). They reveal the same main jet features: the core A0 and two knots positioned on the east and west sides of the core that are associated with K15 and K16, respectively. In addition, the 3\,mm image on October 2, 2016, shows two knots besides A0 and K16, which can be associated with stationary knots A1 and A2 according to their positions.

Fig.~\ref{fig:knots} {\it left} follows the separation of knot K16 from the core. The figure shows also quasi-stationary knots, A1 and A2, detected at some epochs, which are identified with quasi-stationary features A1 and A2, respectively, reported previously by \cite{J01,jo05,J17} based on similarity of their parameters. As can be seen in Fig.~\ref{fig:knots} {\it left}, K16 appears to accelerate beyond A2. In order to investigate whether the break in the motion of K16 is statistically significant, two mathematical representations have been fitted to the measured distance values of K16 from A0 as function of time: a straight line (2 free parameters) and a broken line (with 4 free parameters). While the $\chi^2/N_{dof}=17.1/12$ obtained with the linear fit is statistically acceptable, a much lower value of $\chi^2/N_{dof}=1.4/10$ is obtained for the broken linear fit. 
The $\chi^2$ values are likely underestimated due to the involvement of a correlated systematic error connected with the amplitude calibration, which dominates the uncertainties in the brightness temperature calculations. Nevertheless, the large difference in the computed $\chi^2$ points to a better description of the movement of knot K16 by the broken line. The break occurs on JD~$2457716\pm 26$ when K16 is at the same distance from the core as A2. The apparent speeds of K16 before and after the break are $\beta_{app}=10.8 \pm 2.5$ and $\beta_{app}=25.8 \pm 2.9$, respectively, in units of $c$. The ejection of K16 from the core took place on JD~$2457361 \pm 58$ and the passage of K16 through A1 is estimated on  JD~$2457497 \pm 46$. 

Fig.~\ref{fig:knots} {\it left} also presents the positions of knot K16 with respect to the core in the 3\,mm images, under the assumption that the locations of the
3\,mm and 7\,mm cores are the same. It is possible that the VLBI core at 3\,mm is located closer to the central engine than the 7\,mm core owing to frequency-dependent opacity \citep{Konigl82}. However, positions of K16 at 3\,mm agree very well with the initial (linear) motion derived from the 7\,mm images, implying that any separation between the 3\,mm and 7\,mm cores is significantly less than the resolution of the 7\,mm images. 

Table~\ref{tab:KParm} gives the physical parameters (Lorentz factor, $\Gamma$, Doppler factor, $\delta$, and viewing angle, $\Theta_\circ$) for K15 and K16 using the approach suggested by \cite{jo05}. 
The apparent speed $\beta_{app}$ in the host galaxy frame depends on two parameters, $\Gamma$ and $\Theta_\circ$, according to $\beta_{app}=\beta\sin(\Theta_\circ)/[1-\beta\cos(\Theta_\circ)]$, where $\beta=\sqrt{1-\Gamma^{-2}}$ is the velocity in units of the speed of light. 
A change of the apparent speed can be caused by a variation in either of these parameters. The values of $\Gamma\sim 23$ and $\Theta_\circ\sim0.6^\circ$ of K16
listed in Table~\ref{tab:KParm} are obtained near the core where radiative energy
losses dominate over adiabatic losses \citep{jo05}. Unfortunately, the method cannot be used farther down the jet where adiabatic losses become important. However, the measurement of $\beta_{app}\sim 26$ of K16 observed later implies that the Lorentz factor of K16 becomes $\Gamma\ge 26$. If a minimum increase of Lorentz factor to $\Gamma=26$ is assumed, then $\beta_{app}\sim 26$ requires $\Theta_\circ\sim2.2^\circ$. This is a significant change of the viewing angle with respect to the value derived near the core. The change is comparable with the opening angle of the jet, $\sim 1^\circ$ \citep{J17}, and should be noticeable in projection; 
however, the projected position angle, $\Theta$, of K16 does not change significantly. One
possibility is that the change in jet direction is along the same position angle, as has been found to be the case between the parsec and kiloparsec scale jet of \source\ \citep{Homan2002}. If, instead, the viewing angle of the feature were to remain stable during the epochs discussed here, K16 would need to have accelerated from $\Gamma\sim 23$ to $\Gamma\sim 38$. Although the MOJAVE survey \citep{HOMAN15} and VLBA-BU-BLAZAR program \citep{J17} have found that, statistically, relativistic jets of quasars accelerate (increase in Lorentz factor) gradually on scales from several to 100\,pc, the change of the speed of K16 appears to occur specifically at the location of quasi-stationary feature A2. Note, however, that deceleration of a knot was found previously at the same position \citep[knot B4 in][]{J17}. In any case, a change of knot speeds after passage through the location of A2 supports the hypothesis that an interaction between the moving and quasi-stationary features occurred.

Agreement between the timing of passage of K16 through A1 and the VHE \g-ray event makes the stationary feature especially interesting. The average distance of A1 is $\sim$0.18\,mas downstream of A0, which is translated into a de-projected distance of $\sim 40\,$pc for an average viewing angle of the jet of $1.2^{\circ}$ \citep{J17}. The distance between the black hole and VLBI core at 43 GHz in several blazars is $\sim 10\,$pc,
as reported by, e.g., \cite{fromm15,karaman16}. In the case of \source, this appears as a reasonable assumption, given that the $15\,$GHz VLBI core has been inferred to be located $\sim$18\,pc from the black hole \citep{p12}. Therefore, the distance of A1 from the black hole is $\sim 50\,$pc. Uncertainties in the jet viewing angle, angular distance to A1 (cf. Table~\ref{tab:KParm}), as well as $\sim 50\%$ relative uncertainty in the distance of the 43\,GHz core, translate to a large uncertainty of $\sim 15-20\,$pc of this distance estimate, which is nonetheless incompatible with the typical BLR or dusty torus (DT) distances, which are $\sim$1-3~pc from the central engine \citep[e.g.,][]{Kaspi2000}. 

The same VLBA data at 7\,mm as discussed here have been analyzed by \cite{Park19}. They have also found two moving components, K15 and J15, which in general correspond to K15 and K16, respectively, of the present study. There are some differences in the identification of knots and fitting of polynomials yielding different ejection times of the knots. However, independent of these differences, according to Fig.~4 in \cite{Park19} the time of passage of J15 through stationary feature A1 is JD~$\sim$2457500 (April 21, 2016), which agrees within 1$\sigma$ uncertainty with that of K16 derived above. While \cite{Park19} do not identify the quasi-stationary knots A1 and A2, it is a justifiable assumption that they are present in their modeling as well: e.g., their knot J15b can be identified with the stationary feature A2, and an unidentified knot near the core, seen in the last three epochs in their Fig.~2, has parameters similar to those of feature A1.   

Fig.~\ref{fig:knots} {\it right} presents the light curves of the moving knots, core, and stationary features A1 and A2, along with the Mets\"ahovi light curve of \source\ at 37~GHz and the flux densities of the sum of all components detected at a given epoch. Both the core and 37\,GHz light curves exhibited a significant increase in flux density at the time of ejection of K16. The light curves of K15 and K16 have been used to estimate the timescale of variability, $\tau$, of the knots, which is $\sim 0.1$\,yr, similar for both knots. This value is applied to calculate the Doppler factor, $\delta$, of K15 and K16 given in Table~\ref{tab:KParm}.
Knot K16 possesses higher Doppler and Lorentz factors than K15. Although the knots have similar viewing angles, according to the VLBA and GMVA images they should be located on opposite sides of the projected jet axis. 
The ratio of the maximum flux density of K16 to that of K15, $\sim 2$, agrees with $\delta$ being the main factor for the modulation of the fluxes of knots, with $S_\nu\propto\delta^{3+\alpha}$, where $-\alpha$ is the slope of the optically thin flux density spectrum.

The magnetic field geometry of the jet can be inferred by analyzing the polarization parameters of the core and K16. Fig.~\ref{fig:PolA0} shows the degree, $P$, and electric-vector position angle, EVPA, of linear polarization of the core. The core polarization is moderate around the ejection time of K16, and the EVPA rotates as K16 separates from the core, with a large swing of $\sim$70$^{\circ}$ just after the VHE \g-ray event. Rotation of optical EVPAs during VHE \g-ray events have been reported for a number of blazars, e.g., in S5 0716$+$71 \citep{Ahnen18}. In order to study the magnetic field of K16, profiles of the polarization parameters have been constructed along a line drawn through the center of the knot and perpendicular to the jet. This line moves away from the core with time, as K16 does. An example is shown in Fig.~\ref{fig:PolK16} {\it left}. The profiles of $P$ and the EVPA at some epochs are plotted in Fig.~\ref{fig:PolK16} {\it right}. The profiles show a low degree of polarization in the center of K16, and a significant increase of $P$ closer to the edges of the feature, where the EVPA is oblique or perpendicular to the jet direction. Such polarization properties are typical for a spine-sheath structure of the jet, as discussed, e.g., in \cite{NICK15}. A low polarization degree in the center of K16 indicates a very turbulent magnetic field in the spine. 

%
%
\section{Results and discussion} \label{sec:corstd}
%
%
\subsection{Variability study} \label{sec:varia}
\begin{figure}[th]
\centering
\includegraphics[width=0.48\textwidth]{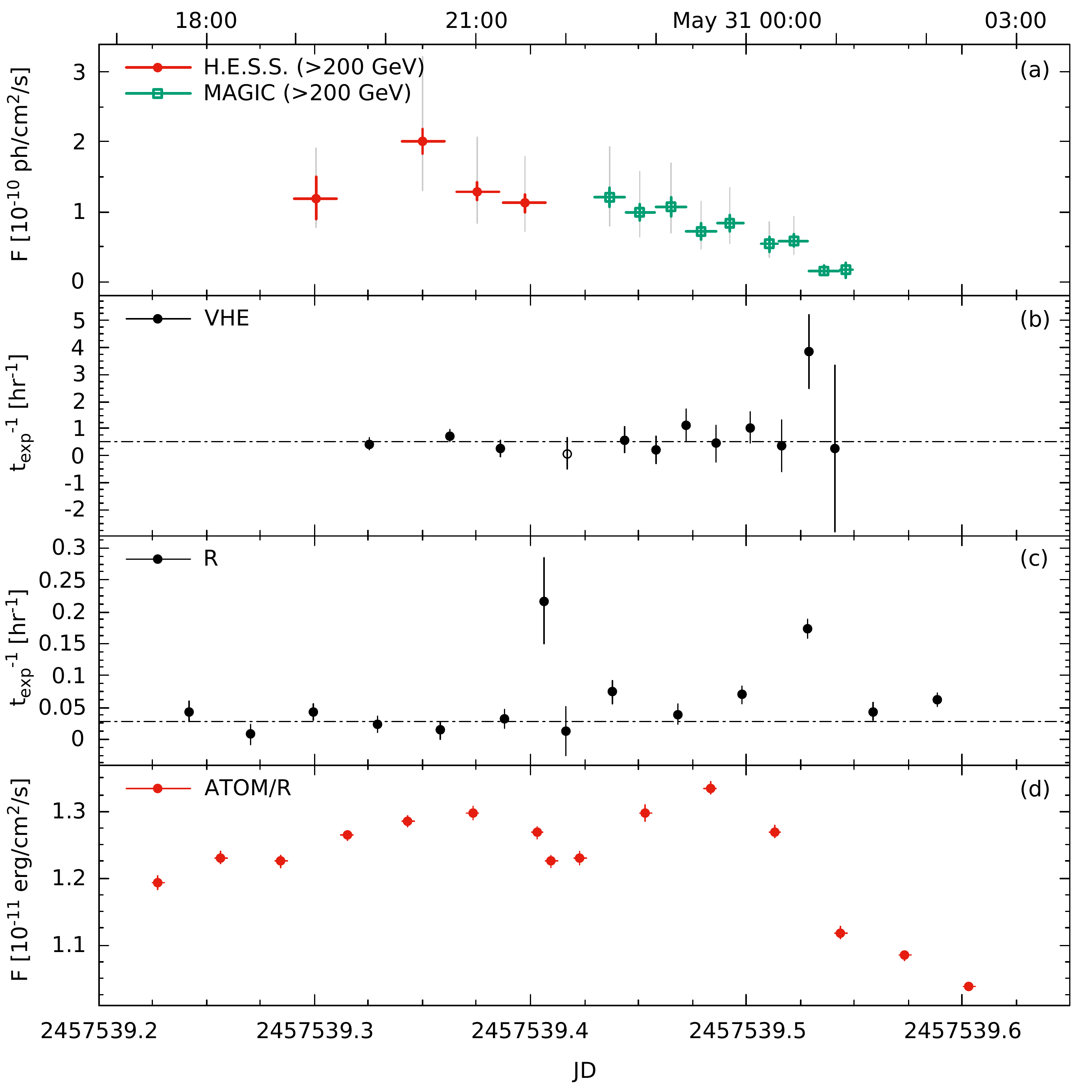} 
\caption{Lightcurves and time scales of \source\ of the flare night, JD~2457539. 
{\bf (a)} VHE \g-ray lightcurve from H.E.S.S. (red points) and MAGIC (green open squares). The binning is 28\,min and 20\,min for the \hess\ and MAGIC lightcurves, respectively. The grey bars mark the systematic uncertainties. 
{\bf (b)} Inverse exponential time scales between subsequent points of the VHE \g-ray lightcurve. Systematic uncertainties have been considered for the open symbol, as it marks the step from H.E.S.S. data points to MAGIC data points. The grey dash-dotted line marks the harmonic mean with $\bar{t}_{\rm exp}^{-1}\sim 0.56\,$hr$^{-1}$. 
{\bf (c)} Inverse exponential time scales between subsequent points of the R-band lightcurve. The grey dash-dotted line marks the harmonic mean with $\bar{t}_{\rm exp}^{-1}\sim 0.03\,$hr$^{-1}$. 
{\bf (d)} Optical R-band lightcurve from ATOM showing individual exposures of 8\,min duration each. 
}
\label{fig:timescales}
\end{figure}
\begin{figure}[th]
\centering 
\includegraphics[width=0.48\textwidth]{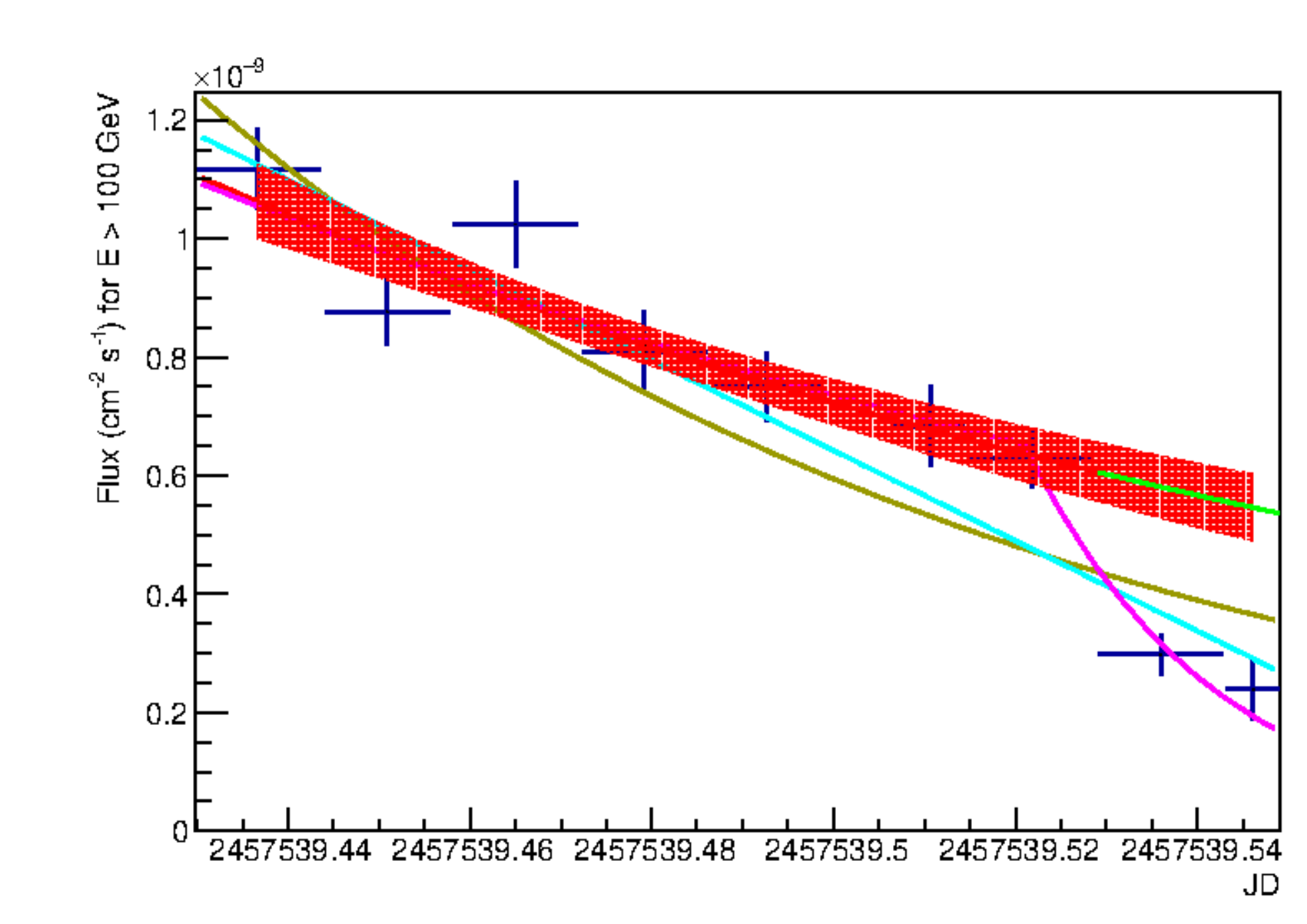}
\caption{Various fits to the MAGIC lightcurve above 100\, GeV on JD~2457539.
Olive green line shows the exponential fit, magenta the broken exponential and cyan the linear fit. 
The exponential fit to the first 7 points of the light curve is shown with red line, with the red shaded region showing the 68\% C.L. uncertainty band and the green line shows its extrapolation to the end of the light curve. 
}
\label{fig:lcfits}
\end{figure}
From the VHE \g-ray lightcurves in Figs.~\ref{fig:mwl_lc_all}(a) and \ref{fig:mwl_lc_57538}(a) the flare duration $t_{\rm dur}$ can be constrained to at most $48\,$hr at these energies. The HE \g-ray spectral index (Fig.~\ref{fig:mwl_lc_all}(c)) and the optical lightcurve (Fig.~\ref{fig:mwl_lc_all}(d)) suggest that the flare may have started even a day earlier implying a maximum duration of $72\,$hr. The lightcurves also suggest a slow rise (up to 2 days) compared to a faster decay (potentially within a single night). 
The low cadence of observations in the VHE \g-ray and optical bands, as well as the small statistics in the HE \g-ray band do not allow one to draw firmer conclusions on the full extend of the flare. 

It is not even clear, if the maximum in the VHE \g-ray lightcurve corresponds to the true peak of the flare or a secondary peak. While the HE \g-ray index indicates that the global peak probably took place within this 1-day bin, which is centered on the \hess\ observation window, any time between the \hess\ observations on JD~2457538 and JD~2457539 is possible. The same is true for the optical lightcurve. The average fluxes of nights JD~2457538 and JD~2457539 are equal. However, the former measurement is only from a single exposure, while in the latter night 15 exposures were conducted and display clear variability. Interestingly, the VHE \g-ray and optical lightcurves of JD~2457539 (Fig.~\ref{fig:mwl_lc_57538}(a) and (c)) show a different evolution. While the VHE \g-ray lightcurve is suggestive of a single peak and a subsequent decay, the optical lightcurve exhibits two peaks with the second one brighter than the first one. The decay after the second peak appears steeper than any other optical flux variation of the night.

The variability in \source\ is further investigated using a phenomenological description assuming exponential variability (a test with a linear variability definition giving fully consistent results is provided in App.~\ref{sec:som2var}) of the emission with the goal to determine the characteristic variability time scales and potential changes thereof.
The exponential variability time scale between subsequent flux points $F_i = F(t_i)$  is defined as

\begin{align}
 t_{\rm exp} = \frac{t_{i+1}-t_{i}}{|\ln{F_{i+1}}-\ln{F_{i}}|} \label{eq:texp}.
\end{align}
The inverse of this time scale is plotted in Fig.~\ref{fig:timescales}(b) for the VHE \g-ray lightcurve. 
The time scales are compatible with the harmonic mean of $\bar{t}_{\rm exp}\sim 1.8\,$hr$\,\sim 108\,$min except for the next to last time step. 
Here, the exponential time scale $t_{\rm exp}= (16\pm 5\stat)\,$min indicates a deviation from the harmonic mean of $2.4\sigma$.

\subsubsection{Assessing the significance of the VHE \g-ray flux drop}
In order to verify the steep decline of the flux in the penultimate time step, further tests have been conducted on the MAGIC lightcurve. The energy threshold of $200\,$GeV used for the lightcurve in Fig.~\ref{fig:timescales}(a) was selected to facilitate a comparison of the  \g-ray emission above the same analysis threshold for both H.E.S.S. and MAGIC. 
However, due to the steepness of the source spectrum the choice of the common threshold is not optimal for the investigations of lightcurve features. 
By applying the new energy threshold of $100\,$GeV, the total excess of 
\g-ray events over the background obtained on the flare night becomes 5.5 times larger and 2.3 times more significant  ($4443 \pm 86$ compared to $805 \pm 36$ excess events).
The observed drop in the flux is by more than a factor of two, therefore it cannot be explained by possible variations in the atmospheric transmission. 
Additionally, the coincident rate of gamma-like events from background control regions does not show a similar drop.

Several fitting functions have been tested on the $>100\,$GeV MAGIC lightcurve (see Fig.~\ref{fig:lcfits}). 
Due to the fast variability, for the fits the average value of the function inside the light curve bin is used rather than the value corresponding to the bin center. 
The lightcurve fitted with an exponential decay in the time range covered by the MAGIC observations results in a flux halving time of $(95\pm7\stat)$\,min. 
However the corresponding $\chi^2/N_{dof}=44.0/7$ results in chance probability of only $2.2\times 10^{-7}$ falsifying the hypothesis of a simple exponential decay. 
Fitting the MAGIC lightcurve with a broken exponential one obtains $\chi^2/N_{dof}=8.7/5$, i.e. the broken exponential is preferred over a simple exponential at the level of 5.5$\sigma$. 
The resulting fit shows the change of the exponential time scale from $(180\pm40\stat)$\,min before the break to $(21\pm7\stat)$\,min. 
It was also tested if the lightcurve can be described with another function, in particular with a linear decline. 
However the resulting $\chi^2/N_{dof}=26.5/7$ disfavours such a fit. 
Importantly, as the above fits are done from the light curves of a single instrument, and due to large amplitude of the observed variability, the uncertainties of the exponential time scales are dominated by the statistical rather then by systematic uncertainties.

An independent method has been used to confirm the significance of the break in the lightcurve. 
A fit of the MAGIC lightcurve has been conducted with a single exponential in a limited range (the first 7 points). 
Such a fit gives a good description of the beginning of the lightcurve with $\chi^2/N_{dof}=7.0/5$. 
Then the fit is extrapolated to the last two points. 
The corresponding $\chi^2/N_{dof}=35.9/2$ is calculated taking into account both the uncertainty of the fit extrapolation and of the last two flux measurements.
Conservatively applying 8 trials (motivated by the fact that the light curve could be broken between each two consecutive points) one obtains a chance probability of $1.3 \times 10^{-7}$ corresponding to $5.1\sigma$, compatible with the preference of the broken exponential over a single one. 

As a final check of the significance of the lightcurve feature a toy MC study has been performed. 
The assumption has been used that the source exhibits an exponential decay following the time scale and the flux normalization of the single exponential fitted to the first 7 points of the LC. 
For each bin the integrated flux from this fit has been calculated, and using the collection area and effective time, a number of \g\ events has been computed. The latter is used as expected value of a Poisson distribution from which random numbers of \g\ events are drawn. 
The number of background events is also computed from Poissonian distribution following the measured background rates. 
The excess is then converted into the flux using the same collection area and effective time. 
The uncertainties of the flux take into account both the Poissonian fluctuations and the uncertainty of the collection area (the uncertainty of the effective time is negligible). 
Then, $2\times 10^{8}$ such light curves have been generated and the same analysis has been repeated as in the second method. 
A fraction of $4.7 \times 10^{-7}$ lightcurves achieve the $\chi^2$ value greater than the one obtained from the data. 
Correcting conservatively for 8 trials the chance probability is $3.8 \times 10^{-6}$, corresponding to $4.5\sigma$. 

\begin{figure*}[th]
\centering 
\includegraphics[width=0.49\textwidth]{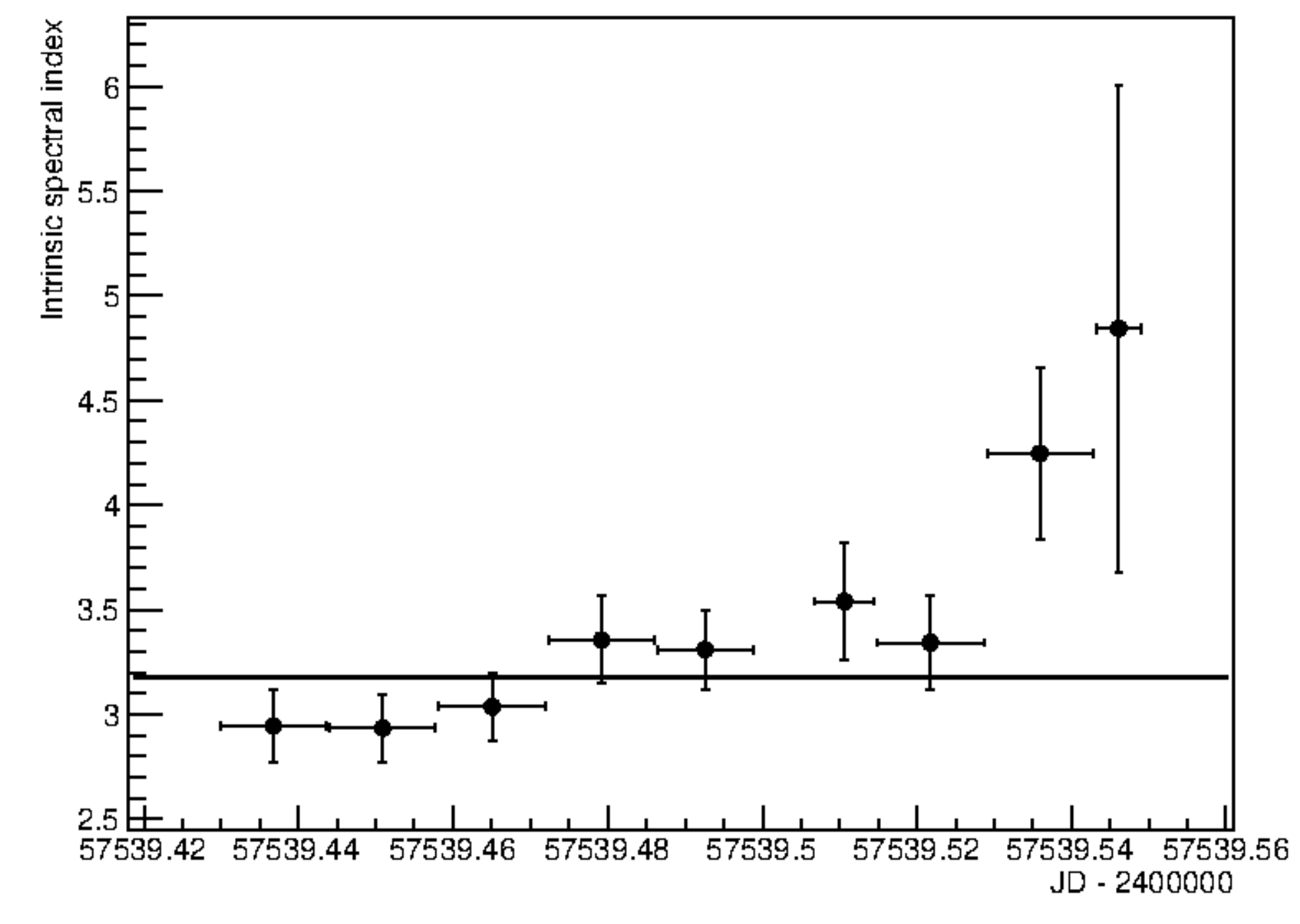}
\includegraphics[width=0.49\textwidth]{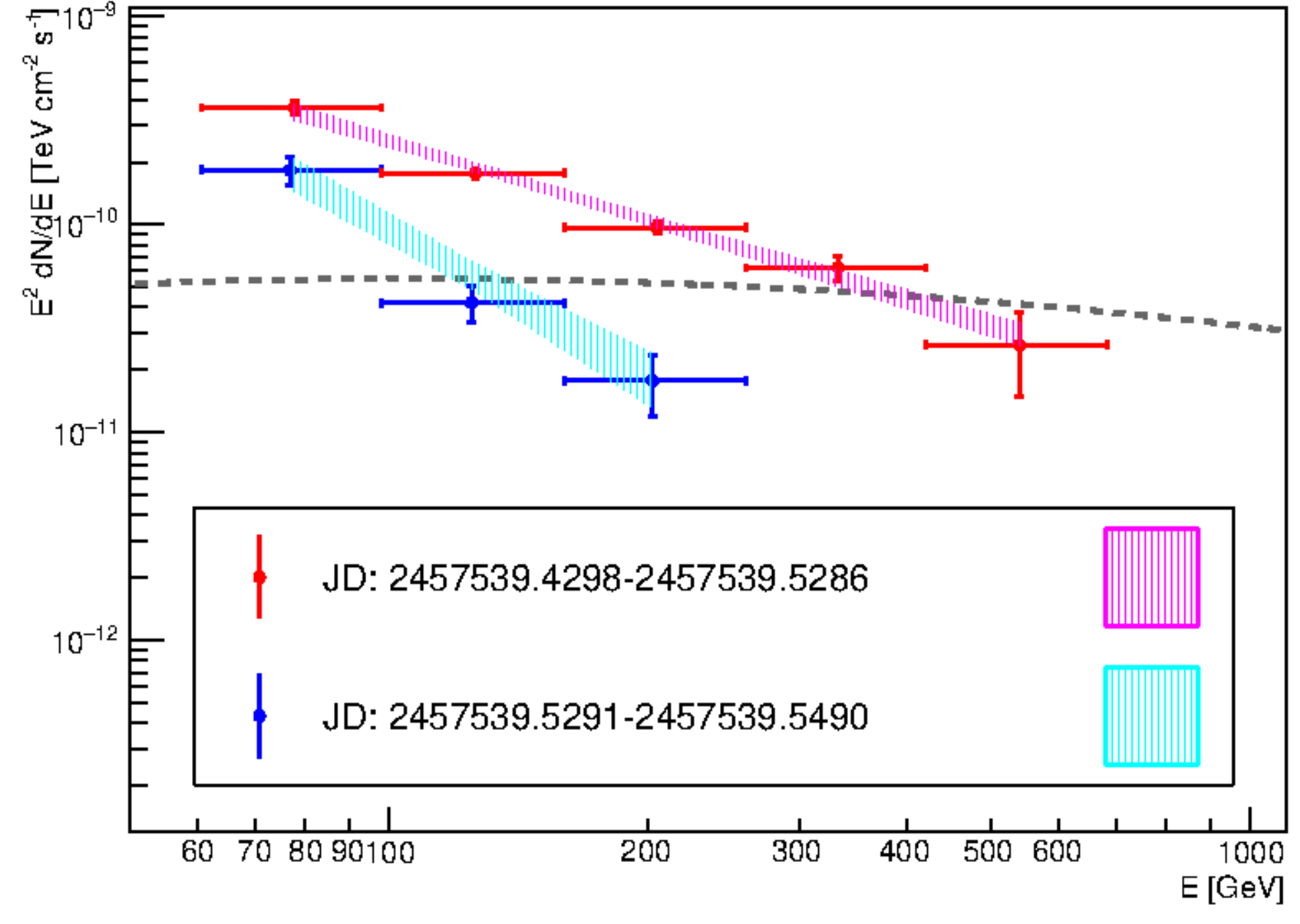}
\caption{{\it Left:} evolution of the intrinsic spectral index as measured by MAGIC on the flare night with the line indicating the average value.
{\it Right:} MAGIC spectra of the source obtained before (red points and magenta shaded region) and after (blue points and cyan shaded region) break in the light curve. The dashed line shows the Crab nebula spectrum for a reference \citep{2015JHEAp...5...30A}. 
}
\label{fig:sed_jump}
\end{figure*}

In order to test if the variability in flux is accompanied with spectral variability the intrinsic (i.e. corrected for EBL absorption) spectral index has been computed for each 20-min long observation run of MAGIC.
The results are shown in the left panel of Fig.~\ref{fig:sed_jump}. 
The first 7 runs, even if a hint of spectral softening can be seen, are still consistent with a constant spectral index ($\chi^2/N_{dof}=7.9/6$).
Including the points after the break into the fit, the fit chance probability gets much lower ($\chi^2/N_{dof}=16.9/8$). 
Additionally, the average spectra from the 7 runs of MAGIC before the break and from the 2 runs after the break have been constructed. 
Both spectra are shown in the right panel of Fig.~\ref{fig:sed_jump}.
Consistent with the run-by-run analysis the spectrum obtained from the end of observations with MAGIC appears softer than from the observations before the break in the light curve. 
The spectral index changes from $3.3\pm0.1_{stat}$ to $4.4\pm0.5_{stat}$. 
A softening of the \g-ray spectrum points to a softening of the underlying electron distribution. 

In conclusion, a simple exponential decay of the VHE \g-ray lightcurve is inconsistent with data observed by MAGIC above $100\,$GeV at a $> 4.5\sigma$ level. Instead, it shows evidence of a faster decline of the flux with a time scale of $\sim 20$\,min and a hint of accompanying spectral softening.

\subsubsection{Variability in the optical lightcurve}
For the optical lightcurve, the variability has also been assessed using the phenomenological function in Eq.~(\ref{eq:texp}) with the results presented in Fig.~\ref{fig:timescales}(c). Note that the linear variability function in App.~\ref{sec:som2var} provides consistent results, too. 
The harmonic mean of the optical exponential time scales is $\bar{t}_{\rm exp}\sim 35\,$hr, about a factor 30 longer than the time scales at VHE \g\ rays during the flare. This slow variability time scale may be influenced by a significant background flux in the optical domain from parts of the jet not participating in the flare or from the accretion disk \citep{dAea11}, as the optical flux rises above the adjacent flux levels by only a factor of $\lesssim 2$, see Fig.~\ref{fig:mwl_lc_all}(d).

There are two outliers in the optical timescales (see Fig.~\ref{fig:timescales}(c)), with the second one being more significant than the first one. The first outlier corresponds to a time scale of $t_{\rm exp}= (5 \pm 1)\,$hr. It deviates by $2.8\sigma$ from the harmonic mean. Note that this deviation is strongly influenced by the higher cadence of observations compared to the rest of the lightcurve. For the second outlier, $t_{\rm exp}= (5.8 \pm 0.5)\,$hr is obtained. Here, the deviation is more than $9\sigma$ from the harmonic mean.

\subsubsection{Comparison between the multiwavelength lightcurves}
Both the VHE \g-ray and optical lightcurves exhibit intranight variability, with evidence for a  significant steepening of the flux decay occurring in both bands almost simultaneously near the end of the flare. 
This implies a change in the physical conditions of the emission region, namely from the interplay of injection, acceleration, and cooling to cooling only \citep[e.g.,][their Fig.~10]{sb10}. 
The change in source behavior is further supported by a hint of spectral softening in the VHE \g-ray band coincident with the drop in the observed lightcurve. 

The total flux variation in the VHE \g-ray lightcurve is about an order of magnitude during the flare night, while the optical variation is only about 10\%. This is also reflected in the variability time scales as mentioned above. However, both differences could be influenced by a significant steady background flux in the optical domain -- such as from other regions of the jet, or the accretion disk -- which would reduce the effect of the variable flux component on the total flux. Furthermore, the details in the lightcurves are different. The VHE \g-ray lightcurve exhibits a single peak followed by the decay extensively discussed above. On the contrary, the optical lightcurve displays two peaks with neither being coincident with the peak in the VHE \g-ray lightcurve. The simultaneous steep flux decay near the end of the observations may therefore point to a common termination of the flare processes.

On timescales of a few days, see Fig.~\ref{fig:mwl_lc_all}, the optical lightcurve matches the flux evolution of the HE \g-ray lightcurve within errors. Both lightcurves exhibit similar flux variations of about a factor two. While this correlation can provide insights for modeling attempts, it is noteworthy that it is not a common feature in flares in \source\ \citep[e.g.,][]{nea12,aMea17,pgn19}.  

While it is the first time that a fast flare is observed in VHE \g\ rays in \source, fast flares have been observed in other energy bands in this source before. 
Notably, $20-30\,$min flux doubling timescales -- corresponding to $30-40\,$min exponential variability timescale as used in the present paper -- were derived in HE \g\ rays during a very bright flare in 2011 \citep{Foschini2013,msb19} with peak fluxes exceeding $10^{-5}\,$ph\,cm$^{-2}$s$^{-1}$.
As the HE \g-ray flux is about an order of magnitude lower during the event discussed here, the characteristics of the flares are different despite the similar variability timescale. 

In previous cases of fast variability in the VHE \g-ray band in blazars, the observations revealed mainly either the smoothly raising or falling component of the flare (see \citealp{pks1222} for the FSRQ  PKS\,1222+216, and \citealp{acc19} for the BL Lac object BL Lacertae), 
a combination of individual flares (see \citealp{aHea07} for the BL Lac object PKS\,2155-304 and \citealp{ic310} for the BL Lac object/radio galaxy IC\,310), 
or nearly symmetric flares, without evidence of a faster drop of the emission (see \citealp{mrk501} for the BL Lac object Mrk\,501). 
Therefore, the sudden decrease of the flux in the lightcurve of \source\ marks a rare occurrence of the cessation of a flare observed at VHE \g\ rays. 
A similar event to the 2016 flare of \source\ was observed in the BL Lac object PKS\,2155-304 on MJD~53945.97 (see Fig.~1 in \citealp{2009A&A...502..749A}) and possibly also on MJD~53944 (see Fig.~8 in \citealp{2010A&A...520A..83H}). 
The observation of such a feature in the FSRQ \source\ is particularly interesting, as the BLR absorption of VHE \g\ rays in this class of objects makes it impossible for the emission to originate in the inner, more compact parts of the jet. 

%
\subsection{The $\gamma$-ray spectrum} \label{sec:gspec}
\begin{figure}[t]
    \centering
    \includegraphics[width=0.98\linewidth]{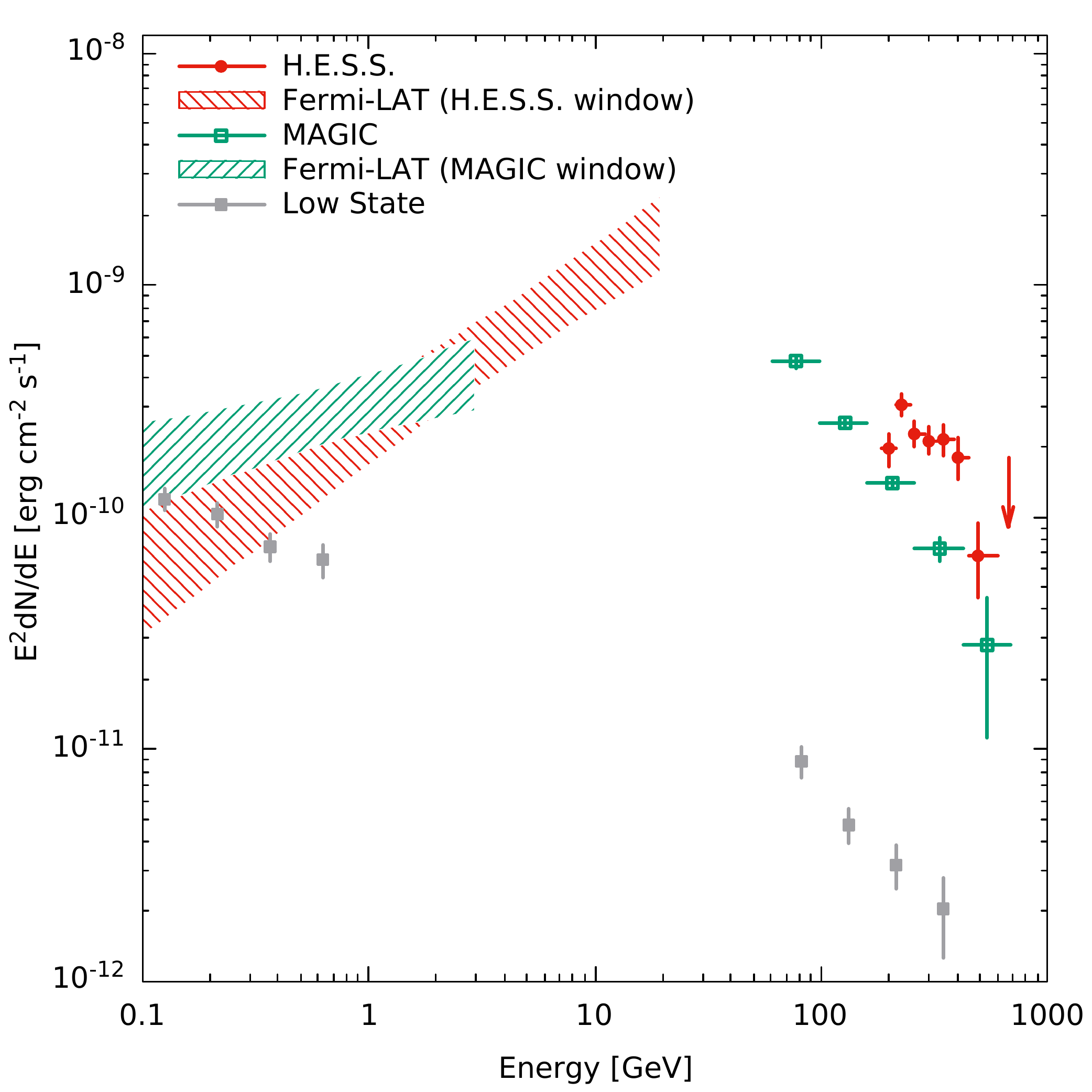}
    \caption{Observed HE and VHE \g-ray spectra of \source\ during the flare night, JD~2457539. The \fermi\ butterflies are integrated over the precise \hess\ (red) and MAGIC (green) observation windows and are plotted until the highest detected photon energy. The VHE \g-ray spectra are corrected for EBL absorption using the model of \cite{frv08}, with the \hess\ spectrum in red and the MAGIC spectrum in green. The grey data points are the low-state spectra from \cite{acc18}. For this spectrum, HE points above 1\,GeV are omitted due to potential bias \citep[for details, see][]{acc18}.}
    \label{fig:spec_gamma}
\end{figure}
The EBL-deabsorbed HE and VHE \g-ray spectra of the peak night (JD~2457539) are plotted in Fig.~\ref{fig:spec_gamma}. The apparent softening of the HE \g-ray spectrum from one observation window to the other, despite being not statistically significant, is probably influenced by the \mbox{(non-)detection} of photons with energies exceeding $10\,$GeV during the observation windows (see Fig. \ref{fig:mwl_lc_57538}(b)). The peak of the $\gamma$-ray SED is located somewhere between $10$ and $60\,$GeV. The grey data points display the low-state spectrum of \source\ \citep{acc18} with a peak position at $100\,$MeV or less. 
During the flare, the peak position shifted by more than a factor $100$ to higher energies. 

The EBL-deabsorbed VHE \g-ray spectra are devoid of curvature. Interestingly, the spectral break between the HE and VHE \g-ray ranges does not change significantly between the observation windows. During the H.E.S.S. window it is $\Delta\Gamma = 1.5\pm0.3\stat$, while during the MAGIC window $\Delta\Gamma = 1.7\pm0.2\stat$. Hence, the cause of the break appears steady during the evolution of the flare. Such a break is absent in the low-state data, with the connection between HE and VHE \g-ray data being smoother than during the flare. 

There are a couple of possible explanations for the break. 
It could correspond to the maximum of the accelerated particle distribution, determined by the maximum efficiency of the acceleration process.
A second possibility could be a break due to Klein-Nishina effects depending on the energies of the particles and the ambient photon field that is inverse Compton scattered to produce the \g\ rays. These possibilities are further discussed in Sec.~\ref{sec:mod}. Here, the focus is on 
a third interpretation, namely absorption of the \g\ rays by a soft photon field located close to the central engine, like the broad-line region (BLR). By assessing the maximum absorption allowed by the data, the distance of the emission region from the black hole can be estimated. The following study is based on \cite{msb19} and \cite{Hea19}.

The detailed \fermi, H.E.S.S. and MAGIC spectra, covering more than 3 decades in energy, enable a simultaneous fit of a set of assumed intrinsic spectra folded with absorption patterns by the BLR.
The latter is derived following the model of \cite{Finke:2016}, which is motivated by reverberation mapping and assumes that accretion disc radiation is absorbed by the BLR clouds and re-emitted as monochromatic lines at fixed distances from the black hole. 

Two geometries of the BLR are implemented in the study. In the \emph{shell} geometry, BLR photons are emitted in infinitesimal thin shells around the black hole, whereas in the \emph{ring} geometry, the BLR photons originate from thin rings orthogonal to the jet axis. 
The model includes emission lines from Ly$\epsilon$ to H$\alpha$ but neglects any contribution from the thermal continuum. Motivated from reverberation mapping, each line has an associated luminosity and is emitted in a shell or a ring at a fixed distance \citep[see Table~5 in][]{Finke:2016}. As input the model requires the black hole mass, $M_{\bullet}$, and the luminosity of the H$\beta$ line, $L(\mathrm{H}\beta)$. For \source, the reverberation-mapping results obtained by \cite{Liu:2006} are used, who find $\log_{10} (M_{\bullet} / M_{\odot}) = 8.2$ with the solar mass $M_{\odot}$, and $L(\mathrm{H}\beta) = 1.77\times10^{43}\,\mathrm{erg}\,\mathrm{s}^{-1}$.
Using the relations summarized in \cite{Finke:2016} between $L(\mathrm{H}\beta)$ and $L(5100\,$\AA), as well as between $L(5100\,$\AA) and the radius of the $\mathrm{H}\beta$, the radius of the Ly$\alpha$ emitting shell or ring is determined, $R_{\mathrm{Ly}\alpha} = 7.69\times10^{16}\,$cm.
Its luminosity is the highest in the model (a factor of 12 higher than $L(\mathrm{H}\beta)$) and therefore responsible for most of the absorption.
The total BLR luminosity in this model is equal to $5.3\times10^{44}\,\mathrm{erg}\,\mathrm{s}^{-1}$ and the Ly$\alpha$ emission accounts for 40\,\% of the total luminosity.
Following \cite{Finke:2016} the optical depths $\tau_{\gamma\gamma}(r,E^\prime)$ are calculated as a function of the distance $r$ of the emission region to the black hole and the $\gamma$-ray energy in the host galaxy frame $E^\prime$. 
At $E^\prime = 50\,$GeV (100\,GeV) the ring geometry results in an optical depth of 1.1 (7.9) when the emission region is located at $0.5R_{\mathrm{Ly}\alpha}$ and decreases to 0.4 (2.0) when the emission region is placed at  $R_{\mathrm{Ly}\alpha}$.

The shell geometry results generally in higher values of the optical depth \citep[compare also Fig.~14 in ][]{Finke:2016}. 
In addition, a BLR model with a ring geometry is tested which has a $L(H\beta)$ luminosity 13 times larger than the value reported by \cite{Liu:2006}. This results in a BLR radius $R_\mathrm{BLR} \sim R_{\mathrm{Ly}\alpha} = 2.6\times10^{17}\,$cm, about 3 times as large as before, which corresponds to the  value used in \cite{aMea14}.
This model is dubbed high luminosity BLR ring geometry.

\begin{figure*}[th]
    \centering
    \includegraphics[width = .49\linewidth]{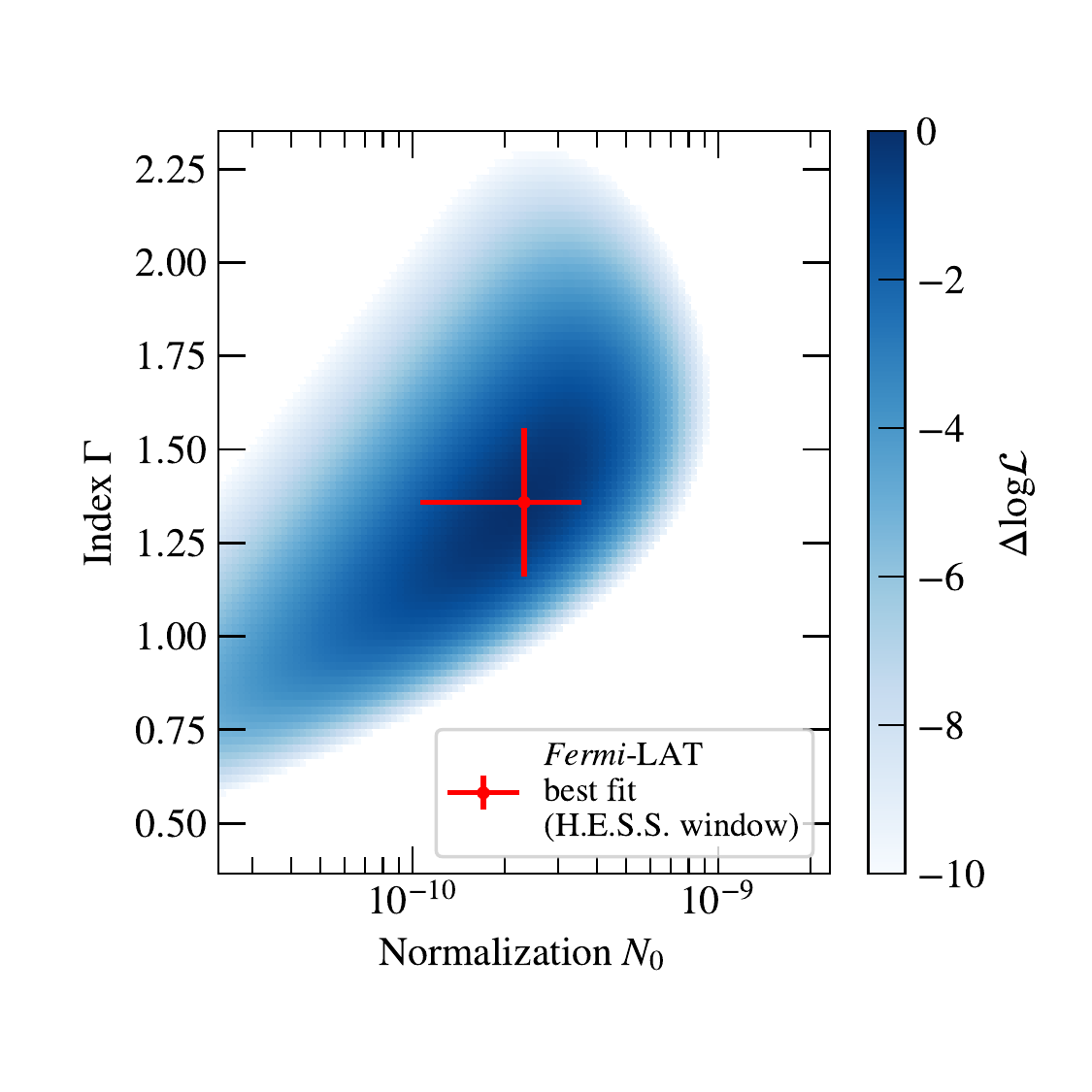}
    \includegraphics[width = .49\linewidth]{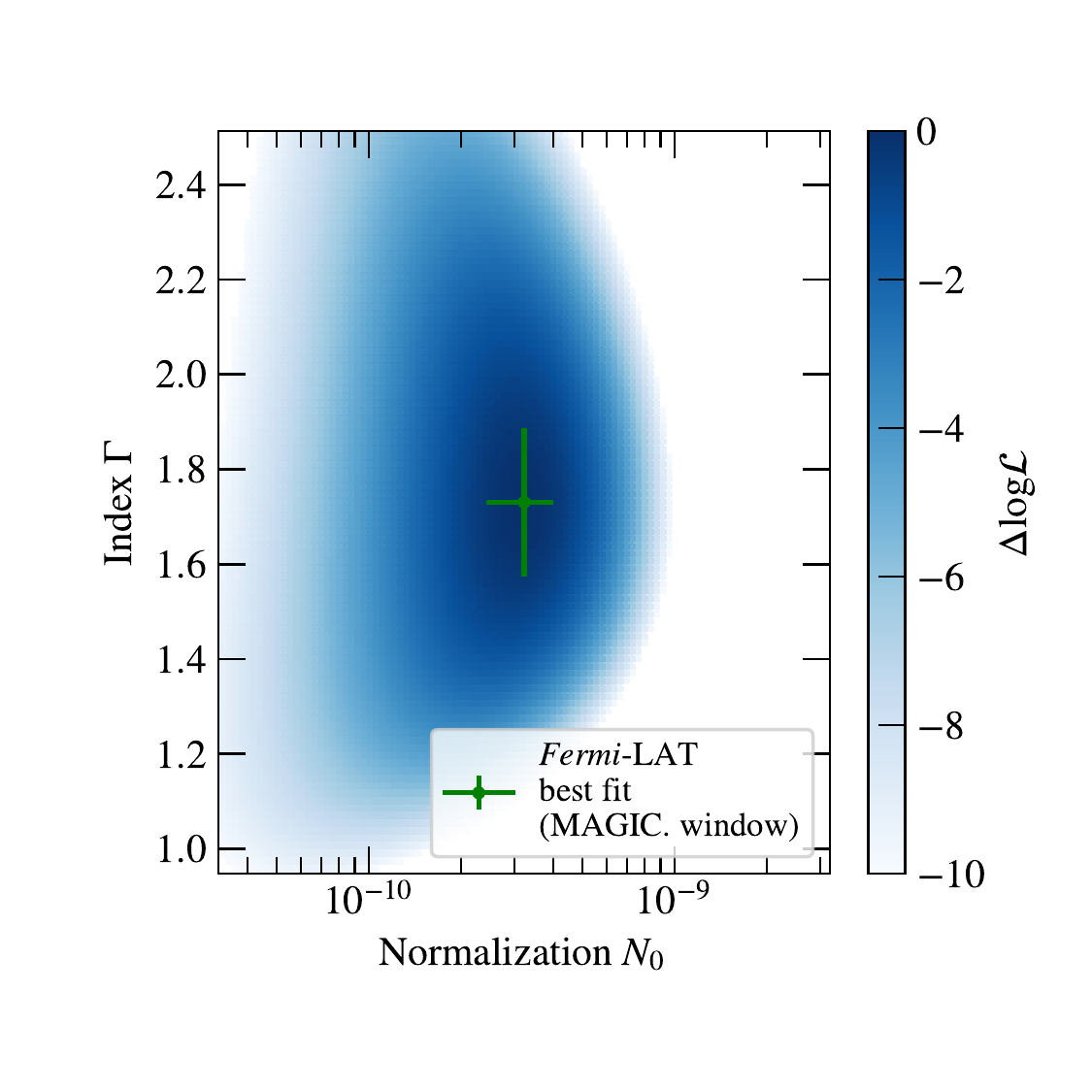}
    \caption{Two dimensional likelihood surface of a power-law fit to the \fermi\ data contemporaneous to the \hess\ observations (left) and MAGIC observations (right). 
    The best-fit values are marked in red for \hess\ and green for MAGIC.
    The likelihood surfaces are used in a simultaneous fit with IACT data to search for BLR absorption features, see main text for further details.}
    \label{fig:fermilikelihoodhess}
\end{figure*}

The distance $r$ from the black hole is constrained by simultaneously fitting an intrinsic spectrum $F(E)$ modified by the BLR absorption $\exp(-\tau_{\gamma\gamma})$ to the \fermi\ and IACT data, where the IACT data are corrected for EBL absorption following \cite{frv08}.\footnote{The EBL absorption in the \fermi\ energy band is negligible; at the energy of the detected highest energy photon it amounts to $\sim5\,\%$.}
\fermi\ data are used which are contemporaneous with the \hess\ and MAGIC observation windows.
Since these windows encompass less than $3\,$hr and \source\ was not in a flaring state in the LAT energy band, the likelihood is extracted as a function of the normalization $N_0$ and spectral index $\Gamma$ and this likelihood surface, $\mathcal{L}_{Fermi}(N_0,\Gamma,\boldsymbol{\theta} | D_{Fermi})$, is used in the combined fit rather than flux points or likelihood curves for each energy bin as done by \cite{msb19}. 
In the likelihood function, $\boldsymbol{\theta}$ denotes nuisance parameters (spectral parameters of the other sources in the ROI) and $D_{Fermi}$ denotes the data.
The likelihood surfaces for the \hess\ and MAGIC observation windows are shown in Fig.~\ref{fig:fermilikelihoodhess}.

Different spectral shapes are tested since intrinsic curvature or a cut-off are degenerate with a cut-off induced by $\gamma$-ray absorption.
Only spectral functions are used that can be described as power laws in the \fermi\ energy band in order to use the extracted likelihood surface, namely a simple power law (PL), a power law with sub-exponential cut-off (EPL), and a smoothly broken power law (BPL). 
It should be noted that for these functions, which are additionally multiplied by the BLR absorption, $\mathcal{L}_{Fermi}$ is only an approximation of the true likelihood value, since $\mathcal{L}_{Fermi}$ does not account for possible  curvature at the highest \fermi\ energies. 
As BLR absorption only sets in above $\sim 20\,$GeV and the curvature of the chosen models also occurs at energies close to or beyond the highest energy photon detected with the LAT, the results are not expected to change if a complete likelihood formulation had been used instead.

From a physical point of view, a break in the spectrum is expected also for the intrinsic emission, as the SED gamma-ray peak in blazars is usually located inside the observed energy range, except for only the most extreme high-synchrotron-peaked blazars \citep[e.g.,][]{singh2019,biteau2020}.
Therefore, the intrinsic spectral shape should be modeled either with a BPL or EPL. 
However, the PL model is retained as a test case to evaluate whether the observed spectral break could in principle be explained with BLR attenuation only. 

For each combination of intrinsic spectrum and assumed BLR geometry (ring, high luminosity ring, or shell), the parameters of the intrinsic spectrum and $r$ are optimized.
For the MAGIC data, the objective function $\chi^2 - 2\ln\mathcal{L}_{Fermi}$ is minimized, where $\chi^2$ is the $\chi^2$ value of the fit to the MAGIC data points which takes into account the full correlation matrix of the flux points. 
For the \hess\ data, on the other hand, it is assumed that the likelihood for a flux value in the $i$th energy bin  can be described with a Gaussian distribution, $\mathcal{L}_{\mathrm{\hess},i}$, as done in \cite{Hea19}. 
Therefore, in this case, the objective function $2(\ln\mathcal{L}_{Fermi} + \sum_i\ln\mathcal{L}_{\mathrm{\hess},i})$ is maximized.
For flux upper limits, one sided Gaussian distributions are used. 
The sum over the likelihoods in each energy bin is then combined with $\mathcal{L}_{Fermi}$.

\begin{figure*}[th]
    \centering
    \includegraphics[width=0.98\linewidth]{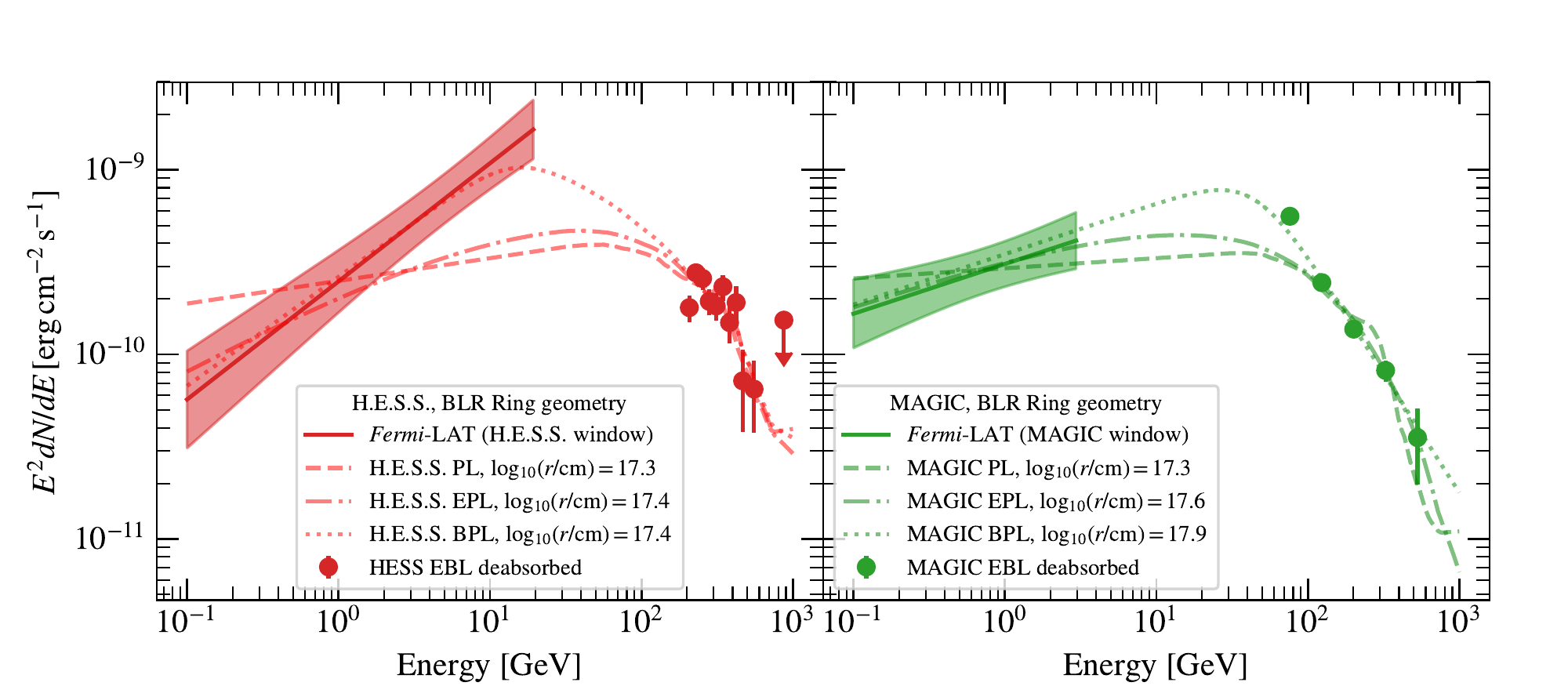}
    \caption{HE and VHE $\gamma$-ray spectra of \source\ during the flare night, JD~2457539. The \fermi\  butterflies are integrated over the precise \hess\ (red, left panel) and MAGIC (green, right panel) observation windows and are plotted up to the highest detected photon energy. The VHE \g-ray spectra are corrected for the EBL absorption with the model of \cite{frv08}, with the \hess\ spectrum in red and the MAGIC spectrum in green. Lines mark the best-fit spectra for combined \fermi\ and \hess\ and MAGIC fits for the BLR ring geometry and different intrinsic spectra: a power law (PL), power law with sub-exponential cut-off (EPL), and a broken power law (BPL). 
   In the PL case, curvature is only provided by the BLR absorption, whereas the EPL and BPL models include intrinsic spectral curvature. The best-fit values of the distance $r$ from the black hole are provided in the legend.}
    \label{fig:fit_blr}
\end{figure*}

The best-fit spectra for all spectral functions and the BLR ring geometry are shown in Fig.~\ref{fig:fit_blr}. The best-fit index for the \hess\ and \fermi\ spectrum for the PL spectrum deviates strongly from the best-fit index in the \fermi\ energy band, indicating the presence of an intrinsic spectral curvature.  
However, due to the broad likelihood profile (see Fig.~\ref{fig:fermilikelihoodhess}), the change in $\mathcal{L}_{Fermi}$ is only $-3.8$ for the PL  spectrum. 

\begin{figure*}[th]
    \centering
    \includegraphics[width=.49\linewidth]{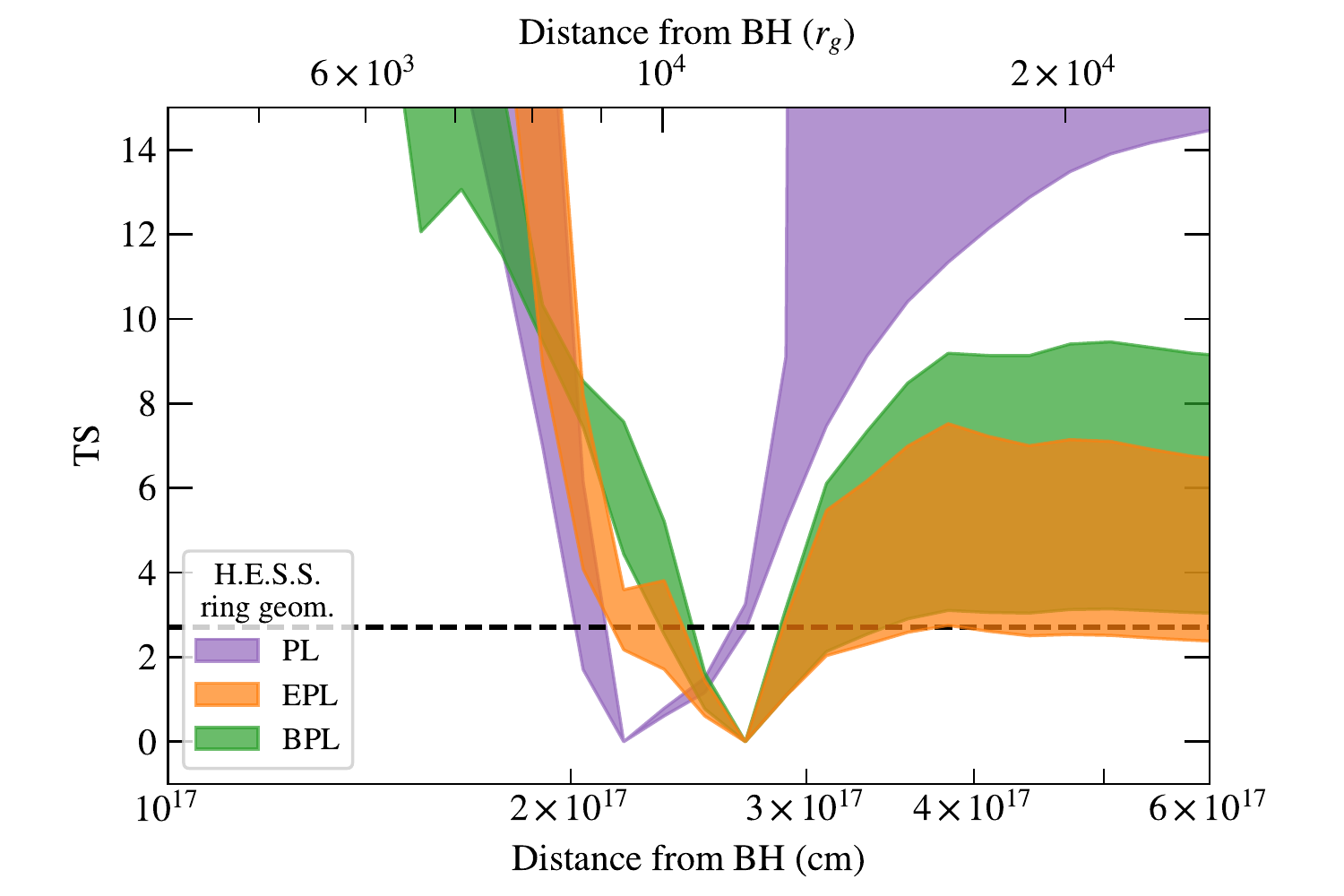}
    \includegraphics[width=.49\linewidth]{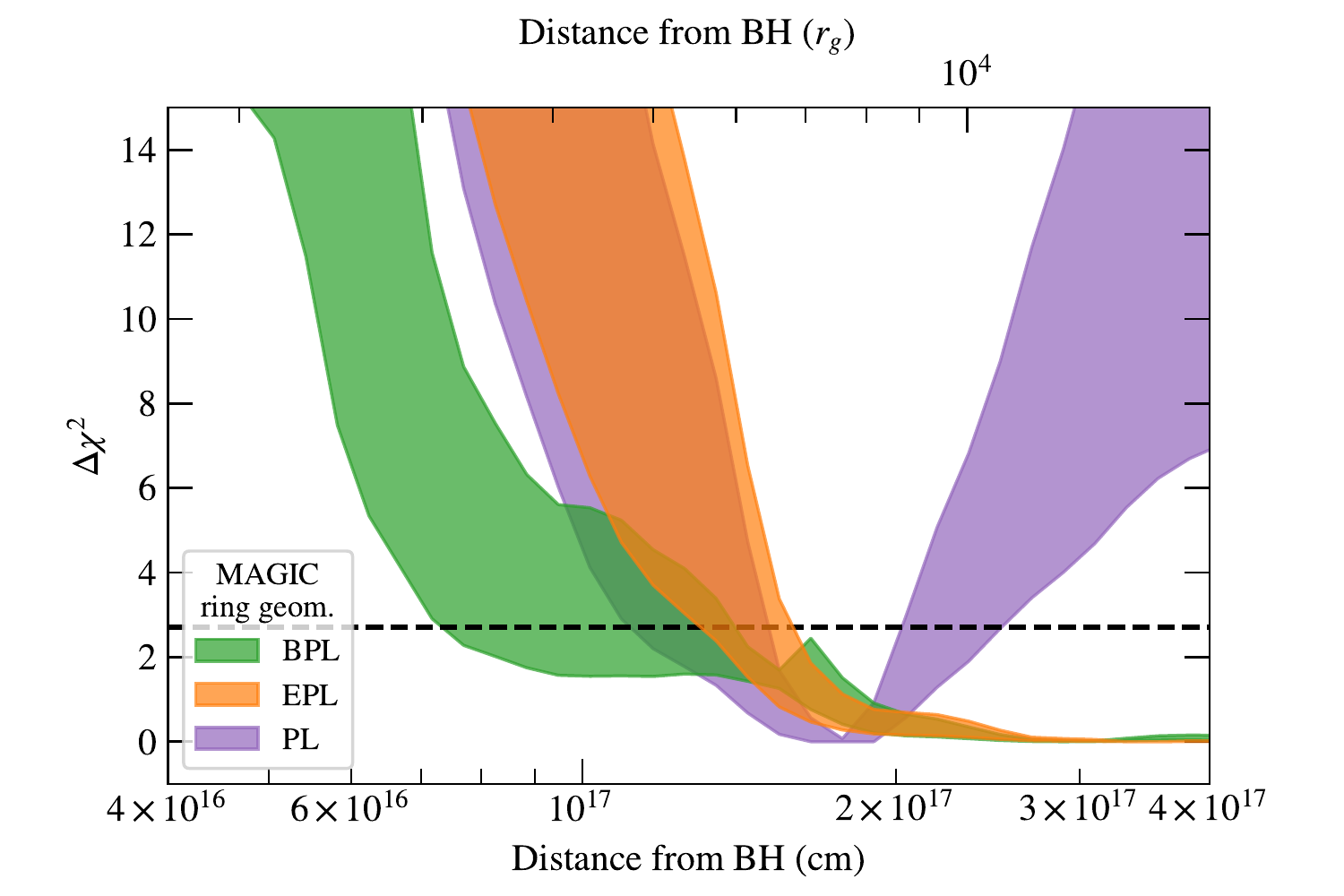}
    \caption{Likelihood and $\Delta\chi^2$ profiles for the distance $r$ in the BLR ring geometry for the combined \emph{Fermi}-LAT and \hess\ (left) and \emph{Fermi}-LAT and MAGIC fits. 
    The spread of the profiles shows the effect of the IACT systematic uncertainties on the measured fluxes. 
    For the \hess\ and \emph{Fermi}-LAT fits, the TS value is defined as twice the difference between the best-fit log-likelihood and the log-likelihood profiled over $r$. The dashed lines show where the TS or $\Delta\chi^2$ values are equal to 2.71, which corresponds to lower limits at 95\,\% confidence.}
    \label{fig:loglike_blr}
\end{figure*}
\begin{table*}[th]
    \centering
    \caption{Limit values $r_\mathrm{lim}$ on the distance of the $\gamma$-ray emitting region to the central black hole for each tested BLR geometry and intrinsic spectrum. The $\min(\log_{10}(r_\mathrm{lim}))$ denote the minimum limit lower limit values found when the analysis is repeated with different energy scales shifted by $(0\,\%,\pm15\,\%)$, which should account for the main contribution of systematic uncertainties.
    Additionally, the number of parameters $n_\mathrm{par}$ and the AIC and $p(\mathrm{AIC})$ values are provided for the case of an un-shifted energy scale.}
    \label{tab:blr-result}
    \begin{tabular}{lccccc}
\hline
\hline
Intr. Spectrum & $\log_{10}(r_\mathrm{lim} / \mathrm{cm})$ & $\min(\log_{10}(r_\mathrm{lim} / \mathrm{cm}))$ & $n_\mathrm{par}$ & AIC & $p(\mathrm{AIC})$\\
\hline
\multicolumn{5}{c}{H.E.S.S., ring geom.}\\
\hline
PL & $17.296$ & $17.296$ & 3 & $-29.44$ & $0.13$ \\
EPL & $17.341$ & $17.327$ & 5 & $-33.59$ & $1.00$ \\
BPL & $17.382$ & $17.366$ & 5 & $-33.45$ & $0.93$ \\
\hline
\multicolumn{5}{c}{MAGIC, ring geom.}\\
\hline
PL & $17.170$ & $17.044$ & 3 & $8.27$ & $1.00$ \\
EPL & $17.192$ & $17.114$ & 5 & $10.50$ & $0.33$ \\
BPL & $17.096$ & $16.862$ & 5 & $10.17$ & $0.39$ \\

\hline
\multicolumn{5}{c}{H.E.S.S., ring geom., high lumi.}\\
\hline
PL & $17.961$ & $17.949$ & 3 & $-25.98$ & $0.06$ \\
EPL & $17.975$ & $17.954$ & 5 & $-27.47$ & $0.13$ \\
BPL & $17.925$ & $17.910$ & 5 & $-31.54$ & $1.00$ \\
\hline
\multicolumn{5}{c}{MAGIC, ring geom., high lumi.}\\
\hline
PL & $17.903$ & $17.835$ & 3 & $7.97$ & $1.00$ \\
EPL & $17.913$ & $17.866$ & 5 & $11.10$ & $0.21$ \\
BPL & $17.764$ & $17.726$ & 5 & $11.57$ & $0.17$ \\

\hline
\multicolumn{5}{c}{H.E.S.S., shell geom.}\\
\hline
PL & $17.604$ & $17.077$ & 3 & $-27.31$ & $0.09$ \\
EPL & $17.621$ & $17.617$ & 5 & $-32.11$ & $1.00$ \\
BPL & $17.632$ & $17.620$ & 5 & $-31.02$ & $0.58$ \\
\hline
\multicolumn{5}{c}{MAGIC, shell geom.}\\
\hline
PL & $17.578$ & $17.528$ & 3 & $8.55$ & $1.00$ \\
EPL & $17.579$ & $17.552$ & 5 & $10.32$ & $0.41$ \\
BPL & $17.572$ & $17.511$ & 5 & $10.14$ & $0.45$ \\

\hline
    \end{tabular}
\end{table*}

In order to judge whether BLR absorption is significantly detected, the likelihood profiles of the distance $r$ are derived for the different intrinsic spectra, BLR geometries, and also by taking into account the systematic uncertainties of the IACT measurements by shifting the IACT flux points to lower and higher flux values following the assessments in Sec.~\ref{sec:ana}. 

The profiles are shown in the panels of  Fig.~\ref{fig:loglike_blr} for the \hess\ (left panel) and MAGIC (right panel) combined fits; the spread of the profiles corresponds to the spread caused by the systematic uncertainties. 
For both MAGIC and \hess, when a simple PL is assumed as the intrinsic function, BLR absorption is required to provide the spectral break in order to describe the data. This results in a clear minimum in the likelihood curves for the PL model.  
The statistical preference of the fit with BLR absorption can be assessed by the $\mathrm{TS}$ and $\Delta\chi^2$ values for the maximum values of $r$ in Fig. 16, since for such large values of $r$ the absorption is negligible in the considered energy bands.
However, as described below, the PL fit is not significantly preferred over the intrinsic spectra with curvature. 
As a result, a detection of BLR absorption cannot be claimed. 
Furthermore, as discussed above, it is expected for an FSRQ like \source\ that the SED peak of the high-energy emission falls within the $\gamma$-ray energy band, which is captured in intrinsic models with curvature.  
The preference of one intrinsic spectral model over another one 
is assessed with the Akaike information criterion \citep{1974ITAC...19..716A}, 

\begin{equation}
    \mathrm{AIC} = 2n_\mathrm{par} - 2\ln\mathcal{L},
\end{equation}
where $n_\mathrm{par}$ is the number of fit parameters and $\mathcal{L}$ is the total summed likelihood ($2\ln\mathcal{L} = -\chi^2 + 2\ln\mathcal{L}_{Fermi}$ for the combined \emph{Fermi}-LAT and MAGIC fit).
Models with an AIC larger than the minimum value, $\mathrm{min}(\mathrm{AIC})$, are less likely to minimize the loss function with a probability $p(\mathrm{AIC}) = \exp[(\mathrm{min}(\mathrm{AIC})- \mathrm{AIC})/2]$ \citep[e.g.,][]{bah11}.
The AIC and $p(\mathrm{AIC})$ values are provided in Table~\ref{tab:blr-result}. 
Even though for the MAGIC spectra the PL fits provide the minimum AIC value, the other spectral models are almost as likely to minimize the loss function, with $p(\mathrm{AIC})$ between 0.2 and 0.5. 
For \hess\, all spectral models seem to prefer a fit including BLR absorption (assuming the ring geometry). 
However, for the EPL and BPL models the preference is not significant when the systematic uncertainties on the observed flux are taken into account. Judging from the AIC, models including curvature appear to be marginally more likely to minimize the cost function compared to a simple power law. 
However, this preference is not at a significant level.

The resulting best-fit spectra and observations are shown in Fig.~\ref{fig:fit_blr}. The best-fit values for the distance $r$ are between $2\,\times10^{17}\,$cm and $8\,\times10^{17}\,$cm, which corresponds to $2.6R_{\mathrm{BLR}} \lesssim r \lesssim 10.4R_{\mathrm{BLR}}$. 
The cut-off caused by the BLR is clearly seen for spectra without intrinsic curvature, which prefer a lower value of $r$, causing a steep and sharp drop in the spectra shown in Fig.~\ref{fig:fit_blr}. 

Obviously, it is not possible to disentangle an intrinsic cut-off from a BLR-induced one, i.e. it is not possible to put an upper limit on the distance of the emission region. 
Even though the broken power law seems to be the only function providing a satisfactory fit to the \fermi\ and \hess\ data, the other spectral shapes cannot be ruled out.
Therefore, only lower limits (at 95\% confidence level) on the distance, $r_\mathrm{lim}$, have been derived. 
Of the tested BLR geometries and parameter space, the BLR ring geometry produces the least $\gamma$-ray attenuation and hence provides the most conservative (least constraining) limits. 
The limit values also demonstrate that the uncertainty introduced by the unknown exact BLR geometry dominates the systematic errors. For the combined \hess\ and \fermi\ fit, the lowest lower limit is $r > 2.0\times10^{17}\mathrm{cm} = 2.6R_{\mathrm{BLR}}$. 
This limit is a factor 3 stronger than previous limits \citep{msb19} derived from \fermi\ data alone. 
Due to the softer \fermi\ and MAGIC spectra, the limit relaxes during the MAGIC observation window to $r > 7.2\times10^{16}\mathrm{cm} =
0.9R_{\mathrm{BLR}}$.

As the emission region is unlikely to move backwards (against the jet flow) between the \hess\ and MAGIC observations, 
the lower limit during the \hess\ observation window provides the minimum distance of the emission region from the black hole. Therefore, the emission region must be located outside of the BLR at a position where the jet has significantly widened. 

It has also been investigated if absorption of the sub-TeV photons in DT radiation can be used to put further constraints on the location of the emission region. 
For specific parameters (DT temperature of 1000K and emission region deep in the DT radiation field) the absorption can considerably affect the observed spectrum above $\sim 400$\,GeV.
Nevertheless the large statistical and systematic uncertainties in the measured flux at the highest energies detected by H.E.S.S. and MAGIC, combined with poorly constrained DT parameters prevent deriving any robust limits on the location of the emission region with this method. 

Finally, the fits with curved models can be used to  estimate the possible position of the inverse Compton peak.
Using the break energy in the BPL model and the BLR ring geometry, the peak is determined to be located at $E_\mathrm{peak} = (24.1\pm23.2_\mathrm{stat})\,\mathrm{GeV}$ for the \hess observation window and 
$(57.2\pm47.5_\mathrm{stat})\,\mathrm{GeV}$ for the  MAGIC window. 
The large errors indicate that the peak can only be weakly constrained due to the large gap in energy between the \fermi\ and IACT observations. 
Within uncertainties, the peak positions are consistent between the two measurements.

%
\subsection{The large-scale jet structure} \label{sec:vlbadis}
\begin{figure}[th]
\centering
\includegraphics[width=0.48\textwidth]{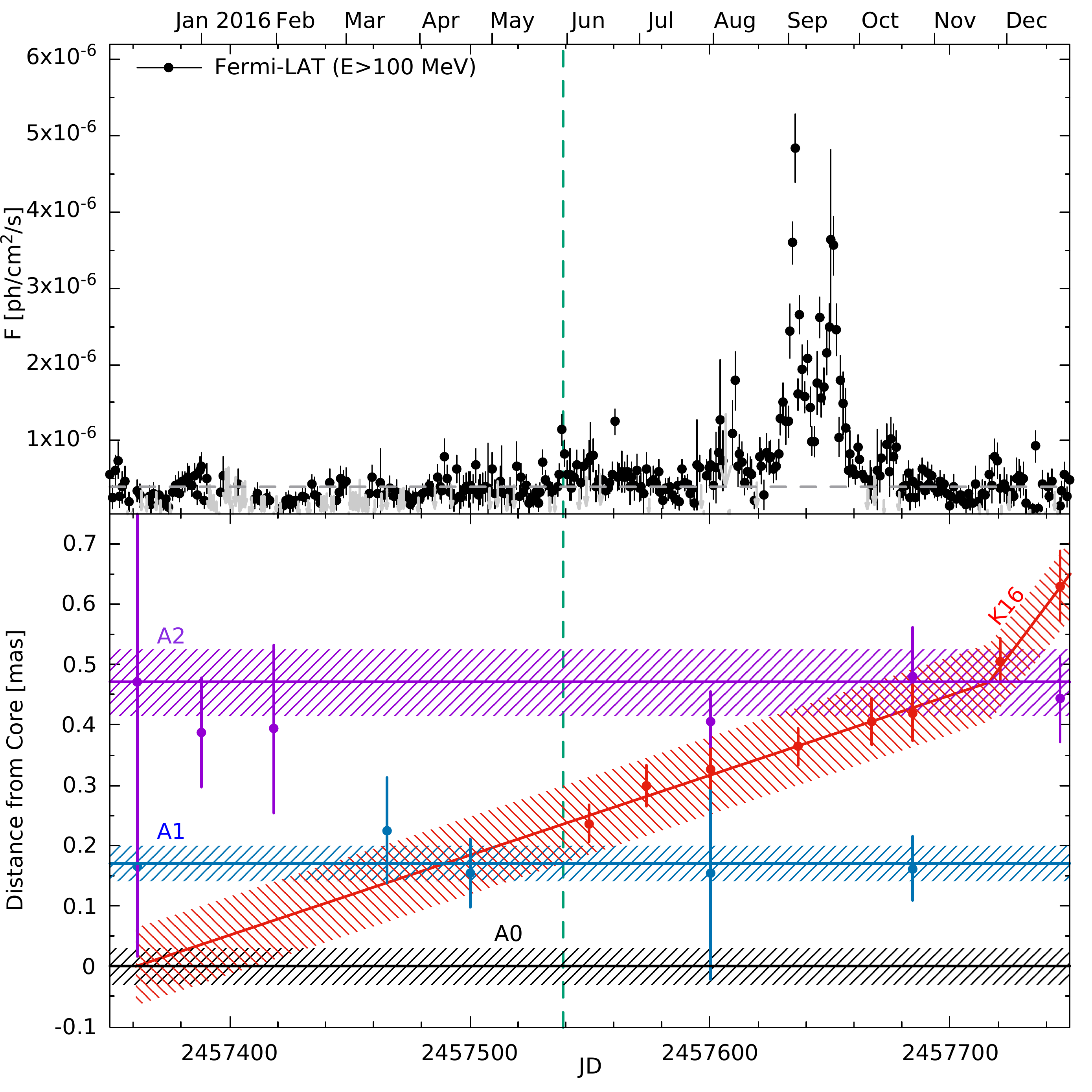}
\caption{{\it Top:} HE \g-ray light curve of \source\ in $24\,$hr bins. Grey arrows are upper limits (95\% confidence level for TS~$<9.0$), and the grey dashed line indicates the 11-year average. 
{\it Bottom:} Separation of knot K16 (red) from the core A0 (black line) according to the 43\,GHz maps. The red line approximates the motion of K16 with the $1\sigma$ uncertainty given by the shaded region. The same is shown for A1 (blue) and A2 (magenta). 
In both panels the green dashed line indicates the time of the VHE \g-ray flare.
} 
\label{fig:sketch}
\end{figure}
In the previous section, it has been established that the emission region must have been located beyond the BLR. Interestingly, in coincidence with the flare a fast radio knot, K16, moved through the jet. A possible connection between K16 and the flare is analyzed here.

The radio data suggest that K16 was ejected from the core between JD~2457303 (October 07, 2015) and JD~2457419 (January 31, 2016). A sketch showing the separation of K16 from A0, is presented in Fig.~\ref{fig:sketch} (bottom panel), along with A1 and A2. 
The parameters of A1 and A2 (see Table~\ref{tab:KParm}) during the current observations are similar to those reported previously \citep{J17}, and the projected position angle of A1 coincides with that of K16 at the time of the VHE \g-ray flare. Note that although the viewing angles of K15 and K16 are similar, according to the images the knots should be moving along trajectories on opposite sides of the jet cross-section resulting in different position angles.
Using the proper motion of K16 determined from the VLBA images (Table~\ref{tab:KParm}), the knot should reach the distance of the A1 stationary feature in $\sim$4.5 month, which yields an epoch of possible interaction between the centroids of K16 and A1 of JD~2457497$\pm$46 (March 3 -- June 3, 2016), taking into account the uncertainties of the proper motion of K16 and the position of A1. 
Interestingly, the 1$\sigma$ uncertainty of the interaction time includes the time of the VHE \g-ray flare suggesting a possible connection of the events.

Inspection of Fig.~\ref{fig:sketch}, which also presents the HE \g-ray lightcurve over the entire time frame, suggests interesting associations between \g-ray events and activity in the parsec-scale jet: 
(1) the passage of K16 through the stationary feature A1 coincides with the VHE \g-ray flare and the beginning of some activity in the HE \g-ray band; and
(2) prominent activity at HE \g-rays begins as K16 starts to interact with the upstream portion of A2. This is interesting as K16 may have changed direction or accelerated after the interaction with A2.  
Unfortunately, the large number of radio knots detected in \source, and the uncertainties in their crossing times make an unambiguous identification difficult. However, each of the four VHE \g-ray flares detected previously in \source\ can be associated with a superluminal knot propagating in the jet \citep{apm10,Hea13,aMea14,l15,aMea17}.  

The analysis of the polarization parameters of the core and K16 (Figs.~\ref{fig:PolA0} and \ref{fig:PolK16}) reveals a highly variable, but low value of polarization $P$ in K16, which varies from $3\%$ to an undetectable level as K16 separates from the core. The core polarization increases after separation of K16 from the core, and its EVPA rotates by $\sim$70$^{\circ}$ just after the TeV detection (see Fig.~\ref{fig:PolA0}).
This implies that the knot moves down a turbulent spine that possesses a weakly ordered magnetic field component oriented perpendicular to the jet (Fig.~\ref{fig:vlba16}). From this orientation, K16 may be interpreted as a transverse shock. 

%
%
\subsection{Modeling constraints} \label{sec:mod}
The most common interpretation of FSRQ \g-ray emission is inverse-Compton scattering of BLR and/or DT photons \citep[see, e.g.,][]{bea13,vdBea19}. 
Here, the lack of absorption by the BLR photons places the emission region at least $2.0\E{17}\,$cm from the black hole outside of the BLR (for the assumed BLR model with $R_{\rm BLR}=7.69 \times 10^{16}$\,cm). A common assumption in blazar modeling is that the emission region fills the entire width of the jet. In turn, the maximum distance $r$ of the emission region from the black hole in a conical jet with opening angle (in units of radians)
$\alpha=0.26/\Gamma_b$ \citep[cf.][]{pu09}, can be calculated as

\begin{align}
    r \approx \frac{t^{\ast}_{\rm VHE}c\delta}{(1+z)\alpha} = 2.1\E{17}\,\mbox{cm}\,\est{t^{\ast}_{\rm VHE}}{20\,\mbox{min}}{}\est{\delta}{45}{}\est{\Gamma_b}{45}{},
\end{align}
employing the observed minimum variability time scale $t^{\ast}_{\rm VHE}\sim 20\,$min.
Ignoring for the moment the possible association of the VHE \g-ray flare with the interaction of the knot K16 with the standing feature A1, the usual approximation $\delta=\Gamma_b$ has been chosen. Even with the rather extreme assumptions on the opening angle and the Doppler factor, the distance is barely compatible with the minimum distance allowed by the absorption study. If the emission region is located beyond this distance, it cannot fill the width of the jet \citep{lea13,wea16}.

While an absorption limit cannot be obtained for the DT, the lower limit derived from the BLR absorption is compatible with the distance of $\sim 50\,$pc from the black hole, where the interaction between the knot K16 and the standing feature A1 took place. 
If the flare was indeed triggered by the interaction of K16 with A1, the emission region is not immersed in neither the  BLR nor the DT photon fields, which are located at most a few parsecs from the black hole. This assumption is used in the following. In turn, from now on the kinematic values derived in Sec.~\ref{sec:radio} are used, namely $\delta=43$ and $\Gamma_b=23$.

During most observed states, the peak position of the inverse-Compton component in the SED of \source\ is located at energies $<100\,$MeV \citep{sea15,b13,bea14,Hea13,acc18}. During this flare, the peak position has  shifted by more than a factor 100 to $\sim 50\,$GeV (Sec.~\ref{sec:gspec}), while the spectral fluxes at $100\,$MeV barely change (c.f. Figs.~\ref{fig:spec_gamma} and \ref{fig:mwl_lc_all}). This implies that the distribution of the flare electrons is either hard or narrow with a high minimum electron Lorentz factor $\gamma_{\rm min}$. 

In case of a hard electron distribution, the break in the \g-ray spectrum could represent the Klein-Nishina break. The scattered photon energy in the Thomson regime is $E_{\gamma}=4\gamma^2 E_{\rm ph}$, where $\gamma$ is the electron Lorentz factor and $E_{\rm ph}$ the soft photon energy. 
In the Klein-Nishina regime, nearly all of the electron energy is transferred to the \g\ ray, i.e. $E_{\gamma} \approx \gamma m_e c^2$. 
At the  Klein-Nishina break, the quantity $4\gamma E_{\rm ph}=m_ec^2$, resulting in $E_{\rm ph}=m_ec^2/(4\gamma)=(m_ec^2)^2/(4E_{\gamma})$. The energy $E_{\rm ph}$ of the soft photon in the comoving frame then is

\begin{align}
    E_{\rm ph} = \frac{\delta (m_e c^2)^2}{4 (1+z) E\as_{\gamma,br}} 
    = 41\,\mbox{eV}\,\est{\delta}{43}{}\est{E\as_{\gamma,br}}{50\,\mbox{GeV}}{-1} \label{eq:KNenergy},
\end{align}
where $E\as_{\gamma,br}$ is the \g-ray energy at the Klein-Nishina break in the observer's frame. The electron Lorentz factor at the break is $\gamma_{br} = m_ec^2/(4E_{\rm ph})= 3200 \est{\delta}{43}{-1}\est{E\as_{\gamma,br}}{50\,\mathrm{GeV}}{}$. 
If the photon is produced inside the jet, it should be a synchrotron photon. 
In this case, the synchrotron component would have extended well into the soft X-ray domain, which would be unusual for FSRQs, where the X-ray domain typically belongs to the inverse-Compton component. Unfortunately, no data are available to probe this possibility.

If the soft photon is provided by external, isotropic photon sources, another transformation to the frame of the photon production region needs to be applied giving 

\begin{align}
E\p\sim E_{\rm ph}/\Gamma_b = 1.8\,\mbox{eV}\est{\delta}{43}{}\est{E\as_{\gamma,br}}{50\,\mathrm{GeV}}{-1}\est{\Gamma_b}{23}{-1} \label{eq:photonenergyA1}
\end{align}
in case this frame is 
that of the host galaxy. This is in the (red) optical domain. 
Such photons could be synchrotron emission from the radio feature A1 within a ring-of-fire scenario \citep{NICK15}.

As the emission region is smaller than the full diameter of the jet, the surrounding jet material could also provide synchrotron seed photons within a spine-sheath scenario \citep{tg08}. In this case, the relative Lorentz factor between the emission region and the surrounding jet material is less than between the emission region and a stationary (in the host-galaxy frame) source such as A1. Assuming a relative Lorentz factor of $\Gamma_{\rm rel}\sim 5$, the required synchrotron photon energy in the sheath is $\sim 8\,$eV.

The observed variability is a combination of particle acceleration and cooling effects, as well as the light-crossing time through the source. Which one is dominating at any given time is difficult to say without proper modeling. Nonetheless, some important clues can be derived if cooling dominates the cessation of the VHE \g-ray and optical flare. 

Assuming first again that only internal processes dominate the cooling (resulting in an SSC situation), the Compton dominance -- the ratio of the \g-ray peak flux to the synchrotron peak flux -- being probably larger than unity\footnote{In \source\ the Compton dominance is typically on the order of 10 \citep[e.g.][]{nea12,bea14,sea15,aMea17}. From Fig.~\ref{fig:spec_gamma} one can deduce that the \g-ray peak flux probably increased by a factor 10, while the R-band flux -- the only representative available for the synchrotron component -- increased by about a factor 2. Even if different parts of the synchrotron components have changed more than that, it is unlikely that they would have risen by at least a factor 10. In turn, it is likely that the Compton dominance is also larger than unity during this flare.} 
dictates that the energy density in the synchrotron photons dominates over the magnetic field energy density. In such a situation, the cooling becomes nonlinear \citep{zs12}, which could help explaining the fast cooling during the flare cessation. While simple estimates indicate that SSC is a viable option, proper constraints cannot be derived as the synchrotron spectrum during the flare is unknown. 

If the cooling is dominated by inverse-Compton scattering of external photon sources, constraints on that photon source can be derived. The inverse-Compton cooling time scale is given by

\begin{align}
    t_{\rm cool} = \frac{3m_ec^2}{4c\sigma_T} \left( u_{\rm tot}F_{\rm KN}\gamma \right)^{-1} \label{eq:tcool},
\end{align}
with $\sigma_T$ the Thomson cross section, $u_{\rm tot}$ the total energy density, and 

\begin{align}
    F_{\rm KN} \approx \begin{cases}
        1, & 4E_{\rm ph}\gamma \ll m_e c^2 \\
        \left( 1+\frac{4E_{\rm ph}\gamma}{m_e c^2} \right)^{-1}, & 4E_{\rm ph}\gamma \gg m_e c^2 
    \end{cases} \label{eq:FKN}
\end{align}
a Klein-Nishina correction factor \citep{mea05}. Assuming that the VHE \g-ray photons with energies $>200\,$GeV are emitted in the (deep) Klein-Nishina domain, the total energy density can be approximated as

\begin{align}
    u_{\rm tot} \approx \frac{3 E_{\rm ph}}{c\sigma_T t_{\rm cool}} 
    = 0.26\,\frac{\mbox{erg}}{\mbox{cm}^{3}} \est{t^{\ast}_{\rm VHE}}{20\,\mbox{min}}{-1} \est{E\as_{\gamma,br}}{50\,\mbox{GeV}}{-1}, \label{eq:utot}
\end{align}
where Eq.~(\ref{eq:KNenergy}) is used, and the cooling time is estimated from the observed time scale  of the flux drop $t\as_{\rm VHE} = t_{\rm cool} (1+z)/\delta$.
Note that Eq.~(\ref{eq:utot}) assumes a narrow soft photon distribution peaking at $E_{\rm ph}$, as no other handle is available for the nature of these photons. While it may be a bad representation of the true soft photon distribution and may influence the results, it should not invalidate the main conclusion.

In case of the ring-of-fire scenario, the total energy density that must be provided by A1 is 

\begin{align}
u\p_{\rm rf}\sim u_{tot}/\Gamma_b^2 = 5\E{-4}\,\frac{\mbox{erg}}{\mbox{cm}^{3}} \est{t^{\ast}_{\rm VHE}}{20\,\mbox{min}}{-1} \est{E\as_{\gamma,br}}{50\,\mbox{GeV}}{-1} \est{\Gamma_b}{23}{-2} \label{eq:utotA1ring}, 
\end{align}
while in the spine-sheath scenario the sheath must provide 

\begin{align}
u\p_{\rm ss}\sim u_{tot}/\Gamma_{\rm rel}^2 = 9\E{-3}\,\frac{\mbox{erg}}{\mbox{cm}^{3}} \est{t^{\ast}_{\rm VHE}}{20\,\mbox{min}}{-1} \est{E\as_{\gamma,br}}{50\,\mbox{GeV}}{-1} \est{\Gamma_{\rm rel}}{5}{-2} \label{eq:utotA1spine}. 
\end{align}
These are high energy densities for a jet region $50\,$pc from the black hole \citep[c.f., the modeling of a pc-scale jet in][which obtains $u\lesssim 10^{-6}$\,erg\,cm$^{-3}$]{zw16}. Such energy densities may leave observable signatures of the jet in the optical or UV domain. 

The energy densities can be transformed into fluxes at Earth and compared to the total observed optical fluxes from the lightcurve in Fig.~\ref{fig:mwl_lc_all}.
The maximum optical flux during the flare is $\sim 1.3\E{-11}\,$erg\,cm$^{-2}$s$^{-1}$.
The synchrotron energy densities are transformed to fluxes according to

\begin{align}
    F_i\as = u\p_{\rm i} \delta_i^4 c \left( \frac{a_{\rm pc}}{2 D_L} \right)^2
    \label{eq:flux},
\end{align}
where $u\p_{\rm i}$ is the result from either Eq.~(\ref{eq:utotA1ring}) or Eq.~(\ref{eq:utotA1spine}), 
$\delta_i$ the respective Doppler factor, $a_{\rm pc}$ is the diameter of A1 in pc, which can be calculated from the angular size of A1 in Table~\ref{tab:KParm}, and $D_L$ the luminosity distance of \source. Inserting $t^{\ast}_{\rm VHE}\approx 20\,\mbox{min}$ and $E\as_{\gamma,br}\approx 50\,\mbox{GeV}$, the total flux in the ring-of-fire scenario is $F_{\rm rf}\as = 8\E{-14}\,$erg\,cm$^{-2}$s$^{-1}$ with $\delta_{\rm rf}=1$ for a stationary source in the host galactic frame. In the spine-sheath scenario one obtains 
$F_{\rm ss}\as = 6\E{-10}\est{\delta_{\rm ss}}{4.5}{4}\,$erg\,cm$^{-2}$s$^{-1}$, 
where $\delta_{\rm ss}$ is the Doppler factor of the sheath 
which has been estimated using the observation angle of the jet, $1.2^{\circ}$ \citep{J17}, and $\Gamma_{\rm rel} = \Gamma_{b}\Gamma_{\rm ss}(1-\beta_b\beta_{\rm ss})$ \citep{gtc05}, where $\beta_b$ and $\beta_{\rm ss}$ are the normalized speeds of K16 and the sheath, respectively, and $\Gamma_{b}$ and $\Gamma_{\rm ss}$ are the corresponding Lorentz factors.

The expected flux in the ring-of-fire scenario is more than two orders of magnitude below the optical flux limit. This scenario is also in line with the fact that the variations of the \g-ray flux on the flare night by over an order of magnitude are accompanied with only much smaller variations in the optical flux.
On the other hand, the flux in the spine-sheath scenario exceeds the optical flux limit by a factor $60$. This disfavors the spine-sheath scenario as a viable option under the assumption that A1 (and thus the sheath) exhibits a homogeneous radiation density. The flux could be substantially reduced, if the sheath Doppler factor $\delta_{\rm ss}$ is lower or the relative Lorentz factor $\Gamma_{\rm rel}$ is higher. Both imply a lower sheath Lorentz factor $\Gamma_{\rm ss}$. However, in order to drop the flux by the factor $60$ requires $\Gamma_{\rm ss}\sim 1.5$. This almost resembles the ring-of-fire scenario.

The estimates in the ring-of-fire scenario can also be used to constrain a model in which the radiation field comes from luminous stars. Such models could naturally explain fast variations of the emission \citep[see e.g.][]{bbs16}. The estimated value of $F_{\rm rf}\as$ however corresponds to the isotropic luminosity of $4\times 10^{43}\,\mathrm{erg\,s^{-1}}$, which is a few orders of magnitude larger than the luminosity of bright stars. Therefore even close to the surface of the star the radiation density would not be enough to cause the cooling break. An alternative argument against such a scenario operating in the case of the flare discussed here, is the connection with the radio component at the distance of $50\,$pc from the base of the jet. The size of the jet at such a distance is orders of magnitude larger than the radius of stars, therefore only a small fraction of the blob filled with relativistic particles would be immersed in the strong radiation field from the star. 

Combining the ring-of-fire scenario with models for fast variability -- such as the Turbulent Extreme Multi-Zone (TEMZ) model of \cite{ALAN14} and/or the magnetic reconnection model of \cite{Giannios09} -- could explain the observations. 
Both models describe the radiation of small emission regions within a larger turbulent zone. 
In the TEMZ model, efficient acceleration of very high energy electrons that cause rapid VHE \g-ray flares results from temporary alignment of the turbulent magnetic field with a direction relative to the shock normal that is conducive to such acceleration \citep{Baring17}. In the magnetic reconnection model of \cite{Giannios09}, regions of plasma containing oppositely directed magnetic fields come into contact with each other, perhaps also as a result of turbulent motions. The magnetic fields reconnect, which creates ``mini-jets'' containing extremely energetic electrons that stream with high bulk Lorentz factors relative to the already highly relativistic ambient jet flow. Both models can produce sporadic VHE \g-ray flares with very short timescales of variability. 

In summary, if the association between the VHE \g-ray flare and the interaction of K16 with A1 is true, the most reasonable options seem to be turbulently-variable SSC or the ring-of-fire scenario. In the former case photons are provided by the flaring component itself, while in the latter case the standing feature A1 provides the target photons for the inverse-Compton scattering process.

%
%
\section{Summary and Conclusions} \label{sec:sumcon}
The dense monitoring of the FSRQ \source\ in the VHE \g-ray band with \hess\ \citep{zea19} and MAGIC \citep{acc18} led to the detection of a bright and short flare at this energy range lasting for two nights. The observed flux surpassed the previous flares by a factor ten. 

In the first night (JD~2457538), \hess\ observed a significant flux increase compared to previous nights, but no variability within the $1.5\,$hr of observation. In the second night (JD~2457539), an even brighter flux was detected with \hess\ including significant variability. This is the first time that VHE \g-ray intranight variability has been detected in \source. From a peak flux of about $80\%$ of the Crab Nebula flux above an energy of $200\,$GeV, the flux decayed throughout the rest of the night as shown through observations with MAGIC. Variability analyses have revealed that the VHE \g-ray lightcurve exhibited a common trend with a variability timescale of $1.5\,$hr. Close to the end of the observations, a deviation from this common trend occurred revealing a variability time scale of about $20\,$min. A detailed analysis indicates that this sudden cessation is significant with more than $4.5\sigma$.

Optical R-band observations with ATOM show a complex lightcurve on JD~2457539, which deviates from the behavior at VHE \g-rays. Unlike the VHE \g-ray lightcurve, the optical one exhibits a double peaked structure. The first peak occurs after the peak in the VHE \g-ray lightcurve, while the second optical peak has no correspondence in the VHE \g\ rays. The common variability time scale in the optical lightcurve is $\sim 15\,$hr. Interestingly, at the same time as the cessation in the VHE \g-ray lightcurve, a similar steepening of the decay also takes place in the optical domain. 
This may imply a common process being responsible for the faster drops of the lightcurves.

Despite the strong flare at VHE \g-ray energies, the HE \g-ray lightcurve ($E>100\,$MeV) was less affected by the outburst exhibiting only a mild flux variation. 
On the other hand, the HE \g-ray spectrum significantly hardened, reaching a value of the photon index of $1.6$, compared to the usual value of $\sim 2.4$. 
This implies a shift of the Compton peak for this flare towards the VHE range.
Hence, the flare mainly influenced the higher energies of the \fermi\ energy range.

The \g-ray spectrum also shows remarkable features. For the first time in \source, the observed VHE \g-ray spectrum is significantly curved. The curvature is fully explained through absorption by the EBL. By combining the VHE with the HE \g-ray spectrum, additional absorption features by the BLR were searched for. The spectra are compatible with negligible absorption by the BLR implying that the emission region is located at a distance from the black hole of at least $r>2.6 R_{\rm BLR}$. 
This is in line with results from other FSRQs \citep{pks1222,aMea14,cea18,Zea19,msb19,Hea19} indicating that the \g-ray emission region is located beyond the BLR.

VLBI observations at $43\,$GHz over the course of a few months around the flare revealed a fast moving knot, K16 ($\delta \sim 43$), that crossed the standing jet feature, A1, around the time of the flare. A1 is located at a de-projected distance of about $50\,$pc from the black hole. 
Hence, if the association of the VHE \g-ray flare with the crossing of a stationary emission feature $50\,$pc from the black hole by a fast, bright knot is true, the production of VHE \g\ rays can occur far down the jet where turbulent plasma crosses a standing shock \citep{bd98,NICK15}. 

The change in the variability time scale during the cessation of the VHE \g-ray and optical flare indicates that the injection and acceleration of particles had stopped. 
In turn, the particle spectrum was only influenced by cooling processes, possibly inverse-Compton cooling on the optical photons of the feature A1, producing the steep decline in the lightcurves \citep{sb10}. 
Hence, the emission region must have been compact \citep{Giannios09,ALAN14}, and probably has not filled the width of the jet. 
All of this is contrary to the standard expectation that emission regions producing such energetic flares should be located within $\sim 1\,$pc of the black hole and fill the width of the jet.

%
%
\begin{appendix}
\section{Linear variability time scale} \label{sec:som2var}
As an additional test to the exponential time scale of variability, the linear variability time scale between subsequent flux points \citep[following][]{zea99} has been employed:

\begin{eqnarray}
 t_{\rm lin} = \frac{F_{i}+F_{i+1}}{2} \frac{t_{i+1}-t_{i}}{|F_{i+1}-F_{i}|} \label{eq:tvar}.
\end{eqnarray}
The time scales defined by Eq.~(\ref{eq:tvar}) corroborate the results obtained with the exponential time scale, Eq.~(\ref{eq:texp}). 
The step with the fastest variability according to the exponential time scale ($t_{\rm exp}= (16\pm 5\stat)\,$min deviating $2.4\sigma$ from the harmonic mean) exhibits a linear time scale of $t_{\rm lin}= (17\pm 5\stat)$\,min deviating $3\sigma$ from the corresponding harmonic mean.

In the optical domain, Eq.~(\ref{eq:tvar}) results in the same values as the exponential time scale. In turn, the same conclusion can be drawn about the deviating time steps from the harmonic mean.

Note that the cadence of observations with H.E.S.S., MAGIC and ATOM is comparable within a factor of 2. Therefore the presented lightcurves in Fig.~\ref{fig:timescales} probe the same frequencies of variability at VHE \g-ray and optical ranges.

\end{appendix}
%

%
%
\begin{acknowledgement}
The authors wish to thank the anonymous referee for a constructive report, which helped to improve the manuscript. \\

The support of the Namibian authorities and of the University of Namibia in facilitating the construction and operation of H.E.S.S. is gratefully acknowledged, as is the support by the German Ministry for Education and Research (BMBF), the Max Planck Society, the German Research Foundation (DFG), the Helmholtz Association, the Alexander von Humboldt Foundation, the French Ministry of Higher Education, Research and Innovation, the Centre National de la Recherche Scientifique (CNRS/IN2P3 and CNRS/INSU), the Commissariat à l’énergie atomique et aux énergies alternatives (CEA), the U.K. Science and Technology Facilities Council (STFC), the Knut and Alice Wallenberg Foundation, the National Science Centre, Poland grant no. 2016/22/M/ST9/00382, the South African Department of Science and Technology and National Research Foundation, the University of Namibia, the National Commission on Research, Science \& Technology of Namibia (NCRST), the Austrian Federal Ministry of Education, Science and Research and the Austrian Science Fund (FWF), the Australian Research Council (ARC), the Japan Society for the Promotion of Science and by the University of Amsterdam. We appreciate the excellent work of the technical support staff in Berlin, Zeuthen, Heidelberg, Palaiseau, Paris, Saclay, Tübingen and in Namibia in the construction and operation of the equipment. This work benefited from services provided by the H.E.S.S. Virtual Organisation, supported by the national resource providers of the EGI Federation. \\

We would like to thank the Instituto de Astrof\'{\i}sica de Canarias for the excellent working conditions at the Observatorio del Roque de los Muchachos in La Palma. The financial support of the German BMBF and MPG; the Italian INFN and INAF; the Swiss National Fund SNF; the ERDF under the Spanish MINECO (FPA2017-87859-P, FPA2017-85668-P, FPA2017-82729-C6-2-R, FPA2017-82729-C6-6-R, FPA2017-82729-C6-5-R, AYA2015-71042-P, AYA2016-76012-C3-1-P, ESP2017-87055-C2-2-P, FPA2017-90566-REDC); the Indian Department of Atomic Energy; the Japanese ICRR, the University of Tokyo, JSPS, and MEXT;  the Bulgarian Ministry of Education and Science, National RI Roadmap Project DO1-153/28.08.2018 and the Academy of Finland grant nr. 320045 is gratefully acknowledged. This work was also supported by the Spanish Centro de Excelencia ``Severo Ochoa'' SEV-2016-0588 and SEV-2015-0548, the Unidad de Excelencia ``Mar\'{\i}a de Maeztu'' MDM-2014-0369 and the "la Caixa" Foundation (fellowship LCF/BQ/PI18/11630012), by the Croatian Science Foundation (HrZZ) Project IP-2016-06-9782 and the University of Rijeka Project 13.12.1.3.02, by the DFG Collaborative Research Centers SFB823/C4 and SFB876/C3, the Polish National Research Centre grant UMO-2016/22/M/ST9/00382 and by the Brazilian MCTIC, CNPq and FAPERJ. \\

The \textit{Fermi} LAT Collaboration acknowledges generous ongoing support
from a number of agencies and institutes that have supported both the
development and the operation of the LAT as well as scientific data analysis.
These include the National Aeronautics and Space Administration and the
Department of Energy in the United States, the Commissariat \`a l'Energie Atomique
and the Centre National de la Recherche Scientifique / Institut National de Physique
Nucl\'eaire et de Physique des Particules in France, the Agenzia Spaziale Italiana
and the Istituto Nazionale di Fisica Nucleare in Italy, the Ministry of Education,
Culture, Sports, Science and Technology (MEXT), High Energy Accelerator Research
Organization (KEK) and Japan Aerospace Exploration Agency (JAXA) in Japan, and
the K.~A.~Wallenberg Foundation, the Swedish Research Council and the
Swedish National Space Board in Sweden. 
Additional support for science analysis during the operations phase is gratefully
acknowledged from the Istituto Nazionale di Astrofisica in Italy and the Centre
National d'\'Etudes Spatiales in France. This work performed in part under DOE
Contract DE-AC02-76SF00515. \\

The VLBA is an instrument of the National Radio Astronomy Observatory. The National Radio Astronomy Observatory is a facility of the National Science Foundation operated by Associated Universities, Inc. The research at Boston University was supported by NASA Fermi Guest Investigator grant 80NSSC17K0649. 
This research has made use of data obtained with the Global Millimeter VLBI Array (GMVA), which consists of telescopes operated by the MPIfR, IRAM, Onsala, Metsahovi, Yebes, the Korean VLBI Network, the Greenland Telescope, the Green Bank Observatory and the VLBA. The GMVA VLBI data were correlated at the correlator of the MPIfR in Bonn, Germany.

\end{acknowledgement}

%
%

%

%
%
\end{document}